\documentclass[10pt]{article}
\pdfoutput=1

\usepackage[utf8]{inputenc}
\usepackage[T2A]{fontenc}

\input{epsf}
\usepackage{epsfig}
\usepackage{amssymb}
\usepackage{amsfonts}
\usepackage{amsbsy}
\usepackage[cmtip,all]{xy}
\usepackage{amsmath}
\usepackage{esint}
\usepackage{collectbox}
\usepackage{amscd}
\usepackage{mathrsfs}
\usepackage{amsmath,amsthm}
\usepackage{enumitem}
\usepackage{dsfont}
\usepackage{soul}
\usepackage{comment}
\usepackage{cite} 
\usepackage[most]{tcolorbox}
\usepackage{stmaryrd}
\usepackage{xcolor}
\definecolor{burgundy}{rgb}{0.5, 0.0, 0.13}
\definecolor{olive}{rgb}{0.50, 0.50, 0.0}
\usepackage[linktocpage=true,colorlinks=true,linkcolor=burgundy,citecolor=black!20!blue,urlcolor=violet]{hyperref}
\usepackage{accents}
\usepackage{xfrac}

\usepackage{tikz}
\usepackage{tikz-cd}
\usetikzlibrary{arrows}
\usetikzlibrary{arrows.meta}
\usetikzlibrary{positioning}
\usetikzlibrary{shapes}
\usetikzlibrary{fit}
\usetikzlibrary{decorations.pathmorphing,decorations.pathreplacing,decorations.markings}
\usetikzlibrary{calc}

\usepackage[font=footnotesize,labelfont=bf]{caption}

\theoremstyle{definition}

\newcommand{\sect}[1]{\setcounter{equation}{0}\section{#1}}

\DeclareMathAlphabet{\mathpzc}{OT1}{pzc}{m}{it}

\def\exp{{\rm exp}}
\def\Ext{{\rm Ext}}
\def\Hom{{\rm Hom}}
\def\I{{\rm i}}
\def\im{\mbox{Im }}
\renewcommand{\Im}{{\rm Im }}
\def\ker{{\rm ker}}
\def\log{{\rm log}}
\def\mod{{\rm mod}}

\def\re{\mbox{Re }}
\renewcommand{\Re}{{\rm Re }}
\def\sdet{{\rm sdet}}
\def\sgn{{\rm sgn\,}}
\def\STr{{\rm STr}}
\def\sym{{\rm Sym}}
\def\tr{{\rm tr\,}}
\def\Tr{{\rm Tr}}
\def\vol{{\rm vol\,}}

\def\la{\langle}
\def\ra{\rangle}
\def\half{\frac{1}{2}}
\def\p{\partial}
\def\prl{\parallel}

\def\rem{$\clubsuit$}

\def\pbar{\bar{\p}}

\def\rem{$\clubsuit$}
\def\bcom{{$\blacktriangleright$}}
\def\ecom{{$\blacktriangleleft$}}

\def\tM{\tilde{M}}
\def\tOmega{\tilde{\Omega}}
\def\tp{\tilde \phi}

\def\sx{{\bf x}}
\def\sy{{\bf y}}
\def\vx{{\vec{x}}}
\def\vy{{\vec{y}}}

\def\ba{\bar{a}}
\def\bb{\bar{b}}
\def\bc{\bar{c}}
\def\bd{\bar{d}}
\def\bz{\bar{z}}
\def\bZ{\bar{Z}}
\def\bW{\bar{W}}
\def\bD{\bar{D}}
\def\bA{\bar{A}}
\def\bB{\bar{B}}
\def\bR{\bar{R}}
\def\bS{\bar{S}}
\def\bT{\bar{T}}
\def\bU{\bar{U}}
\def\bPi{\bar{\Pi}}
\def\bOmega{\bar{\Omega}}
\def\bpartial{\bar{\partial}}
\def\bj{{\bar{j}}}
\def\bi{{\bar{i}}}
\def\bk{{\bar{k}}}
\def\bl{{\bar{l}}}
\def\bm{{\bar{m}}}
\def\btheta{\bar{\theta}}
\def\bpsi{\bar{\psi}}
\def\bF{\bar{F}}
\def\bs{\bar{s}}
\def\bt{\bar{t}}
\def\bv{\bar{v}}
\def\bx{\bar{x}}
\def\by{\bar{y}}
\def\bz{\bar{z}}
\def\btau{\bar{\tau}}
\def\bA{\bar{A}}
\def\bB{\bar{B}}
\def\bC{\bar{C}}
\def\bX{\bar{X}}
\def\bY{\bar{Y}}
\def\bZ{\bar{Z}}
\def\bnabla{{\bar \nabla}}

\def\bDelta{\boldsymbol{\Delta}}

\def\CA{{\cal A}}
\def\CB{{\cal B}}
\def\CC {{\cal C}}
\def\CD {{\cal D}}
\def\CE {{\cal E}}
\def\CF {{\cal F}}
\def\CG {{\cal G}}
\def\CH {{\cal H}}
\def\CI {{\cal I}}
\def\CJ {{\cal J}}
\def\CK {{\cal K}}
\def\CL {{\cal L}}
\def\CM {{\cal M}}
\def\CN {{\cal N}}
\def\CO {{\cal O}}
\def\CP {{\cal P}}
\def\CR {{\cal R}}
\def\CV {{\cal V}}
\def\CW {{\cal W}}
\def\CX {{\cal X}}
\def\CO {{\cal O}}
\def\CZ {{\cal Z}}
\def\CE {{\cal E}}
\def\CG {{\cal G}}
\def\CH {{\cal H}}
\def\CI {{{\cal I}}}
\def\CB {{\cal B}}
\def\CQ {{\cal Q}}
\def\CS {{\cal S}}
\def\CT {{\cal T}}
\def\CU {{\cal U}}
\def\CX {{\cal X}}
\def\CY {{\cal Y}}
\def\CZ{{\cal Z}}

\def\IA{\mathbb{A}}
\def\IB{\mathbb{B}}
\def\IC{\mathbb{C}}
\def\ID{\mathbb{D}}
\def\IE{\mathbb{E}}
\def\IF{\mathbb{Z}}
\def\IG{\mathbb{G}}
\def\IH{\mathbb{H}}
\def\II{\mathbb{I}}
\def\IJ{\mathbb{J}}
\def\IK{\mathbb{K}}
\def\IL{\mathbb{L}}
\def\IM{{\cal E}} 
\def\IN{\mathbb{N}}
\def\IO{\mathbb{O}}
\def\IP{\mathbb{P}}
\def\IQ{\mathbb{Q}} 
\def\IR{{\mathbb{R}}}
\def\IS{{\mathbb{S}}}
\def\IT{{\mathbb{T}}}
\def\IV{{\mathbb{V}}}
\def\IW{{\mathbb{W}}}
\def\IZ{{\mathbb{Z}}}

\def\fa{\mathfrak{a}}
\def\fb{\mathfrak{b}}
\def\fc{\mathfrak{c}}
\def\fd{\mathfrak{d}}
\def\fD{\mathfrak{D}}
\def\fe{\mathfrak{e}}
\def\ff{\mathfrak{f}}
\def\lieg{\mathfrak{g}}
\def\fg{\mathfrak{g}}
\def\lieh{\mathfrak{h}}
\def\fh{\mathfrak{h}}
\def\fj{\mathfrak{j}}
\def\fk{\mathfrak{k}}
\def\fl{\mathfrak{l}}
\def\fm{\mathfrak{m}}
\def\fn{\mathfrak{n}}
\def\fo{\mathfrak{o}}
\def\fp{\mathfrak{p}}
\def\fq{\mathfrak{q}}
\def\fr{\mathfrak{r}}
\def\fs{\mathfrak{s}}
\def\ft{\mathfrak{t}}
\def\liet{\mathfrak{t}}
\def\fp{\mathfrak{p}}
\def\fq{\mathfrak{q}}
\def\fr{\mathfrak{r}}
\def\fs{\mathfrak{s}}
\def\ft{\mathfrak{t}}
\def\fu{\mathfrak{u}}
\def\fv{\mathfrak{v}}
\def\fw{\mathfrak{w}}
\def\fx{\mathfrak{x}}
\def\fy{\mathfrak{y}}
\def\fz{\mathfrak{z}}
\def\fA{\mathfrak{A}}
\def\fB{\mathfrak{B}}
\def\fC{\mathfrak{C}}
\def\fE{\mathfrak{E}}
\def\fH{\mathfrak{H}}
\def\fG{\mathfrak{G}}
\def\fI{\mathfrak{I}}
\def\fJ{\mathfrak{J}}
\def\fK{\mathfrak{K}}
\def\fL{\mathfrak{L}}
\def\fM{\mathfrak{M}}
\def\fN{\mathfrak{N}}
\def\fR{\mathfrak{R}}
\def\fS{\mathfrak{S}}
\def\fT{\mathfrak{T}}
\def\fP{\mathfrak{P}}
\def\fV{\mathfrak{V}}
\def\fQ{\mathfrak{Q}}
\def\fX{\mathfrak{X}}
\def\fY{\mathfrak{Y}}

\def\bPhi{{\boldsymbol{\Phi}}}
\def\bSigma{{\boldsymbol{\Sigma}}}
\def\bPsi{{\boldsymbol{\Psi}}}
\def\bphi{{\boldsymbol{\phi}}}
\def\bpsi{{\boldsymbol{\psi}}}
\def\bvphi{{\boldsymbol{\varphi}}}

\def\lm{\limits}
\def\nn{\nonumber}
\def\nb{\nabla}

\hoffset= - 1.0in

\numberwithin{equation}{section}

\tcbset{boxrule=0.6pt,colback=white!97!green}

\DeclareSymbolFont{bbsymbol}{U}{bbold}{m}{n}
\DeclareMathSymbol{\bbzero}{\mathbin}{bbsymbol}{"30}
\DeclareMathSymbol{\bbone}{\mathbin}{bbsymbol}{"31}
\DeclareMathSymbol{\bbtwo}{\mathbin}{bbsymbol}{"32}
\DeclareMathSymbol{\bbthree}{\mathbin}{bbsymbol}{"33}
\DeclareMathSymbol{\bbfour}{\mathbin}{bbsymbol}{"34}
\DeclareMathSymbol{\bbfive}{\mathbin}{bbsymbol}{"35}
\DeclareMathSymbol{\bbsix}{\mathbin}{bbsymbol}{"36}
\DeclareMathSymbol{\bbseven}{\mathbin}{bbsymbol}{"37}
\DeclareMathSymbol{\bbeight}{\mathbin}{bbsymbol}{"38}
\DeclareMathSymbol{\bbnine}{\mathbin}{bbsymbol}{"39}

\newcommand\Kappa{\mathrm{K}}
\def\myY{\mathsf{Y}}
\def\myblue{white!40!blue}
\def\mygray{white!20!gray}
\newcommand\sqbox[1]{{
	\setbox0=\hbox{\mbox{$\Box$}}
	\setbox1=\hbox{\mbox{\raisebox{0.35ex}{\tiny #1}}}
	\mbox{\raisebox{-0.2ex}{\rlap{\hbox to \wd0{\hss{\box1}\hss}}\box0}}
}}

\begin{document}

\hfill MIPT/TH-12/25

\hfill ITEP/TH-15/25

\hfill IITP/TH-13/25

\vskip 1.5in
\begin{center}
	
    {\bf\Large On geometric bases for A-polynomials II: $\mathfrak{su}_3$ and Kuperberg bracket}
	
	\vskip 0.2in
	\renewcommand{\thefootnote}{\fnsymbol{footnote}}
	{Dmitry Galakhov
		\footnote[2]{e-mail: d.galakhov.pion@gmail.com, galakhov@itep.ru} and  Alexei Morozov
		\footnote[3]{e-mail: morozov@itep.ru}}\\
	
	\vskip 0.2in
	\renewcommand{\thefootnote}{\roman{footnote}}
	{\small{
			\textit{
				MIPT, 141701, Dolgoprudny, Russia}
			\vskip 0 cm
			\textit{
				NRC ``Kurchatov Institute'', 123182, Moscow, Russia}
			\vskip 0 cm
			\textit{
				IITP RAS, 127051, Moscow, Russia}
			\vskip 0 cm
			\textit{
				ITEP, Moscow, Russia}
	}}
\end{center}

\vskip 0.2in
\baselineskip 16pt

\centerline{ABSTRACT}

\bigskip

{\footnotesize
    We continue the study of quantum A-polynomials -- equations for knot polynomials with respect to their coloring
    (representation-dependence) --  as the relations between different links, obtained by hanging additional ``simple''
    components on the original knot.
    Depending on the choice of this ``decoration'', the knot polynomial is either multiplied by a number
    or decomposes into a sum over ``surrounding'' representations by a cabling procedure.
    What happens is that these two of decorations, when complicated enough, become dependent --
    and this provides an equation.
    Remarkably it can be made independent of the representation.
    However, the equivalence of links is not a topological property --
    it follows from the properties of $R$-matrices, and strongly depends on the choice
    the gauge group and particular links.
    The relatively well studied part of the story concerns $\mathfrak{su}_2$, where $R$-matrices can be chosen in an
    especially convenient Kauffman form, what makes the derivation of equations rather geometrical.
    To make these geometric methods somewhat simpler we suggest to use an arcade formalism/representation of the braid group to simplify decorating links universally.
    Here we attempt to extend this technique to the next case, $\mathfrak{su}_3$, where the Kauffman rule is substituted by a more involved Kuperberg rule, still remains more geometric
    than generic analysis of MOY-diagrams, needed for higher ranks.
    Already in this case we encounter a classification problem for possible ``decorations'' and
    emergence of two-lined Young diagrams in enumeration of representations.

}

\bigskip

\bigskip

\tableofcontents

\bigskip

\section{Introduction}

``Quantum A-polynomials'' \cite{garoufalidis2004characteristic,Gukov:2003na,Aganagic:2012jb,Garoufalidis:2016zhf} is a nickname for somewhat mysterious finite-difference
equations with respect to the coloring parameter $R$, which are satisfied by HOMFLY-PT knot polynomials $H_K(R)$ -- averages of Wilson loop operators in 3d Chern-Simons theory.
In the classical limit $q\to 1$ they convert into ordinary A-polynomials,
which have a pure topological interpretation as complex curves associated to a character variety of the knot complement \cite{cooper1994plane, yoshida1991ideal, neumann1985volumes, Gukov:2003na, 2019arXiv190301732D, Kashaev:1996kc, murakami2001colored, Gukov:2006ze, A=B, Arthamonov:2013rfa}.
They have close relatives, named quantum C-polynomials \cite{GarC,2009.11641}.
Here we continue the program of \cite{Galakhov:2024eco} in constructing derivation mechanisms for (quantum) A-polynomials by manipulations with auxiliary links obtained by decorating the original knot $K$ with additional ``simply colored'' strands.
The idea is that different decorations are not independent -- and this is ``the same''
dependence as that between different colorings.
This follows from the fact that the change of colorings can be provided by {\it cabling}
\cite{adams2004knot, cabling} the original knot with specific links -- and in this way {\it some} map between colorings and links can be established.
The problem is to tame the dependencies on the decorations and to find its true origins.
It is not a pure topological equivalence -- at some stage the true Reshetikhin-Turaev-style (RT)
calculation \cite{RT1,RT2,1112.2654} of {\it some} link transformations looks unavoidable.
The task is to minimize this need and, hopefully, to localize it in some clever way.
We are approaching this goal in small steps, looking through particular examples
and trying to find something in common.

In this particular paper we suggest to formalize the work with links by the use of planar {\it arcades}
on 2-dimensional knot diagrams, which provide a sophisticated, still handable representation of
the braid group -- in this way shifting the problem of link dependencies to ``just'' the closures of braids.
The arcade formalism would allow us to ``planarize'' arbitrary links in a universal way.
The planarized link is still captured by a link diagram yet drawn not in a presence of the knot question, rather on a plain with punctures.
We believe these diagrammatic language is somewhat simpler and allows one to simplify further, untangle links by applying \emph{brackets}.

On the other hand, we try to extend the formalism from the case of the gauge algebra $\fs\fu_2$, strongly simplified
by the use of the Kauffman bracket \cite{kauffman2001knots} (special choice of $R$-matrices, very good for planarity studies),
to other $\fs\fu_{n\geq 3}$.
In this case links are actually substituted by MOY diagrams \cite{MOY}, which seems to be a severe complication.
However, this is not necessarily the case, because for the RT formalism this is not a very big difference,
as we demonstrate with the help of Kuperberg's generalization \cite{kuperberg1996spiders} of the Kauffman bracket for $\fs\fu_3$.

This note is organized as follows.
In Sec.~\ref{sec:link_bas} a link planarization method and its application to a derivation of A-polynomials is discussed.
In Sec.~\ref{sec:arcade} a braid group action on sets of arcades is introduced.
This representation is used for algorithmic calculation of arbitrary link planarization.
In Sec.~\ref{sec:RT} a brief reminder on elements of the Reshetikhin-Turaev formalism and properties, brackets of quantum algebras $U_q(\fs\fu_n)$ is presented.
Sec.~\ref{sec:Kauff_ex} is devoted to tests of the proposed arcade formalism by calculations of known cases of A-polynomials for some simple knots.
Eventually, in Sec.~\ref{sec:su(3)} we discuss A-polynomials for the gauge algebra $\fs\fu_3$ and their derivation from link planarization together with the Kuperberg bracket for an explicit example of the simplest trefoil knot.

\section{Towards link bases}\label{sec:link_bas}

\subsection{A-polynomials as relations among links}

HOMFLY-PT polynomials \cite{HOMFLY,PT,Mironov:2011ym} for knot $K$ depend on the choice of the gauge group $G$ and its irreducible representation $R$.
Here we denote them as $\Psi_K(R)$.
$\Psi_K(R)$ is believed to represent a Chern-Simons average with gauge group $G$ of Wilson line $\Tr_R{\rm Pexp}\oint_K A$.
The Wilson line imposes singular boundary conditions for fields as one approaches the knot strand, so it is natural to treat the respective average as a Chern-Simons partition function on $S^3\setminus K$ where we cut from from $S^3$ the knot singularity together with a small knot tubular neighborhood.
The path integral on a manifold $M$ with a boundary $\p M$ and prescribed boundary conditions is a wave function -- an element of the Hilbert space $\mathscr{H}(\p M)$.

Probably, the most canonical representation of the knot complement as a manifold $M$ with boundary $\p M$ is performed for the figure-eight knot, depicted in Fig.~\ref{fig:inversion}.
The complement of the knot tubular neighborhood (a) is cut in two tetrahedra with carved out vertices (b) in such a way that two unidentified tetrahedron edges depicted by green and gray colors are oriented as it is depicted in figure (a).
The tetrahedra are used to construct Thurston's hyperbolic metric on the complement of hyperbolic knot $4_1$ in a canonical way \cite{thurston2022geometry, kim2018octahedral}. 
The surface of the knot tubular neighborhood is covered by blue and red triangles as it is depicted in a zoomed in picture of knot neighborhood in (c).
Eventually, these triangles form a closed 2d surface (d) that is a topological torus (e).

\begin{figure}[ht!]
	\centering
	\begin{tikzpicture}
		\node(A) {$\begin{array}{c}
				\begin{tikzpicture}[scale=0.8]
					\draw[white,line width = 3mm] (-0.5,0) to[out=90,in=270] (1,1.5) to[out=90,in=0] (0,2);
					\draw[line width = 2.5mm] (-0.5,0) to[out=90,in=270] (1,1.5) to[out=90,in=0] (0,2);
					\draw[white, line width = 2.2mm] (-0.5,0) to[out=90,in=270] (1,1.5) to[out=90,in=0] (0,2);
					\node[right] at (1,1.5) {$\scriptstyle 12',10''$};
					\draw[white,line width = 3mm] (0,1) to[out=180,in=90] (-1.5,0) to[out=270,in=180] (-0.5,-0.7) to[out=0,in=270] (0.5,0);
					\draw[line width = 2.5mm] (0,1.2) to[out=180,in=90] (-1.5,0) to[out=270,in=180] (-0.5,-0.7) to[out=0,in=270] (0.5,0);
					\draw[white, line width = 2.2mm] (0,1.2) to[out=180,in=90] (-1.5,0) to[out=270,in=180] (-0.5,-0.7) to[out=0,in=270] (0.5,0);
					\node[left] at (-1.5,0) {$\scriptstyle 2',1''$};
					\draw[white,line width = 3mm] (0,2) to[out=180,in=90] (-1,1.5) to[out=270,in=90] (0.5,0);
					\draw[line width = 2.5mm] (0,2) to[out=180,in=90] (-1,1.5) to[out=270,in=90] (0.5,0);
					\draw[white, line width = 2.2mm] (0,2) to[out=180,in=90] (-1,1.5) to[out=270,in=90] (0.5,0);
					\node[left] at (-1,1.5) {$\scriptstyle 8',9''$};
					\begin{scope}[xscale=-1]
						\draw[white,line width = 3mm] (0,1.2) to[out=180,in=90] (-1.5,0) to[out=270,in=180] (-0.5,-0.7) to[out=0,in=270] (0.5,0);
						\draw[line width = 2.5mm] (0,1.2) to[out=180,in=90] (-1.5,0) to[out=270,in=180] (-0.5,-0.7) to[out=0,in=270] (0.5,0);
						\draw[white, line width = 2.2mm] (0,1.2) to[out=180,in=90] (-1.5,0) to[out=270,in=180] (-0.5,-0.7) to[out=0,in=270] (0.5,0);
						\node[right] at (-1.5,0) {$\scriptstyle 6',4''$};
					\end{scope}
					\draw[thick, gray, postaction={decorate},decoration={markings, mark= at position 0.7 with {\arrow{stealth}}}] (-0.5,0) -- (0.5,0);
					\draw[thick, black!40!green, postaction={decorate},decoration={markings, mark= at position 0.7 with {\arrow{stealth}}}] (0,1.2) -- (0,2);
					\node[left] at (-0.5,0) {$\scriptstyle 11$};
					\node[right] at (0.5,0) {$\scriptstyle 3$};
					\node[above right] at (0,1.2) {$\scriptstyle 5$};
					\node[above] at (0,2) {$\scriptstyle 7$};
					\begin{scope}[shift={(0,2)}]
						\begin{scope}[rotate=90]
							\begin{scope}[yscale=0.5]
								\draw[fill=white] ([shift={(0:0.16)}]0,0) arc (0:180:0.16);
								\draw[fill=white, dashed] ([shift={(180:0.16)}]0,0) arc (180:360:0.16);
							\end{scope}
						\end{scope}
					\end{scope}
					\begin{scope}[shift={(0,1.2)}]
						\begin{scope}[rotate=90]
							\begin{scope}[yscale=0.5]
								\draw[fill=white] ([shift={(0:0.16)}]0,0) arc (0:180:0.16);
								\draw[fill=white, dashed] ([shift={(180:0.16)}]0,0) arc (180:360:0.16);
							\end{scope}
						\end{scope}
					\end{scope}
					\begin{scope}[shift={(-0.5,0)}]
						\begin{scope}[rotate=180]
							\begin{scope}[yscale=0.5]
								\draw[fill=white] ([shift={(0:0.16)}]0,0) arc (0:180:0.16);
								\draw[fill=white, dashed] ([shift={(180:0.16)}]0,0) arc (180:360:0.16);
							\end{scope}
						\end{scope}
					\end{scope}
					\begin{scope}[shift={(0.5,0)}]
						\begin{scope}[rotate=180]
							\begin{scope}[yscale=0.5]
								\draw[fill=white] ([shift={(0:0.16)}]0,0) arc (0:180:0.16);
								\draw[fill=white, dashed] ([shift={(180:0.16)}]0,0) arc (180:360:0.16);
							\end{scope}
						\end{scope}
					\end{scope}
					\begin{scope}[shift={(-1.5,0)}]
						\begin{scope}[rotate=180]
							\begin{scope}[yscale=0.5]
								\draw[fill=white] ([shift={(0:0.16)}]0,0) arc (0:180:0.16);
								\draw[fill=white, dashed] ([shift={(180:0.16)}]0,0) arc (180:360:0.16);
							\end{scope}
						\end{scope}
					\end{scope}
					\begin{scope}[shift={(1.5,0)}]
						\begin{scope}[rotate=180]
							\begin{scope}[yscale=0.5]
								\draw[fill=white] ([shift={(0:0.16)}]0,0) arc (0:180:0.16);
								\draw[fill=white, dashed] ([shift={(180:0.16)}]0,0) arc (180:360:0.16);
							\end{scope}
						\end{scope}
					\end{scope}
					\draw[-stealth] (0.3,2) -- (-0.3,2);
				\end{tikzpicture}
			\end{array}$};
		\node at (3.5,-1.5) {$\begin{array}{c}
				\begin{tikzpicture}[scale=0.8]
					\draw[fill=white!90!red,white!90!red] (0,0.2) -- (2,0.2) to[out=180,in=120] (1,-0.2) -- (0,-0.2) to[out=180,in=180] (0,0.2);
					\draw[fill=white!90!blue,white!90!blue] (1,-0.2) to[out=120,in=180] (2,0.2) to[out=0,in=0] (2,-0.2) -- (1,-0.2);
					\draw (0,0.2) -- (2,0.2) (0,-0.2) -- (2,-0.2) (2,0.2) to[out=0,in=0] (2,-0.2) to[out=180,in=180] (2,0.2) (0,0.2) to[out=180,in=180] (0,-0.2) (1,-0.2) to[out=120,in=180] (2,0.2);
					\draw[dashed] (0,0.2) to[out=0,in=0] (0,-0.2) (0,0.2) to[out=0,in=60] (1,-0.2);
					\node[above] at (1,0.2) {$\scriptstyle a'$};
					\node[below] at (1,-0.2) {$\scriptstyle d''$};
					\node[left] at (-0.1,0) {$\scriptstyle b$};
					\node[right] at (2.1,0) {$\scriptstyle c$};
					\draw[-stealth] (0,0.6) -- (2,0.6);
					\node[above] at (1,0.6) {\tiny orientation};
				\end{tikzpicture}
			\end{array}$};
		\node at (4,0.5) {$\begin{array}{c}
				\begin{tikzpicture}[rotate=20,scale=0.8]
					\foreach \p in {0, 1, 2, 3} 
					{\begin{scope}[rotate = 90 * \p]
							\draw[fill, white!90!green] (0,0) -- (0.8,0) -- (0.8,0.2) -- (0.2,0.8) -- (0,0.8) -- cycle;
					\end{scope}}
					\draw[fill=white!90!red] (0,-0.6) to node[left,pos=0.5]{\tiny 1} (-0.2,-0.8) to node[below,pos=0.5]{\tiny 3} (0.2,-0.8) to node[above,pos=0.5] {\tiny 2} (0,-0.6);
					\begin{scope}[rotate=90]
						\draw[fill=white!90!red] (0,-0.6) to node[below,pos=0.5]{\tiny 12} (-0.2,-0.8) to node[ right,pos=0.5]{\tiny 11} (0.2,-0.8) to node[left,pos=0.5] {\tiny 10} (0,-0.6);
					\end{scope}
					\begin{scope}[rotate=180]
						\draw[fill=white!90!red] (0,-0.6) to node[right,pos=0.5]{\tiny 8} (-0.2,-0.8) to node[above,pos=0.5]{\tiny 9} (0.2,-0.8) to node[below,pos=0.5] {\tiny 7} (0,-0.6);
					\end{scope}
					\begin{scope}[rotate=270]
						\draw[fill=white!90!red] (0,-0.6) to node[above,pos=0.5]{\tiny 5} (-0.2,-0.8) to node[left,pos=0.5]{\tiny 4} (0.2,-0.8) to node[right,pos=0.5] {\tiny 6} (0,-0.6);
					\end{scope}
					\draw[black!40!green, postaction={decorate},decoration={markings, mark= at position 0.7 with {\arrow{stealth}}}] (-1,0) -- (0,1);
					\draw[black!40!green, postaction={decorate},decoration={markings, mark= at position 0.7 with {\arrow{stealth}}}] (0,-1) -- (0,1);
					\draw[black!40!green, postaction={decorate},decoration={markings, mark= at position 0.7 with {\arrow{stealth}}}] (-1,0) -- (-0.1,0) (0.1,0) -- (1,0);
					\draw[gray, postaction={decorate},decoration={markings, mark= at position 0.7 with {\arrow{stealth}}}] (-1,0) -- (0,-1);
					\draw[gray, postaction={decorate},decoration={markings, mark= at position 0.7 with {\arrow{stealth}}}] (1,0) -- (0,-1);
					\draw[gray, postaction={decorate},decoration={markings, mark= at position 0.7 with {\arrow{stealth}}}] (1,0) -- (0,1);
				\end{tikzpicture}
			\end{array}$};
		\node at (6,0.5) {$\begin{array}{c}
				\begin{tikzpicture}[rotate=20,scale=0.8]
					\foreach \p in {0, 1, 2, 3} 
					{\begin{scope}[rotate = 90 * \p]
							\draw[fill, white!90!green] (0,0) -- (0.8,0) -- (0.8,0.2) -- (0.2,0.8) -- (0,0.8) -- cycle;
					\end{scope}}
					\draw[fill=white!90!blue] (0,-0.6) to node[left,pos=0.5]{\tiny 4} (-0.2,-0.8) to node[below,pos=0.5]{\tiny 11} (0.2,-0.8) to node[above,pos=0.5] {\tiny 6} (0,-0.6);
					\begin{scope}[rotate=90]
						\draw[fill=white!90!blue] (0,-0.6) to node[below,pos=0.5]{\tiny 8} (-0.2,-0.8) to node[ right,pos=0.5]{\tiny 3} (0.2,-0.8) to node[left,pos=0.5] {\tiny 9} (0,-0.6);
					\end{scope}
					\begin{scope}[rotate=180]
						\draw[fill=white!90!blue] (0,-0.6) to node[right,pos=0.5]{\tiny 12} (-0.2,-0.8) to node[above,pos=0.5]{\tiny 10} (0.2,-0.8) to node[below,pos=0.5] {\tiny 7} (0,-0.6);
					\end{scope}
					\begin{scope}[rotate=270]
						\draw[fill=white!90!blue] (0,-0.6) to node[above,pos=0.5]{\tiny 5} (-0.2,-0.8) to node[left,pos=0.5]{\tiny 1} (0.2,-0.8) to node[right,pos=0.5] {\tiny 2} (0,-0.6);
					\end{scope}
					\draw[black!40!green, postaction={decorate},decoration={markings, mark= at position 0.7 with {\arrow{stealth}}}] (-1,0) -- (0,1);
					\draw[black!40!green, postaction={decorate},decoration={markings, mark= at position 0.7 with {\arrow{stealth}}}] (0,-1) -- (0,1);
					\draw[black!40!green, postaction={decorate},decoration={markings, mark= at position 0.7 with {\arrow{stealth}}}] (-1,0) -- (-0.1,0) (0.1,0) -- (1,0);
					\draw[gray, postaction={decorate},decoration={markings, mark= at position 0.7 with {\arrowreversed{stealth}}}] (-1,0) -- (0,-1);
					\draw[gray, postaction={decorate},decoration={markings, mark= at position 0.7 with {\arrowreversed{stealth}}}] (1,0) -- (0,-1);
					\draw[gray, postaction={decorate},decoration={markings, mark= at position 0.7 with {\arrowreversed{stealth}}}] (1,0) -- (0,1);	   
				\end{tikzpicture}
			\end{array}$};
		\node at (10,0.5) {\scalebox{0.7}{$\begin{array}{c}
					\begin{tikzpicture}
						\draw[fill=white!90!blue] (3,0) to (4,1) to node[pos=0.2,right]{\scriptsize 3} (4,0) to node[pos=0.5,below]{\scriptsize 8} (3,0);
						\draw[fill=white!90!red] (3,1) to node[pos=0.5,above]{\scriptsize 8} (4,1) to node[pos=0.8,right]{\scriptsize 9} (3,0) to (3,1);
						\begin{scope}[shift={(-1,0)}]
							\draw[fill=white!90!blue] (3,0) to (4,1) to node[pos=0.2,right]{\scriptsize 7} (4,0) to node[pos=0.5,below]{\scriptsize 12} (3,0);
							\draw[fill=white!90!red] (3,1) to node[pos=0.5,above]{\scriptsize 12} (4,1) to node[pos=0.8,right]{\scriptsize 10} (3,0) to (3,1);
						\end{scope}
						\begin{scope}[shift={(-2,0)}]
							\draw[fill=white!90!blue] (3,0) to (4,1) to node[pos=0.2,right]{\scriptsize 11} (4,0) to node[pos=0.5,below]{\scriptsize 6} (3,0);
							\draw[fill=white!90!red] (3,1) to node[pos=0.5,above]{\scriptsize 6} (4,1) to node[pos=0.8,right]{\scriptsize 4} (3,0) to (3,1);
						\end{scope}
						\begin{scope}[shift={(-3,0)}]
							\draw[fill=white!90!blue] (3,0) to (4,1) to node[pos=0.2,right]{\scriptsize 5} (4,0) to node[pos=0.5,below]{\scriptsize 2} (3,0);
							\draw[fill=white!90!red] (3,1) to node[pos=0.5,above]{\scriptsize 2} (4,1) to node[pos=0.8,right]{\scriptsize 1} (3,0) to node[pos=0.8,right]{\scriptsize 3} (3,1);
						\end{scope}
						\draw[thick,-stealth,gray,decorate,decoration={snake,amplitude=1pt, segment length=10pt}] (0,2) -- (0,1);
						\draw[thick,-stealth,gray,decorate,decoration={snake,amplitude=1pt, segment length=10pt}] (4,2) -- (4,1);
						\draw[thick,-stealth,gray,decorate,decoration={snake,amplitude=1pt, segment length=10pt}] (2,1) -- (2,2);
						\draw[thick, black!40!green, postaction={decorate},decoration={markings, mark= at position 0.7 with {\arrow{stealth}}}] (1,1) -- (1,2);
						\draw[thick, black!40!green, postaction={decorate},decoration={markings, mark= at position 0.7 with {\arrow{stealth}}}] (3,2) -- (3,1);
					\end{tikzpicture}
				\end{array}$}};
		\node(B) at (9,-1.5) {$\begin{array}{c}
				\begin{tikzpicture}[scale=0.4]
					\draw[thick,fill=white!90!green] (-2.9,0.) to[out=-89.8399,in=100.017] (-2.87723,-0.26343) to[out=-79.9828,in=118.828] (-2.70107,-0.76265) to[out=-61.1725,in=134.258] (-2.40139,-1.16537) to[out=-45.742,in=146.399] (-2.02811,-1.47359) to[out=-33.6013,in=156.101] (-1.60625,-1.70413) to[out=-23.8987,in=164.195] (-1.14839,-1.86896) to[out=-15.8051,in=171.311] (-0.66453,-1.97387) to[out=-8.68872,in=184.42] (0.33791,-2.01131) to[out=4.41993,in=198.572] (1.30922,-1.81925) to[out=18.5724,in=207.161] (1.75584,-1.63275) to[out=27.1611,in=217.636] (2.16277,-1.37722) to[out=37.6355,in=230.875] (2.51464,-1.03837) to[out=50.8749,in=247.59] (2.7788,-0.60076) to[out=67.5902,in=269.84] (2.9,0.)
					(2.9,0.) to[out=90.1601,in=292.698] (2.7788,0.60076) to[out=112.698,in=309.358] (2.51464,1.03837) to[out=129.358,in=322.547] (2.16277,1.37722) to[out=142.547,in=332.985] (1.75584,1.63275) to[out=152.985,in=341.551] (1.30922,1.81925) to[out=161.551,in=355.683] (0.33791,2.01131) to[out=175.683,in=8.79496] (-0.66453,1.97387) to[out=-171.205,in=15.9226] (-1.14839,1.86896) to[out=-164.077,in=24.036] (-1.60625,1.70413) to[out=-155.964,in=33.7698] (-2.02811,1.47359) to[out=-146.23,in=45.956] (-2.40139,1.16537) to[out=-134.044,in=61.4422] (-2.70107,0.76265) to[out=-118.558,in=80.2966] (-2.87723,0.26343) to[out=-99.7034,in=89.8399] (-2.9,0.);
					\draw[thick,fill=white]
					(-1.11956,0.09908) to[out=-49.9491,in=153.423] (-0.92834,-0.03418) to[out=-26.5774,in=161.919] (-0.72701,-0.11594) to[out=-18.081,in=169.27] (-0.47349,-0.18059) to[out=-10.7297,in=175.99] (-0.18692,-0.21756) to[out=-4.01049,in=184.113] (0.18692,-0.21756) to[out=4.11276,in=190.839] (0.47349,-0.18059) to[out=10.8386,in=198.203] (0.72701,-0.11594) to[out=18.2034,in=206.723] (0.92834,-0.03418) to[out=26.7226,in=229.949] (1.11956,0.09908)
					(0.97323,0.01035) to[out=150.737,in=339.671] (0.78968,0.09405) to[out=159.671,in=347.286] (0.54899,0.16486) to[out=167.286,in=354.139] (0.2697,0.21034) to[out=174.139,in=357.416] (0.12181,0.22137) to[out=177.416,in=1.0626] (-0.04693,0.22375) to[out=-178.937,in=5.96464] (-0.2697,0.21034) to[out=-174.035,in=12.8258] (-0.54899,0.16486) to[out=-167.174,in=20.4567] (-0.78968,0.09405) to[out=-159.543,in=29.2627] (-0.97323,0.01035);
					\draw[thick]
					(-2.,0.74437) to[out=-110.513,in=111.519] (-1.99235,0.25547) to[out=-68.4811,in=143.432] (-1.59517,-0.22917) to[out=-36.5678,in=164.608] (-0.84954,-0.57841) to[out=-15.3925,in=173.316] (-0.39091,-0.66766) to[out=-6.68404,in=182.646] (0.15119,-0.68718) to[out=2.64617,in=195.535] (0.84954,-0.57841) to[out=15.5348,in=216.766] (1.59517,-0.22917) to[out=36.7659,in=248.787] (1.99235,0.25547) to[out=68.7869,in=290.513] (2.,0.74437)
					(2.,0.74437) to[out=110.827,in=311.265] (1.85392,0.98337) to[out=131.265,in=335.001] (1.41484,1.29737) to[out=155.001,in=348.618] (0.84207,1.47882) to[out=168.618,in=357.467] (0.21479,1.55296) to[out=177.467,in=5.21746] (-0.42348,1.53888) to[out=-174.783,in=15.2103] (-1.03897,1.4325) to[out=-164.79,in=31.5016] (-1.57881,1.20973) to[out=-148.498,in=43.8516] (-1.79277,1.04724) to[out=-136.148,in=69.173] (-2.,0.74437);
					\node[below left] at (-0.84954,-0.57841) {$\scriptstyle B$};
					\draw[thick]
					(1.27781,-1.82961) to[out=18.1969,in=232.273] (1.36305,-1.76682) to[out=52.2727,in=255.341] (1.42518,-1.63128) to[out=75.3413,in=270.783] (1.4499,-1.39568) to[out=90.783,in=277.359] (1.4373,-1.22373) to[out=97.3594,in=285.102] (1.38531,-0.96416) to[out=105.102,in=295.592] (1.25201,-0.60221) to[out=115.592,in=310.13] (1.05809,-0.29197) to[out=130.13,in=326.502] (0.91736,-0.16139) to[out=146.502,in=17.7616] (0.72219,-0.11751);
					\draw[dashed, thick]
					(0.72219,-0.11751) to[out=-161.803,in=69.0245] (0.59176,-0.26297) to[out=-110.975,in=86.938] (0.55164,-0.47351) to[out=-93.062,in=100.849] (0.58046,-0.83433) to[out=-79.151,in=109.283] (0.66132,-1.13397) to[out=-70.717,in=115.835] (0.75151,-1.35224) to[out=-64.1654,in=122.344] (0.84463,-1.5209) to[out=-57.6557,in=135.66] (0.99829,-1.71626) to[out=-44.3405,in=148.487] (1.09512,-1.79371) to[out=-31.5133,in=197.762] (1.27781,-1.82961);
					\node[above right] at (1.4499,-1.39568) {$\scriptstyle A$};
				\end{tikzpicture}
			\end{array}$};
		\draw[stealth-stealth] (A.south) to[out=330,in=210] node[below,pos=0.5] {Inversion} (B.west);
		\node at (-2,0) {(a)};
		\node at (7,1.2) {(b)};
		\node at (5,-1.2) {(c)};
		\node at (12,1) {(d)};
		\node at (11,-1.5) {(e)};
	\end{tikzpicture}
	\caption{Inverting complements of knot tubular neighborhoods}\label{fig:inversion}
\end{figure}

In this note we will not be interested in geometry of the knot complement, so we apply the inversion transform of Fig.~\ref{fig:inversion} to arbitrary knot $K$ and treat $M=S^3/K$ as a manifold with as arbitrarily complicated local geometry as possible, yet $\p M$ is a topological torus $T^2$.
It is well-known that the Hilbert space of the Chern-Simons theory on torus $T^2$ is spanned by Weyl-Kac characters \cite{Moore:1988qv}, so the representation label $R$ we have assigned to our wave functions -- HOMFLY-PT polynomials -- plays the role of a ``coordinate'' on the gauge connection moduli space on a torus.
Surely, one should take certain precautions with this statement as for HOMFLY-PT polynomials the coupling constant parameter $q$ might be thought as a generic complex parameter that would lead to a complex in general Chern-Simons level and complexification of the gauge group.
And the theory with the complex level and non-compact gauge group has its own peculiarities in quantization \cite{witten1991quantization, Galakhov:2014aha, Galakhov:2015gza, Dimofte:2009yn, Dimofte:2016pua, Freed:2021anp}.

Having chosen on the torus boundary A- and B-cycles as it is depicted in Fig.~\ref{fig:inversion} we could construct our first operators acting on $\Psi_K(R)$.
These operators are called $\mu_Q$ (meridian) and $\lambda_Q$ (longitude) and correspond to Wilson lines colored with irrep $Q$ winding along A- and B-cycles respectively.
Let us assume that these operators sink inside the torus body in Fig.~\ref{fig:inversion} and consider further inversion.
In the inverted form these operators represent two types of links of a new strand of irrep $Q$ with the initial strand of rep $R$ (see Fig.~\ref{fig:sinking}): a \emph{Hopf link} for $\mu$ and a \emph{satellite link} for $\lambda$.

Operator monomials also correspond to various links.
To obtain a link corresponding to an operator words one should sink elementary links corresponding to operators in the order opposite to as they appear in the word.

\begin{figure}[ht!]
	\centering
	\begin{tikzpicture}
		\node(A) at (0,0) {$\begin{array}{c}
				\begin{tikzpicture}[scale=0.7]
					\draw[white,line width = 1.5mm] (-0.5,0) to[out=90,in=270] (1,1.5) to[out=90,in=0] (0,2);
					\draw[burgundy, line width =  1mm] (-0.5,0) to[out=90,in=270] (1,1.5) to[out=90,in=0] (0,2);
					\begin{scope}[yscale=0.5]
						\draw[thick] ([shift={(0:0.3)}]-1.5,0) arc (0:180:0.3);
					\end{scope}
					\draw[white,line width = 1.5mm] (0,1) to[out=180,in=90] (-1.5,0) to[out=270,in=180] (-0.5,-0.7) to[out=0,in=270] (0.5,0);
					\draw[burgundy, line width =  1mm] (0,1.2) to[out=180,in=90] (-1.5,0) to[out=270,in=180] (-0.5,-0.7) to[out=0,in=270] (0.5,0);
					\draw[white,line width = 1.5mm] (0,2) to[out=180,in=90] (-1,1.5) to[out=270,in=90] (0.5,0);
					\draw[burgundy, line width =  1mm] (0,2) to[out=180,in=90] (-1,1.5) to[out=270,in=90] (0.5,0);
					\begin{scope}[xscale=-1]
						\draw[white,line width = 1.5mm] (0,1.2) to[out=180,in=90] (-1.5,0) to[out=270,in=180] (-0.5,-0.7) to[out=0,in=270] (0.5,0);
						\draw[burgundy, line width =  1mm] (0,1.2) to[out=180,in=90] (-1.5,0) to[out=270,in=180] (-0.5,-0.7) to[out=0,in=270] (0.5,0);
					\end{scope}
					\begin{scope}[yscale=0.5]
						\draw[white, line width = 0.7mm] ([shift={(180:0.3)}]-1.5,0) arc (180:360:0.3);
						\draw[thick] ([shift={(180:0.3)}]-1.5,0) arc (180:360:0.3);
					\end{scope}
				\end{tikzpicture}
			\end{array}$};
		\node(B) at (7,0) {$\begin{array}{c}
				\begin{tikzpicture}[scale=0.7]
					\draw[white,line width = 1.5mm] (-0.5,0) to[out=90,in=270] (1,1.5) to[out=90,in=0] (0,2);
					\draw[burgundy, line width =  1mm] (-0.5,0) to[out=90,in=270] (1,1.5) to[out=90,in=0] (0,2);
					\begin{scope}[shift={(0.2,-0.2)}]
						\draw[thick, white,line width = 0.7mm] (-0.5,0) to[out=90,in=270] (1,1.5) to[out=90,in=0] (0,2);
						\draw[thick] (-0.5,0) to[out=90,in=270] (1,1.5) to[out=90,in=0] (0,2);
					\end{scope}
					\draw[white,line width = 1.5mm] (0,1.2) to[out=180,in=90] (-1.5,0) to[out=270,in=180] (-0.5,-0.7) to[out=0,in=270] (0.5,0);
					\draw[burgundy, line width =  1mm] (0,1.2) to[out=180,in=90] (-1.5,0) to[out=270,in=180] (-0.5,-0.7) to[out=0,in=270] (0.5,0);
					\begin{scope}[shift={(0.2,-0.2)}]
						\draw[thick, white,line width = 0.7mm] (0,1.2) to[out=180,in=90] (-1.5,0) to[out=270,in=180] (-0.5,-0.7) to[out=0,in=270] (0.5,0);
						\draw[thick] (0,1.2) to[out=180,in=90] (-1.5,0) to[out=270,in=180] (-0.5,-0.7) to[out=0,in=270] (0.5,0);
					\end{scope}
					\draw[white,line width = 1.5mm] (0,2) to[out=180,in=90] (-1,1.5) to[out=270,in=90] (0.5,0);
					\draw[burgundy, line width =  1mm] (0,2) to[out=180,in=90] (-1,1.5) to[out=270,in=90] (0.5,0);
					\begin{scope}[shift={(0.2,-0.2)}]
						\draw[thick, white,line width = 0.7mm] (0,2) to[out=180,in=90] (-1,1.5) to[out=270,in=90] (0.5,0);
						\draw[thick] (0,2) to[out=180,in=90] (-1,1.5) to[out=270,in=90] (0.5,0);
					\end{scope}
					\begin{scope}[xscale=-1]
						\draw[white,line width = 1.5mm] (0,1.2) to[out=180,in=90] (-1.5,0) to[out=270,in=180] (-0.5,-0.7) to[out=0,in=270] (0.5,0);
						\draw[burgundy, line width =  1mm] (0,1.2) to[out=180,in=90] (-1.5,0) to[out=270,in=180] (-0.5,-0.7) to[out=0,in=270] (0.5,0);
					\end{scope}
					\begin{scope}[shift={(0.2,-0.2)}]
						\begin{scope}[xscale=-1]
							\draw[thick, white,line width = 0.7mm] (0,1.2) to[out=180,in=90] (-1.5,0) to[out=270,in=180] (-0.5,-0.7) to[out=0,in=270] (0.5,0);
							\draw[thick] (0,1.2) to[out=180,in=90] (-1.5,0) to[out=270,in=180] (-0.5,-0.7) to[out=0,in=270] (0.5,0);
						\end{scope}
					\end{scope}
				\end{tikzpicture}
			\end{array}$};
		\node[left] at (A.west) {$\begin{array}{c}
				A\mbox{-cycle Wilson}\\
				\mbox{loop operator }\mu
		\end{array}$: };
	\node[left] at (B.west) {$\begin{array}{c}
			B\mbox{-cycle Wison}\\
			\mbox{loop operator }\lambda
		\end{array}$: };
	\node(A) at (0,-2.5) {$\begin{array}{c}
			\begin{tikzpicture}[scale=0.7]
				\begin{scope}[yscale=0.5]
					\draw[thick, gray] ([shift={(0:0.5)}]-1.5,0.5) arc (0:180:0.5);
				\end{scope}
				\draw[white,line width = 1.5mm] (-0.5,0) to[out=90,in=270] (1,1.5) to[out=90,in=0] (0,2);
				\draw[burgundy, line width =  1mm] (-0.5,0) to[out=90,in=270] (1,1.5) to[out=90,in=0] (0,2);
				\begin{scope}[shift={(0.2,-0.2)}]
					\draw[thick, white,line width = 0.7mm] (-0.5,0) to[out=90,in=270] (1,1.5) to[out=90,in=0] (0,2);
					\draw[thick,\myblue] (-0.5,0) to[out=90,in=270] (1,1.5) to[out=90,in=0] (0,2);
				\end{scope}
				\draw[white,line width = 1.5mm] (0,1.2) to[out=180,in=90] (-1.5,0) to[out=270,in=180] (-0.5,-0.7) to[out=0,in=270] (0.5,0);
				\draw[burgundy, line width =  1mm] (0,1.2) to[out=180,in=90] (-1.5,0) to[out=270,in=180] (-0.5,-0.7) to[out=0,in=270] (0.5,0);
				\begin{scope}[shift={(0.2,-0.2)}]
					\draw[thick, white,line width = 0.7mm] (0,1.2) to[out=180,in=90] (-1.5,0) to[out=270,in=180] (-0.5,-0.7) to[out=0,in=270] (0.5,0);
					\draw[thick,\myblue] (0,1.2) to[out=180,in=90] (-1.5,0) to[out=270,in=180] (-0.5,-0.7) to[out=0,in=270] (0.5,0);
				\end{scope}
				\draw[white,line width = 1.5mm] (0,2) to[out=180,in=90] (-1,1.5) to[out=270,in=90] (0.5,0);
				\draw[burgundy, line width =  1mm] (0,2) to[out=180,in=90] (-1,1.5) to[out=270,in=90] (0.5,0);
				\begin{scope}[shift={(0.2,-0.2)}]
					\draw[thick, white,line width = 0.7mm] (0,2) to[out=180,in=90] (-1,1.5) to[out=270,in=90] (0.5,0);
					\draw[thick,\myblue] (0,2) to[out=180,in=90] (-1,1.5) to[out=270,in=90] (0.5,0);
				\end{scope}
				\begin{scope}[xscale=-1]
					\draw[white,line width = 1.5mm] (0,1.2) to[out=180,in=90] (-1.5,0) to[out=270,in=180] (-0.5,-0.7) to[out=0,in=270] (0.5,0);
					\draw[burgundy, line width =  1mm] (0,1.2) to[out=180,in=90] (-1.5,0) to[out=270,in=180] (-0.5,-0.7) to[out=0,in=270] (0.5,0);
				\end{scope}
				\begin{scope}[shift={(0.2,-0.2)}]
					\begin{scope}[xscale=-1]
						\draw[thick, white,line width = 0.7mm] (0,1.2) to[out=180,in=90] (-1.5,0) to[out=270,in=180] (-0.5,-0.7) to[out=0,in=270] (0.5,0);
						\draw[thick,\myblue] (0,1.2) to[out=180,in=90] (-1.5,0) to[out=270,in=180] (-0.5,-0.7) to[out=0,in=270] (0.5,0);
					\end{scope}
				\end{scope}
				\begin{scope}[yscale=0.5]
					\draw[white, line width = 0.7mm] ([shift={(180:0.5)}]-1.5,0.5) arc (180:360:0.5);
					\draw[thick, gray] ([shift={(180:0.5)}]-1.5,0.5) arc (180:360:0.5);
				\end{scope}
			\end{tikzpicture}
		\end{array}$};
	\node(B) at (7,-2.5) {$\begin{array}{c}
			\begin{tikzpicture}[scale=0.7]
				\begin{scope}[yscale=0.5]
					\draw[thick, gray] ([shift={(0:0.3)}]-1.5,0.5) arc (0:180:0.3);
				\end{scope}
				\draw[white,line width = 1.5mm] (-0.5,0) to[out=90,in=270] (1,1.5) to[out=90,in=0] (0,2);
				\draw[burgundy, line width =  1mm] (-0.5,0) to[out=90,in=270] (1,1.5) to[out=90,in=0] (0,2);
				\begin{scope}[shift={(0.3,-0.3)}]
					\draw[thick, white,line width = 0.7mm] (-0.5,0) to[out=90,in=270] (1,1.5) to[out=90,in=0] (0,2);
					\draw[thick,\myblue] (-0.5,0) to[out=90,in=270] (1,1.5) to[out=90,in=0] (0,2);
				\end{scope}
				\draw[white,line width = 1.5mm] (0,1.2) to[out=180,in=90] (-1.5,0) to[out=270,in=180] (-0.5,-0.7) to[out=0,in=270] (0.5,0);
				\draw[burgundy, line width =  1mm] (0,1.2) to[out=180,in=90] (-1.5,0) to[out=270,in=180] (-0.5,-0.7) to[out=0,in=270] (0.5,0);
				\begin{scope}[yscale=0.5]
					\draw[white, line width = 0.7mm] ([shift={(180:0.3)}]-1.5,0.5) arc (180:360:0.3);
					\draw[thick, gray] ([shift={(180:0.3)}]-1.5,0.5) arc (180:360:0.3);
				\end{scope}
				\begin{scope}[shift={(0.3,-0.3)}]
					\draw[thick, white,line width = 0.7mm] (0,1.2) to[out=180,in=90] (-1.5,0) to[out=270,in=180] (-0.5,-0.7) to[out=0,in=270] (0.5,0);
					\draw[thick,\myblue] (0,1.2) to[out=180,in=90] (-1.5,0) to[out=270,in=180] (-0.5,-0.7) to[out=0,in=270] (0.5,0);
				\end{scope}
				\draw[white,line width = 1.5mm] (0,2) to[out=180,in=90] (-1,1.5) to[out=270,in=90] (0.5,0);
				\draw[burgundy, line width =  1mm] (0,2) to[out=180,in=90] (-1,1.5) to[out=270,in=90] (0.5,0);
				\begin{scope}[shift={(0.3,-0.3)}]
					\draw[thick, white,line width = 0.7mm] (0,2) to[out=180,in=90] (-1,1.5) to[out=270,in=90] (0.5,0);
					\draw[thick,\myblue] (0,2) to[out=180,in=90] (-1,1.5) to[out=270,in=90] (0.5,0);
				\end{scope}
				\begin{scope}[xscale=-1]
					\draw[white,line width = 1.5mm] (0,1.2) to[out=180,in=90] (-1.5,0) to[out=270,in=180] (-0.5,-0.7) to[out=0,in=270] (0.5,0);
					\draw[burgundy, line width =  1mm] (0,1.2) to[out=180,in=90] (-1.5,0) to[out=270,in=180] (-0.5,-0.7) to[out=0,in=270] (0.5,0);
				\end{scope}
				\begin{scope}[shift={(0.3,-0.3)}]
					\begin{scope}[xscale=-1]
						\draw[thick, white,line width = 0.7mm] (0,1.2) to[out=180,in=90] (-1.5,0) to[out=270,in=180] (-0.5,-0.7) to[out=0,in=270] (0.5,0);
						\draw[thick,\myblue] (0,1.2) to[out=180,in=90] (-1.5,0) to[out=270,in=180] (-0.5,-0.7) to[out=0,in=270] (0.5,0);
					\end{scope}
				\end{scope}
			\end{tikzpicture}
		\end{array}$};
	\node[left] at (A.west) {$\begin{array}{c}
			\mbox{Operator}\\
			\mbox{monomial }{\color{\myblue}\lambda}\cdot {\color{gray}\mu}
		\end{array}$: };
	\node[left] at (B.west) {$\begin{array}{c}
			\mbox{Operator}\\
			\mbox{monomial }{\color{gray}\mu}\cdot {\color{\myblue}\lambda}
		\end{array}$: };
	\end{tikzpicture}
	\caption{Sinking operators inside the torus body produces links after inversion}\label{fig:sinking}
\end{figure}

Using Reshetikhin-Turaev (RT) formalism for computing HOMFLY-PT polynomials one could push further the action of $\mu$ and $\lambda$ on $\Psi$'s.
Link $\mu$ passes through any knot diagram intersection, in other words $\mu$ commutes with the R-matrix in the RT formalism.
Also in the RT formalism one is able to reconstruct the action of $U_q(\fg)$ from the R-matrix elements, so $\mu$ commutes with the whole $U_q(\fg)$, in other words, $\mu$ correspond to a quantum Casimir element of $U_q(\fg)$.
The other operator $\lambda$ represents a tensor multiplication in the cabling formalism language \cite{adams2004knot,cabling}:
\begin{equation}
	\begin{aligned}
	&\mu_Q\Psi_K(R)=c_Q(R)\times \Psi_K(R)\,,\\
	&\lambda_Q\Psi_K(R)=\Psi_K(R\otimes Q)=\sum\lm_{P\vdash R\otimes Q}\Psi_K(P)\,.
	\end{aligned}
\end{equation}

Here we call quantum A-polynomials a set of ${\rm rk}\;G$ polynomial difference equations for the knot wave function:
\begin{tcolorbox}
    \begin{equation}\label{Apoly}
	    \CA^{(k)}_K\left(\mu_{Q_1},\mu_{Q_2},\ldots;\lambda_{Q_1},\lambda_{Q_2},\ldots\right)\Psi_K(R)=0,\quad k=1,\ldots,{\rm rk}\,G\,.
	\end{equation}
\end{tcolorbox}

We \emph{do hope} the A-polynomials \eqref{Apoly} can be derived as \emph{relations among links}.
We expect that a set of all expectation values for links entangling the base knot $K$ is overdetermined.
Relations allow one to keep only a finite number of such expectation values inexpressible as linear combinations of others -- a link basis.
Under such assumptions the derivation of A-polynomials becomes straightforward.
It is sufficient to consider different monomials of operators $\mu_Q$ and $\lambda_Q$ (these can be represented as links) in the amount exceeding the size of the link basis, then the linear algebra guarantees an existence of a relation among them.

We would like to divide relations among links in two large groups we discuss subsequently:
\begin{itemize}
	\item Topological moves (Reidemeister moves) allowing one to \emph{planarize} an arbitrary link.
	\item Moves shifting the topology of planarized links.
\end{itemize}

The first type of moves is independent of the gauge group and its representations.
It is completely algorithmic and depends only on a braid word in the braid representation of the base knot $K$.
On the contrary the second type of moves incorporates all the information about the theory in question and remains more an art rather than a well-defined algorithmic procedure.

Therefore, unfortunately, we must admit that this procedure of constructing link bases and deriving A-polynomials is still under construction.
Here we concentrate more on successful examples.
Some version of an algorithm could be presented in the case of gauge group $SU(2)$, or $SL(2,\IC)$, and in a quasi-classical limit when $q\to 1$ and representation $R$ spin tends to infinity as the link basis in this case is related to bases of CG chords \cite{Galakhov:2024eco} or of Reeb chords in the knot contact homology theory \cite{2002math.....10124E,ekholm2013knot,Aganagic:2013jpa,Aganagic:2012jb,1210.4803}.

\subsection{Link planarization via topological moves}

Here we think of knots and links in the framework of Reshetikhin-Turaev formalism where the link of interest is represented as a closure of a braid.
Closure and intersection elements of the link diagram are substituted by respective $U_q(\fg)$ \footnote{Quantizing parameter $q$ is related canonically to the Chern-Simons theory level $\kappa$ as $q=e^{\frac{2\pi\sqrt{-1}}{\kappa+c_{\fg}}}$, where $c_{\fg}$ is the dual Coxeter number of $\fg$.} braided tensor category functors.
Physically the braid is considered as an evolution of punctures in the WZW model on a plane $\IC$ with an extra time dimension $\IR_t$ so that the evolution trajectories form the braid in the 3d space $\IC\times \IR_t$. 
And closure elements -- ``caps'' and ``cups'' -- are described by (de)fusion.

Link planarization proves to be a rather helpful trick in classifying link topological classes entangled with the base knot $K$.
By link planarization we imply a simple topological squeeze of a link to a single narrow time layer around some fixed point of $\IR_t$.
After this move we look at the planarized link from above and represent it as link diagram with self-intersections on a plane with punctures corresponding to the base knot strands:
\begin{equation}\label{planar}
    \mbox{(a):}\begin{array}{c}
	\begin{tikzpicture}[scale=0.5]
	    \foreach \i in {0, 1, 2} {
		\begin{scope}[shift={(0,\i)}]
		    \draw[thick, burgundy] (0.5,0) to[out=90,in=270] (-0.5,1);
		    \draw[] (0.7,0) to[out=90,in=270] (-0.3,1);
		    \draw[white, line width = 0.5mm] (-0.5,0) to[out=90,in=270] (0.5,1) (-0.3,0) to[out=90,in=270] (0.7,1);
		    \draw[thick, burgundy] (-0.5,0) to[out=90,in=270] (0.5,1);
		    \draw[] (0.5,1) (-0.3,0) to[out=90,in=270] (0.7,1);
		\end{scope}
		\draw[thick, burgundy] (0.5,3) to[out=90, in=180] (1,3.5) to[out=0,in=90] (1.5,3) -- (1.5,0) to[out=270,in=0] (1,-0.5) to[out=180,in=270] (0.5,0);
		\begin{scope}[xscale=-1]
		\draw[thick, burgundy] (0.5,3) to[out=90, in=180] (1,3.5) to[out=0,in=90] (1.5,3) -- (1.5,0) to[out=270,in=0] (1,-0.5) to[out=180,in=270] (0.5,0);
		\end{scope}
		\begin{scope}[shift={(0.2,0)}]
		\draw[white, line width =0.6mm] (0.5,3) to[out=90, in=180] (1,3.5) to[out=0,in=90] (1.5,3) -- (1.5,0) to[out=270,in=0] (1,-0.5) to[out=180,in=270] (0.5,0);
		\draw[] (0.5,3) to[out=90, in=180] (1,3.5) to[out=0,in=90] (1.5,3) -- (1.5,0) to[out=270,in=0] (1,-0.5) to[out=180,in=270] (0.5,0);
		\begin{scope}[xscale=-1]
		\draw[white, line width =0.6mm] (0.5,3) to[out=90, in=180] (1,3.5) to[out=0,in=90] (1.5,3) -- (1.5,0) to[out=270,in=0] (1,-0.5) to[out=180,in=270] (0.5,0);
		\draw[] (0.5,3) to[out=90, in=180] (1,3.5) to[out=0,in=90] (1.5,3) -- (1.5,0) to[out=270,in=0] (1,-0.5) to[out=180,in=270] (0.5,0);
		\end{scope}
		\end{scope}
	    }
	\end{tikzpicture}
    \end{array}
\;\longrightarrow\;
\mbox{(b):}\begin{array}{c}
		\begin{tikzpicture}
		    \draw[-stealth] (-1,0) -- (-1,1.5);
		    \node[left] at (-1,1.5) {$\scriptstyle t$};
		    \foreach \i in {0, 1, 2}
		    {
		    \begin{scope}[shift = {(0,0.5*\i)}]
		    \draw[burgundy, thick] (0.25,0) to[out=90,in=270] (-0.25,0.5);
		    \draw[white, line width = 1mm] (-0.25,0) to[out=90,in=270] (0.25,0.5);
		    \draw[burgundy, thick] (-0.25,0) to[out=90,in=270] (0.25,0.5);
		    \end{scope}
		    }
		    \begin{scope}[shift={(0,1.5)}]
		    \begin{scope}[scale=0.25,yscale=0.7]
			\draw[white,line width = 0.6mm] (-1.4,0) to[out=270,in=180] (-1,-0.4) to[out=0,in=180] (1,0.8) to[out=0,in=90] (1.8,-0.5);
			\draw[] (-1.4,0) to[out=270,in=180] (-1,-0.4) to[out=0,in=180] (1,0.8) to[out=0,in=90] (1.8,-0.5);
			\draw[] (-1,-0.8) to[out=0,in=270] (0.5,0);
			\begin{scope}[yscale=-1]
				\draw[white,line width = 0.6mm] (-1.4,0) to[out=270,in=180] (-1,-0.4) to[out=0,in=180] (1,0.8) to[out=0,in=90] (1.8,-0.5);
				\draw[] (-1.4,0) to[out=270,in=180] (-1,-0.4) to[out=0,in=180] (1,0.8) to[out=0,in=90] (1.8,-0.5);
			\end{scope}
			\draw[] (1.8,0.5) to[out=90,in=0] (0,1) to[out=180,in=90] (-1.8,0) to[out=270,in=180] (-1,-0.8);
			\draw[white,line width = 0.6mm] (0.5,0) to[out=90,in=0] (-1,1) to[out=180,in=90] (-2.2,0) to[out=270,in=180] (0,-1) to[out=0,in=270] (1.8,-0.5);
			\draw[] (0.5,0) to[out=90,in=0] (-1,1) to[out=180,in=90] (-2.2,0) to[out=270,in=180] (0,-1) to[out=0,in=270] (1.8,-0.5);
		    \end{scope}
		\end{scope}
		\draw[white, line width = 1.2mm] (-0.25,1.5) to[out=90,in=0] (-0.5,1.75) (0.25,1.5) to[out=90,in=180] (0.5,1.75);
		    \draw[burgundy, thick] (-0.25,1.5) to[out=90,in=0] (-0.5,1.75) to[out=180,in=90] (-0.75,1.5) -- (-0.75,0) to[out=270,in=180] (-0.5,-0.25) to[out=0,in=270] (-0.25,0);
		    \begin{scope}[xscale=-1]
\draw[burgundy, thick] (-0.25,1.5) to[out=90,in=0] (-0.5,1.75) to[out=180,in=90] (-0.75,1.5) -- (-0.75,0) to[out=270,in=180] (-0.5,-0.25) to[out=0,in=270] (-0.25,0);
		\end{scope}
		\end{tikzpicture}
	\end{array}\;\longrightarrow\;
\mbox{(c):}\begin{array}{c}
		\begin{tikzpicture}[scale=0.6]
		    \begin{scope}[shift={(-1,0)}]
			\draw[ultra thick, orange] (-0.1,-0.1) -- (0.1,0.1) (0.1,-0.1) -- (-0.1,0.1); 
		    \end{scope}
		    \begin{scope}[shift={(1,0)}]
			\draw[ultra thick, orange] (-0.1,-0.1) -- (0.1,0.1) (0.1,-0.1) -- (-0.1,0.1); 
		    \end{scope}
		    \begin{scope}[shift={(2.5,0)}]
			\draw[ultra thick, orange] (-0.1,-0.1) -- (0.1,0.1) (0.1,-0.1) -- (-0.1,0.1); 
		    \end{scope}
		    \begin{scope}[shift={(-3,0)}]
			\draw[ultra thick, orange] (-0.1,-0.1) -- (0.1,0.1) (0.1,-0.1) -- (-0.1,0.1); 
		    \end{scope}
			\draw[] (-1.4,0) to[out=270,in=180] (-1,-0.4) to[out=0,in=180] (1,0.8) to[out=0,in=90] (1.8,-0.5);
			\draw[] (-1,-0.8) to[out=0,in=270] (0.5,0);
			\begin{scope}[yscale=-1]
				\draw[white,line width = 1.5mm] (-1.4,0) to[out=270,in=180] (-1,-0.4) to[out=0,in=180] (1,0.8) to[out=0,in=90] (1.8,-0.5);
				\draw[] (-1.4,0) to[out=270,in=180] (-1,-0.4) to[out=0,in=180] (1,0.8) to[out=0,in=90] (1.8,-0.5);
			\end{scope}
			\draw[] (1.8,0.5) to[out=90,in=0] (0,1) to[out=180,in=90] (-1.8,0) to[out=270,in=180] (-1,-0.8);
			\draw[white,line width = 1.5mm] (0.5,0) to[out=90,in=0] (-1,1) to[out=180,in=90] (-2.2,0) to[out=270,in=180] (0,-1) to[out=0,in=270] (1.8,-0.5);
			\draw[] (0.5,0) to[out=90,in=0] (-1,1) to[out=180,in=90] (-2.2,0) to[out=270,in=180] (0,-1) to[out=0,in=270] (1.8,-0.5);
		\end{tikzpicture}
	\end{array}\,.
\end{equation}

The link planarization procedure inquires \emph{topological} transport of link strands through intersections from areas below and above time slice $t$:
\begin{equation}\label{braid}
	\begin{array}{c}
		\begin{tikzpicture}
			\draw[line width = 0.5mm] (0.2,1) -- (0.2,-0.5) to[out=270,in=0] (0,-0.7);
			\draw[line width = 2mm,white] (-0.5,1) -- (-0.5,0.5) to[out=270,in=90] (0.5,-0.5) -- (0.5,-1);
			\draw[line width = 1.2mm,black] (-0.5,1) -- (-0.5,0.5) to[out=270,in=90] (0.5,-0.5) -- (0.5,-1);
			\draw[line width = 1mm,black!40!cyan] (-0.5,1) -- (-0.5,0.5) to[out=270,in=90] (0.5,-0.5) -- (0.5,-1);
			\draw[line width = 2mm,white] (0.5,1) -- (0.5,0.5) to[out=270,in=90] (-0.5,-0.5) -- (-0.5,-1);
			\draw[line width = 1.2mm,black] (0.5,1) -- (0.5,0.5) to[out=270,in=90] (-0.5,-0.5) -- (-0.5,-1);
			\draw[line width = 1mm,black!10!orange] (0.5,1) -- (0.5,0.5) to[out=270,in=90] (-0.5,-0.5) -- (-0.5,-1);
			\draw[line width = 1.5mm,white] (-0.2,1) -- (-0.2,-0.5) to[out=270,in=180] (0,-0.7);
			\draw[line width = 0.5mm] (-0.2,1) -- (-0.2,-0.5) to[out=270,in=180] (0,-0.7);
			\begin{scope}[shift={(-1,0)}]
				\draw[-stealth] (0,-1) -- (0,1);
				\node[left] at (0,1) {$\scriptstyle t$};
			\end{scope}
		\end{tikzpicture}\\
		\hline
		t=0:\,\begin{array}{c}
			\begin{tikzpicture}
				\draw[fill=black!10!orange] (-0.5,0) circle (0.1);
				\draw[fill=black!40!cyan] (0.5,0) circle (0.1);
				\draw[line width = 0.5mm] (0,-0.5) -- (0,0.5);
			\end{tikzpicture}
		\end{array}
	\end{array}\;\overset{R}{\longrightarrow}\;\begin{array}{c}
		\begin{tikzpicture}
			\draw[line width = 0.5mm] (0.2,1) to[out=270,in=90] (-0.7,0.8);
			\draw[line width = 2mm,white] (-0.5,1) -- (-0.5,0.5) to[out=270,in=90] (0.5,-0.5) -- (0.5,-1);
			\draw[line width = 1.2mm,black] (-0.5,1) -- (-0.5,0.5) to[out=270,in=90] (0.5,-0.5) -- (0.5,-1);
			\draw[line width = 1mm,black!40!cyan] (-0.5,1) -- (-0.5,0.5) to[out=270,in=90] (0.5,-0.5) -- (0.5,-1);
			\draw[line width = 1.5mm,white]  (-0.7,0.8) to[out=270,in=90] (0.7,0.6);
			\draw[line width = 0.5mm]  (-0.7,0.8) to[out=270,in=90] (0.7,0.6);
			\draw[line width = 2mm,white] (0.5,1) -- (0.5,0.5) to[out=270,in=90] (-0.5,-0.5) -- (-0.5,-1);
			\draw[line width = 1.2mm,black] (0.5,1) -- (0.5,0.5) to[out=270,in=90] (-0.5,-0.5) -- (-0.5,-1);
			\draw[line width = 1mm,black!10!orange] (0.5,1) -- (0.5,0.5) to[out=270,in=90] (-0.5,-0.5) -- (-0.5,-1);
			\draw[line width = 1.5mm,white]  (0.7,0.6) to[out=270,in=270] (-0.2,1);
			\draw[line width = 0.5mm]  (0.7,0.6) to[out=270,in=270] (-0.2,1);
			\begin{scope}[shift={(-1,0)}]
				\draw[-stealth] (0,-1) -- (0,1);
				\node[left] at (0,1) {$\scriptstyle t$};
			\end{scope}
		\end{tikzpicture}\\
		\hline
		t=1:\,\begin{array}{c}
			\begin{tikzpicture}
				\draw[fill=black!40!cyan] (-0.5,0) circle (0.1);
				\draw[fill=black!10!orange] (0.5,0) circle (0.1);
				\draw[line width = 0.5mm] (0,-0.5) to[out=90,in=270] (0.7,0) to[out=90,in=0] (0.5,0.3) to[out=180,in=0] (-0.5,-0.3) to[out=180,in=270] (-0.7,0) to[out=90,in=270] (0,0.5);
			\end{tikzpicture}
		\end{array}
	\end{array},\quad
	\begin{array}{c}
		\begin{tikzpicture}[xscale=-1]
			\draw[line width = 0.5mm] (0.2,1) -- (0.2,-0.5) to[out=270,in=0] (0,-0.7);
			\draw[line width = 2mm,white] (-0.5,1) -- (-0.5,0.5) to[out=270,in=90] (0.5,-0.5) -- (0.5,-1);
			\draw[line width = 1.2mm,black] (-0.5,1) -- (-0.5,0.5) to[out=270,in=90] (0.5,-0.5) -- (0.5,-1);
			\draw[line width = 1.2mm,black!10!orange] (-0.5,1) -- (-0.5,0.5) to[out=270,in=90] (0.5,-0.5) -- (0.5,-1);
			\draw[line width = 2mm,white] (0.5,1) -- (0.5,0.5) to[out=270,in=90] (-0.5,-0.5) -- (-0.5,-1);
			\draw[line width = 1.2mm,black] (0.5,1) -- (0.5,0.5) to[out=270,in=90] (-0.5,-0.5) -- (-0.5,-1);
			\draw[line width = 1mm,black!40!cyan] (0.5,1) -- (0.5,0.5) to[out=270,in=90] (-0.5,-0.5) -- (-0.5,-1);
			\draw[line width = 1.5mm,white] (-0.2,1) -- (-0.2,-0.5) to[out=270,in=180] (0,-0.7);
			\draw[line width = 0.5mm] (-0.2,1) -- (-0.2,-0.5) to[out=270,in=180] (0,-0.7);
			\begin{scope}[shift={(1,0)}]
				\draw[-stealth] (0,-1) -- (0,1);
				\node[left] at (0,1) {$\scriptstyle t$};
			\end{scope}
		\end{tikzpicture}\\
		\hline
		t=0:\,\begin{array}{c}
			\begin{tikzpicture}
				\draw[fill=black!10!orange] (-0.5,0) circle (0.1);
				\draw[fill=black!40!cyan] (0.5,0) circle (0.1);
				\draw[line width = 0.5mm] (0,-0.5) -- (0,0.5);
			\end{tikzpicture}
		\end{array}
	\end{array}\;\overset{R^{-1}}{\longrightarrow}\;\begin{array}{c}
		\begin{tikzpicture}[xscale=-1]
			\draw[line width = 0.5mm] (0.2,1) to[out=270,in=90] (-0.7,0.8);
			\draw[line width = 2mm,white] (-0.5,1) -- (-0.5,0.5) to[out=270,in=90] (0.5,-0.5) -- (0.5,-1);
			\draw[line width = 1.2mm,black] (-0.5,1) -- (-0.5,0.5) to[out=270,in=90] (0.5,-0.5) -- (0.5,-1);
			\draw[line width = 1mm,black!10!orange] (-0.5,1) -- (-0.5,0.5) to[out=270,in=90] (0.5,-0.5) -- (0.5,-1);
			\draw[line width = 1.5mm,white]  (-0.7,0.8) to[out=270,in=90] (0.7,0.6);
			\draw[line width = 0.5mm]  (-0.7,0.8) to[out=270,in=90] (0.7,0.6);
			\draw[line width = 2mm,white] (0.5,1) -- (0.5,0.5) to[out=270,in=90] (-0.5,-0.5) -- (-0.5,-1);
			\draw[line width = 1.2mm,black] (0.5,1) -- (0.5,0.5) to[out=270,in=90] (-0.5,-0.5) -- (-0.5,-1);
			\draw[line width = 1mm,black!40!cyan] (0.5,1) -- (0.5,0.5) to[out=270,in=90] (-0.5,-0.5) -- (-0.5,-1);
			\draw[line width = 1.5mm,white]  (0.7,0.6) to[out=270,in=270] (-0.2,1);
			\draw[line width = 0.5mm]  (0.7,0.6) to[out=270,in=270] (-0.2,1);
			\begin{scope}[shift={(1,0)}]
				\draw[-stealth] (0,-1) -- (0,1);
				\node[left] at (0,1) {$\scriptstyle t$};
			\end{scope}
		\end{tikzpicture}\\
		\hline
		t=1:\,\begin{array}{c}
			\begin{tikzpicture}[xscale=-1]
				\draw[fill=black!10!orange] (-0.5,0) circle (0.1);
				\draw[fill=black!40!cyan] (0.5,0) circle (0.1);
				\draw[line width = 0.5mm] (0,-0.5) to[out=90,in=270] (0.7,0) to[out=90,in=0] (0.5,0.3) to[out=180,in=0] (-0.5,-0.3) to[out=180,in=270] (-0.7,0) to[out=90,in=270] (0,0.5);
			\end{tikzpicture}
		\end{array}
	\end{array}\,.
\end{equation}
If a link is located both below and above $t$ we cut it in open strands along slice $t$ and bring pieces to $t$ separately remembering which strand goes above, which goes below, then close disjoint strands in a link diagram with self-intersections.
It turns out to be an algorithmic procedure in terms of a braid group representation on arcades we describe separately in Sec.~\ref{sec:arcade}.

We find that the mechanism of link planarization is rather useful for our purposes.
However before implementing it at full capacity a comment should made.
Among \emph{all} links (abbreviated to \emph{$a$-links} in this section) one should distinguish a subclass of \emph{boundary} links (or \emph{b-links}).
A generic $a$-link is a Wilson loop located in the bulk of knot complement $M=S^3\setminus K$, however some of the links could be brought by homotopy to boundary $\p M$ and pulled out of $M$.
Conversely, as we mentioned earlier Wilson loop operators acting on the Hilbert space $\mathscr{H}(\p M)$ could be sunk inside the bulk $M$ to become $a$-links.
We call $b$-links those $a$-links that are transportable homotopically to the boundary $\p M$ and, therefore, correspond to some operators on $\mathscr{H}$ (see Fig.\ref{fig:ab-links}).

\begin{figure}[ht!]
    \centering
    \begin{tikzpicture}
	\draw[fill=white!80!green] (-1,0.2)  -- (0,0.2) to[out=0,in=180] (0.5,0) -- (4,0) to[out=0,in=270] (4.5,0.5) to[out=90,in=0] (4,1) -- (1.3,1) to[out=180,in=0] (0.4,0.8) --(-0.4,0.8) to[out=180,in=180] (-1,0.2);
	\draw[dashed] (-1,0.2) to[out=0,in=0] (-0.4,0.8);
	\begin{scope}[shift={(0.3,0)}]
	    \draw[thick] (-1,0.2) to[out=180,in=180] (-0.4,0.8);
	    \draw[thick, dashed] (-1,0.2) to[out=0,in=0] (-0.4,0.8);
	\end{scope}
	\begin{scope}[shift = {(1.5,0.5)}]
	\begin{scope}[scale=0.4]
	    \draw[fill=white] (-0.5,0) to[out=30,in=150] (0.5,0) to[out=210,in=330] cycle;
	    \draw (-0.5,0) -- (-0.673205,0.1) (0.5,0) -- (0.673205,0.1);
	\end{scope}
	\end{scope}
	\begin{scope}[shift = {(2.5,0.5)}]
	    \begin{scope}[yscale=0.5]
	    \draw[thick] (0,0) circle (0.5);
	\end{scope}
	\begin{scope}[scale=0.4]
	    \draw[fill=white] (-0.5,0) to[out=30,in=150] (0.5,0) to[out=210,in=330] cycle;
	    \draw (-0.5,0) -- (-0.673205,0.1) (0.5,0) -- (0.673205,0.1);
	\end{scope}
	\end{scope}
	\begin{scope}[shift = {(3.5,0.5)}]
	\begin{scope}[scale=0.4]
	    \draw[fill=white] (-0.5,0) to[out=30,in=150] (0.5,0) to[out=210,in=330] cycle;
	    \draw (-0.5,0) -- (-0.673205,0.1) (0.5,0) -- (0.673205,0.1);
	\end{scope}
	\end{scope}
	\node[left] at (-1,0.5) {$\scriptstyle\p M$};
	\node[right] at (4.5,0.5) {$\scriptstyle M$};
	\node[below] at (-0.7,0.2) {\small $b$-link};
	\node[below] at (2.5,0) {\small $a$-link};
    \end{tikzpicture}
    \caption{Schematic depiction of $a$-link and $b$-link analogies on a 2d $M$ (a Riemann surface) with 1d $\p M$
    	(a circle at the bottle neck on the left).
    	An example of the $b$-link is homotopic to the boundary, and $a$-link is not.}\label{fig:ab-links}
\end{figure}

We would like to introduce notations for planarized links.
Suppose there are some planarized link diagrams $L_i$ drawn with possible self-intersections on a plane with punctures as in \eqref{planar}(c).
We will denote a HOMFLY-PT polynomial of knot $K$ with inserted planar link diagram $L_i$ as in \eqref{planar}(b) into a chosen time slice as a correlator $\langle L_i\rangle_K$ (if the context permits subscript $K$ will be omitted).
To a word $\langle L_1 L_2\ldots L_n\rangle_K$ we associate a link consisting of $L_i$ beaded on the braid from top to bottom as pancakes on a fork:
\begin{equation}
\langle L_iL_j\rangle_K:=\begin{array}{c}
		\begin{tikzpicture}[scale=0.7]
			\draw[thick] (-1.25,0) to[out=270,in=270] (-0.75,0) (-0.25,0) to[out=270,in=270] (0.25,0) (0.75,0) to[out=270,in=270] (1.25,0);
			\begin{scope}[shift={(0,2.5)}]
				\draw[thick] (-1.25,0) to[out=90,in=90] (-0.75,0) (-0.25,0) to[out=90,in=90] (0.25,0) (0.75,0) to[out=90,in=90] (1.25,0);
			\end{scope}
			\draw[thick] (-1.25,1) -- (-1.25,2.5) (-0.75,1) -- (-0.75,2.5) (-0.25,1) -- (-0.25,2.5) (0.25,1) -- (0.25,2.5) (0.75,1) -- (0.75,2.5) (1.25,1) -- (1.25,2.5);
			\draw (-1.5,0) -- (1.5,0) -- (1.5,1) -- (-1.5,1) -- cycle;
			\node at (0,0.5) {braid};
			\begin{scope}[shift={(0,1.5)}]
				\draw[fill=white](-1.5,0) to[out=10,in=170] (1.5,0) to[out=190,in=350] (-1.5,0);
				\node[left] at (-1.5,0) {$\scriptstyle L_j$};
			\end{scope}
			\begin{scope}[shift={(0,2)}]
				\draw[fill=white](-1.5,0) to[out=10,in=170] (1.5,0) to[out=190,in=350] (-1.5,0);
				\node[left] at (-1.5,0) {$\scriptstyle L_i$};
			\end{scope}
		\end{tikzpicture}
	\end{array}\,.
\end{equation}
It is natural to extend this formalism to linear word combinations:
\begin{equation}
    \langle c_1 L_1+c_2 L_2\rangle_K=c_1\langle L_1\rangle_K+c_2\langle L_2\rangle_K\,.
\end{equation}

In general, such correlators correspond to substantial $a$-links that are not $b$-links!
Therefore we call $L_i$ in brackets \emph{planarized link symbols}\footnote{
	It would be interesting to investigate how the set of planarized links with introduced structures behave under topological moves on the base knot $K$, if an ambient isotopy of knots $K\sim K'$ induces any morphism on the bracket $\langle\cdot\rangle_K\to \langle\cdot\rangle_{K'}$.
	Since the planarized links are graphs essentially, this morphisms might lead to a construction of graph valued link invariants \cite{akimova2020labels,ManNik}.
	We would like to thank Vassily Manturov for point out this possible direction to us.
	}, so they are not confused with operators $\mu_Q$ and $\lambda_Q$ corresponding to $b$-links.
In particular, for link symbols if two bracketed relations are found $\langle P_1(L_i)\rangle_K=0$ and $\langle P_2(L_i)\rangle_K=0$, in general, $\langle P_1(L_i)P_2(L_i)\rangle_K\neq 0$ due to quantum corrections in the correlators!

However, there is a very special planarized link diagram represented by a simple ring without self-intersections around any puncture.
This is a Hopf $b$-link $\mu_Q$.
As a $b$-link it can be brought to the boundary and treated as an operator acting on the HOMFLY-PT polynomial -- a partition function on  a manifold $M$ with boundary -- as a wave function in $\mathscr{H}(\p M)$:
\begin{equation}
    \langle A\; \mu_Q\; B\rangle_K=\mu_Q\;\langle AB\rangle_K\,.
\end{equation}

Also we should mention another treatment of planarized link symbols in the limit $q\to 1$ and large reps coloring knot $K$: as surface operators obtained as monodromies of light fields in a WZW or Liouville conformal block of heavy fields \cite{Galakhov:2014aha,Terashima:2013fg,Drukker:2009id,Kozcaz:2010af,Galakhov:2013jma,Maruyoshi:2010iu,Marshakov:2010fx}.
In principle, one might expect that closed monodromy paths could be cut open what would lead to CG/Reeb chord bases for planarized links \cite{Galakhov:2024eco}.
Yet generalizing these bases from the case of $\fs\fu_2$ to generic $\fs\fu_n$ similar to an appropriate generalization of surface operators \cite{Gaiotto:2012rg,Galakhov:2014xba,Longhi:2016rjt,Neitzke:2021gxr} would require further research, so we leave this discussion for elsewhere.


\subsection{Shifting planarized link topology, brackets and closures}

The first natural candidate for relations modifying the topology of planarized links is skein relation.
A distinguishing feature of these relations is that information about the theory, parameters of $U_q(\fs\fu_n)$, appears manifestly, for example:
\begin{equation}\label{skein}
	\begin{aligned}
	&q^{\frac{1}{n}}\begin{array}{c}
		\begin{tikzpicture}[scale=0.6]
			\draw (-0.173205,0.1) to[out=150, in=0] (-0.5,0.2) to[out=180,in=90] (-0.8,0); 
			\draw (0.173205,0.1) to[out=30, in=0] (0.5,0.2) to[out=0,in=90] (0.8,0); 
			\draw[white,line width = 1.5mm] (-0.5,0) to[out=90,in=270] (1,1.5) to[out=90,in=0] (0,2);
			\draw[burgundy, line width =  0.85mm] (-0.5,0) to[out=90,in=270] (1,1.5) to[out=90,in=0] (0,2);
			\draw[white,line width = 1.5mm] (0,1) to[out=180,in=90] (-1.5,0) to[out=270,in=180] (-0.5,-0.7) to[out=0,in=270] (0.5,0);
			\draw[burgundy, line width =  0.85mm] (0,1.2) to[out=180,in=90] (-1.5,0) to[out=270,in=180] (-0.5,-0.7) to[out=0,in=270] (0.5,0);
			\draw[white,line width = 1.5mm] (0,2) to[out=180,in=90] (-1,1.5) to[out=270,in=90] (0.5,0);
			\draw[burgundy, line width =  0.85mm] (0,2) to[out=180,in=90] (-1,1.5) to[out=270,in=90] (0.5,0);
			\begin{scope}[xscale=-1]
				\draw[white,line width = 1.5mm] (0,1.2) to[out=180,in=90] (-1.5,0) to[out=270,in=180] (-0.5,-0.7) to[out=0,in=270] (0.5,0);
				\draw[burgundy, line width =  0.85mm] (0,1.2) to[out=180,in=90] (-1.5,0) to[out=270,in=180] (-0.5,-0.7) to[out=0,in=270] (0.5,0);
			\end{scope}
			\draw[white, line width = 0.9mm] (-0.173205,0.1) -- (0,0) to[out=330, in=180] (0.5,-0.2) to[out=0,in=270] (0.8,0);
			\draw (-0.173205,0.1) -- (0,0) to[out=330, in=180] (0.5,-0.2) to[out=0,in=270] (0.8,0); 
			\draw[white, line width = 0.9mm] (-0.8,0) to[out=270,in=180] (-0.5,-0.2) to[out=0,in=210] (0,0) -- (0.173205,0.1);
			\draw[postaction={decorate},decoration={markings, mark= at position 0.7 with {\arrow{stealth}}}] (-0.8,0) to[out=270,in=180] (-0.5,-0.2) to[out=0,in=210] (0,0) -- (0.173205,0.1);
			\node[burgundy, left] at (-1,1.5) {$\scriptstyle R$};
		\end{tikzpicture}
	\end{array}-q^{-\frac{1}{n}}\begin{array}{c}
		\begin{tikzpicture}[scale=0.6]
			\draw (-0.173205,0.1) to[out=150, in=0] (-0.5,0.2) to[out=180,in=90] (-0.8,0); 
			\draw (0.173205,0.1) to[out=30, in=0] (0.5,0.2) to[out=0,in=90] (0.8,0); 
			\draw[white,line width = 1.5mm] (-0.5,0) to[out=90,in=270] (1,1.5) to[out=90,in=0] (0,2);
			\draw[burgundy, line width =  0.85mm] (-0.5,0) to[out=90,in=270] (1,1.5) to[out=90,in=0] (0,2);
			\draw[white,line width = 1.5mm] (0,1) to[out=180,in=90] (-1.5,0) to[out=270,in=180] (-0.5,-0.7) to[out=0,in=270] (0.5,0);
			\draw[burgundy, line width =  0.85mm] (0,1.2) to[out=180,in=90] (-1.5,0) to[out=270,in=180] (-0.5,-0.7) to[out=0,in=270] (0.5,0);
			\draw[white,line width = 1.5mm] (0,2) to[out=180,in=90] (-1,1.5) to[out=270,in=90] (0.5,0);
			\draw[burgundy, line width =  0.85mm] (0,2) to[out=180,in=90] (-1,1.5) to[out=270,in=90] (0.5,0);
			\begin{scope}[xscale=-1]
				\draw[white,line width = 1.5mm] (0,1.2) to[out=180,in=90] (-1.5,0) to[out=270,in=180] (-0.5,-0.7) to[out=0,in=270] (0.5,0);
				\draw[burgundy, line width =  0.85mm] (0,1.2) to[out=180,in=90] (-1.5,0) to[out=270,in=180] (-0.5,-0.7) to[out=0,in=270] (0.5,0);
			\end{scope}
			\begin{scope}[xscale=-1]
				\draw[white, line width = 0.9mm] (-0.173205,0.1) -- (0,0) to[out=330, in=180] (0.5,-0.2) to[out=0,in=270] (0.8,0);
				\draw (-0.173205,0.1) -- (0,0) to[out=330, in=180] (0.5,-0.2) to[out=0,in=270] (0.8,0); 
				\draw[white, line width = 0.9mm] (-0.8,0) to[out=270,in=180] (-0.5,-0.2) to[out=0,in=210] (0,0) -- (0.173205,0.1);
				\draw[postaction={decorate},decoration={markings, mark= at position 0.7 with {\arrow{stealth}}}] (-0.8,0) to[out=270,in=180] (-0.5,-0.2) to[out=0,in=210] (0,0) -- (0.173205,0.1);
			\end{scope}
			\node[burgundy, left] at (-1,1.5) {$\scriptstyle R$};
		\end{tikzpicture}
	\end{array}=(q-q^{-1})\begin{array}{c}
		\begin{tikzpicture}[scale=0.6]
			\begin{scope}[yscale=0.5]
				\draw ([shift={(0:0.3)}]-0.5,0) arc (0:180:0.3);
				\draw ([shift={(0:0.3)}]0.5,0) arc (0:180:0.3);
			\end{scope}
			\draw[white,line width = 1.5mm] (-0.5,0) to[out=90,in=270] (1,1.5) to[out=90,in=0] (0,2);
			\draw[burgundy, line width =  0.85mm] (-0.5,0) to[out=90,in=270] (1,1.5) to[out=90,in=0] (0,2);
			\draw[white,line width = 1.5mm] (0,1) to[out=180,in=90] (-1.5,0) to[out=270,in=180] (-0.5,-0.7) to[out=0,in=270] (0.5,0);
			\draw[burgundy, line width =  0.85mm] (0,1.2) to[out=180,in=90] (-1.5,0) to[out=270,in=180] (-0.5,-0.7) to[out=0,in=270] (0.5,0);
			\draw[white,line width = 1.5mm] (0,2) to[out=180,in=90] (-1,1.5) to[out=270,in=90] (0.5,0);
			\draw[burgundy, line width =  0.85mm] (0,2) to[out=180,in=90] (-1,1.5) to[out=270,in=90] (0.5,0);
			\begin{scope}[xscale=-1]
				\draw[white,line width = 1.5mm] (0,1.2) to[out=180,in=90] (-1.5,0) to[out=270,in=180] (-0.5,-0.7) to[out=0,in=270] (0.5,0);
				\draw[burgundy, line width =  0.85mm] (0,1.2) to[out=180,in=90] (-1.5,0) to[out=270,in=180] (-0.5,-0.7) to[out=0,in=270] (0.5,0);
			\end{scope}
			\begin{scope}[yscale=0.5]
				\draw[white, line width = 1mm] ([shift={(180:0.3)}]-0.5,0) arc (180:360:0.3);
				\draw[postaction={decorate},decoration={markings, mark= at position 0.7 with {\arrow{stealth}}}] ([shift={(180:0.3)}]-0.5,0) arc (180:360:0.3);
				\begin{scope}[xscale=-1]
					\draw[white, line width = 1mm] ([shift={(180:0.3)}]-0.5,0) arc (180:360:0.3);
					\draw[postaction={decorate},decoration={markings, mark= at position 0.7 with {\arrow{stealth}}}] ([shift={(180:0.3)}]-0.5,0) arc (180:360:0.3);
				\end{scope}
			\end{scope}
			\node[burgundy, left] at (-1,1.5) {$\scriptstyle R$};
		\end{tikzpicture}
	\end{array},\quad \forall R\,,\\
	&q^{\frac{1}{n}}\begin{array}{c}
		\begin{tikzpicture}[scale=0.5,yscale=-1]
			\begin{scope}[shift={(-1,0)}]
				\draw[ultra thick, orange] (-0.1,-0.1) -- (0.1,0.1) (-0.1,0.1) -- (0.1,-0.1);
			\end{scope}
			\begin{scope}[shift={(1,0)}]
				\draw[ultra thick, orange] (-0.1,-0.1) -- (0.1,0.1) (-0.1,0.1) -- (0.1,-0.1);
			\end{scope}
			\draw(-1,-0.4) to[out=0,in=180] (1,0.4);
			\draw[white,line width = 1.5mm] (-1,0.4) to[out=0,in=180] (1,-0.4);
			\draw (-1,0.4) to[out=0,in=180] (1,-0.4);
			\draw (-1,0.4) to[out=180,in=90] (-1.4,0) to[out=270,in=180] (-1,-0.4);
			\begin{scope}[xscale=-1]
				\draw (-1,0.4) to[out=180,in=90] (-1.4,0) to[out=270,in=180] (-1,-0.4);
			\end{scope}
		\end{tikzpicture}
	\end{array}-q^{-\frac{1}{n}}\begin{array}{c}
	\begin{tikzpicture}[scale=0.5]
		\begin{scope}[shift={(-1,0)}]
			\draw[ultra thick, orange] (-0.1,-0.1) -- (0.1,0.1) (-0.1,0.1) -- (0.1,-0.1);
		\end{scope}
		\begin{scope}[shift={(1,0)}]
			\draw[ultra thick, orange] (-0.1,-0.1) -- (0.1,0.1) (-0.1,0.1) -- (0.1,-0.1);
		\end{scope}
		\draw (-1,-0.4) to[out=0,in=180] (1,0.4);
		\draw[white,line width = 1.5mm] (-1,0.4) to[out=0,in=180] (1,-0.4);
		\draw (-1,0.4) to[out=0,in=180] (1,-0.4);
		\draw (-1,0.4) to[out=180,in=90] (-1.4,0) to[out=270,in=180] (-1,-0.4);
		\begin{scope}[xscale=-1]
			\draw (-1,0.4) to[out=180,in=90] (-1.4,0) to[out=270,in=180] (-1,-0.4);
		\end{scope}
	\end{tikzpicture}
	\end{array}=\left(q-q^{-1}\right)\begin{array}{c}
		\begin{tikzpicture}[scale=0.5]
			\begin{scope}[shift={(-1,0)}]
				\draw[ultra thick, orange] (-0.1,-0.1) -- (0.1,0.1) (-0.1,0.1) -- (0.1,-0.1);
			\end{scope}
			\begin{scope}[shift={(1,0)}]
				\draw[ultra thick, orange] (-0.1,-0.1) -- (0.1,0.1) (-0.1,0.1) -- (0.1,-0.1);
			\end{scope}
			\draw (-0.3,0) to[out=90,in=0] (-1,0.4) to[out=180,in=90] (-1.4,0) to[out=270,in=180] (-1,-0.4) to[out=0,in=270] (-0.3,0);
			\begin{scope}[xscale=-1]
				\draw (-0.3,0) to[out=90,in=0] (-1,0.4) to[out=180,in=90] (-1.4,0) to[out=270,in=180] (-1,-0.4) to[out=0,in=270] (-0.3,0);
			\end{scope}
		\end{tikzpicture}
	\end{array}\,.
	\end{aligned}
\end{equation}

Having a link diagram of some small representation on a plane with punctures one could reduce the diagram further by applying brackets for the respective algebra $U_q(\fg)$.
The ``bracket'' in this context implies that the R-matrix could be explicitly decomposed in terms of invariant tensors admitting graphical calculus.
The resulting graphical calculus is known as MOY diagrams \cite{MOY}.
It is well-known that brackets for $\fs\fu_n$ lead to skein relations like \eqref{skein}.

Due to a natural identification of links with MOY diagrams we call MOY diagrams drawn on a plane with punctures as in \eqref{planar} links as well.

Another type of moves we included in this group could be described as hopping link strands over closing elements.
These moves are topological in general if the link is considered as a collection of strands in a 3d space, however the topology of the very planarized link is shifted drastically:
\begin{equation}\label{cupmove}
    \begin{array}{c}
	\begin{tikzpicture}
	    \draw (-0.4,0) to[out=90,in=180] (0,0.2) to[out=0,in=90] (0.4,0);
	    \draw[white, line width=1mm] (-0.25, 0.25) -- (-0.25,-0.25) to[out=270,in=270] (-1.25,-0.25) -- (-1.25,0.25);
	    \draw[thick, burgundy] (-0.25, 0.4) -- (-0.25,-0.25) to[out=270,in=270] (-1.25,-0.25) -- (-1.25,0.4);
	    \begin{scope}[xscale=-1]
	    \draw[white, line width=1mm] (-0.25, 0.4) -- (-0.25,-0.25) to[out=270,in=270] (-1.25,-0.25) -- (-1.25,0.4);
	    \draw[thick, burgundy] (-0.25, 0.4) -- (-0.25,-0.25) to[out=270,in=270] (-1.25,-0.25) -- (-1.25,0.4);
	    \end{scope}
	    \draw[white, line width =1mm] (-0.4,0) to[out=270,in=180] (0,-0.2) to[out=0,in=270] (0.4,0);
	    \draw (-0.4,0) to[out=270,in=180] (0,-0.2) to[out=0,in=270] (0.4,0);
	\end{tikzpicture}
    \end{array}\;\overset{\mbox{\tiny 1st Reid.}}{\sim}\;\begin{array}{c}
	\begin{tikzpicture}
	    \draw (-0.5,0.144338) to[out=30,in=90] (0.4,0) (-1,0.144338) to[out=150,in=90] (-1.4,0);
	    \begin{scope}[xscale=-1]
	    \draw[white, line width=1mm] (-0.25, 0.25) -- (-0.25,-0.25) to[out=270,in=270] (-1.25,-0.25) -- (-1.25,0.25);
	    \draw[thick, burgundy] (-0.25, 0.4) -- (-0.25,-0.25) to[out=270,in=270] (-1.25,-0.25) -- (-1.25,0.4);
	    \end{scope}
	    \draw[white, line width =1mm] (-0.5,-0.144338) to[out=330,in=270] (0.4,0);
	    \draw (-0.5,-0.144338) to[out=330,in=270] (0.4,0);
	    \draw[white, line width=1mm] (-0.25, 0.25) -- (-0.25,-0.25) to[out=270,in=270] (-1.25,-0.25) -- (-1.25,0.25);
	    \draw[thick, burgundy] (-0.25, 0.4) -- (-0.25,-0.25) to[out=270,in=270] (-1.25,-0.25) -- (-1.25,0.4);
	    \draw[white, line width =1mm] (-1,-0.144338) to[out=210,in=270] (-1.4,0);
	    \draw  (-1,-0.144338) to[out=210,in=270] (-1.4,0);
	    \draw (-1,0.144338) -- (-0.5,-0.144338);
	    \draw[white,line width =1mm] (-1,-0.144338) -- (-0.5,0.144338);
	    \draw (-1,-0.144338) -- (-0.5,0.144338);
	\end{tikzpicture}
    \end{array}, \quad \begin{array}{c}\begin{tikzpicture}[scale=0.5]
    \foreach \i in {0,...,3} 
    {
    	\begin{scope}[shift={(\i,0)}]
    		\draw[orange, ultra thick] (-0.1,-0.1) -- (0.1, 0.1) (0.1,-0.1) -- (-0.1, 0.1);
    	\end{scope}
    }
    \draw (0.5,0) to[out=90,in=180] (1.5,0.5) to[out=0,in=90] (2.5,0);
    \begin{scope}[yscale=-1]
    	\draw (0.5,0) to[out=90,in=180] (1.5,0.5) to[out=0,in=90] (2.5,0);
    \end{scope}
    \end{tikzpicture}
\end{array}\sim \begin{array}{c}
\begin{tikzpicture}[scale=0.5]
\foreach \i in {0,...,3} 
{
	\begin{scope}[shift={(\i,0)}]
		\draw[orange, ultra thick] (-0.1,-0.1) -- (0.1, 0.1) (0.1,-0.1) -- (-0.1, 0.1);
	\end{scope}
}
\draw (0,-0.5) to[out=180,in=270] (-0.5,0) to[out=90,in=180] (0,0.3) -- (1,0.3) to[out=0, in=180] (2,-0.5) to[out=0,in=270] (2.5,0) to[out=90,in=0] (2,0.6) -- (0.7,0.6);
\draw[line width=1.2mm,white] (0.7,0.6) to[out=180,in=0] (0,-0.5);
\draw (0.7,0.6) to[out=180,in=0] (0,-0.5);
\end{tikzpicture}
\end{array}\,.
\end{equation}
These types of moves prove to be rather useful for a derivation of relations among links themselves.
In these moves the information on how knot $K$ is constructed from a vanilla braid via closures is contained.
So  it is natural to guess that they are mostly responsible for the knot $K$ dependence appearing in A-polynomials \eqref{Apoly}.


\section{Arcade representation of braid group}\label{sec:arcade}

Here we introduce a notion of \emph{arcades}.
Consider a set of $k$ punctures.
We call an arcade a triplet of sets of dots ($\bf D$), upper arcs ($\bf U$) and lower arcs ($\bf L$).
Dots $\bf D$ are collected in $k+1$ ordered sets of dots located to the left, to the right and in between gaps among punctures.
The upper and lower arcs are labeled by unordered doublets of dots $x\cap y$ respective arcs connect.
For example, the following arcade encodes a diagram:
\begin{equation}\label{arc_examp}
	\begin{aligned}
	&\left(\begin{array}{c}
		\bf D \\ \bf U \\ \bf L
	\end{array}\right)=\left(\begin{array}{c}
	\{1,2\}, \{3\}, \{4,5\}, \{6\}\\
	3 \cap 4\\
	1\cap 3,  5 \cap 6 
	\end{array}\right)\quad\to\quad\begin{array}{c}
		\begin{tikzpicture}
			\foreach \i in {1,...,3}
			{\begin{scope}[shift={(\i,0)}]
					\draw[ultra thick, orange] (-0.05,-0.05) -- (0.05,0.05) (-0.05,0.05) -- (0.05,-0.05);
			\end{scope}}
			\foreach \a/\b in {1.91667/0.416667}
			{\draw[thick] ([shift={(0:\b)}]\a,0) arc (0:180:\b);}
			\foreach \a/\b in {0.916667/0.583333,3.08333/0.416667}
			{\draw[thick] ([shift={(180:\b)}]\a,0) arc (180:360:\b);}
			\foreach \a/\b in {1/0.333333,2/0.666667,3/1.5,4/2.33333,5/2.66667,6/3.5}
			{
				\draw[\myblue, fill=white] (\b,0) circle (0.1);
				\node at (\b,0) {\tiny \a};
			}
		\end{tikzpicture}
	\end{array}
	\end{aligned}
\end{equation}
In our arcades arcs {\bf do not intersect}.

Apparently, there are certain equivalence moves on arcades mapping arcade diagrams into topologically equivalent ones:
\begin{enumerate}
	\item Stray dot: if there is a dot that does not belong to any arc it can be deleted.
	\item Inflation: if there are two arcs of the same type connected to the same dot, for example, $a\cap b$ and $b\cap c$, then the common dot $b$ can be deleted and arcs merge into a single one $a\cap c$.
	\item Swap: if there is an arc connecting two consequent point in the same gap, these points are merged and the arc is deleted.
\end{enumerate}
We will be interested in arcades \emph{modulo} these equivalence moves.

The \emph{braid group} action on punctures descends naturally to apparent morphisms on arcades.
Let us introduce the braid group generators as $R_k^+$ and $R_k^-$ braiding the $k^{\rm th}$ and the $(k+1)^{\rm th}$ punctures counterclockwise and clockwise respectively.
The corresponding arcade morphisms read:
\begin{equation}\label{arc_braid}
	\begin{aligned}
		&\begin{array}{c}
			\begin{tikzpicture}[scale=1.3]
				\begin{scope}[shift={(-0.5,0)}]
					\draw[ultra thick, black!20!orange] (-0.05,-0.05) -- (0.05,0.05) (-0.05,0.05) -- (0.05,-0.05);
				\end{scope}
				\begin{scope}[shift={(0.5,0)}]
					\begin{scope}[rotate=45]
					\draw[ultra thick, black!40!cyan] (-0.05,-0.05) -- (0.05,0.05) (-0.05,0.05) -- (0.05,-0.05);
					\end{scope}
				\end{scope}
				\draw[thick] (-1.25,-0.75) -- (-1.25,0.75) (-1,-0.75) -- (-1,0.75) (-0.75,-0.75) -- (-0.75,0.75);
				\draw[ultra thick, black!40!red] (-0.25,-0.75) -- (-0.25,0.75); 
				\draw[ultra thick, black!40!green] (0,-0.75) -- (0,0.75);
				\draw[ultra thick, black!40!blue] (0.25,-0.75) -- (0.25,0.75);
				\draw[thick] (1.25,-0.75) -- (1.25,0.75) (1,-0.75) -- (1,0.75) (0.75,-0.75) -- (0.75,0.75);
				\foreach \x/\t in {-1.25/$a_1$, -1/$a_2$, -0.75/$a_3$, -0.25/$b_1$, 0/$b_2$, 0.25/$b_3$, 0.75/$c_1$, 1/$c_2$, 1.25/$c_3$}
				{
					\draw[fill=white, draw=\myblue] (\x,0) circle (0.1);
					\node at (\x,0) {\scalebox{0.5}{\t}};
				}
			\end{tikzpicture}
		\end{array}\;\underset{\begin{array}{c}
			\begin{tikzpicture}[scale=0.5]
				\draw[ultra thick, black!40!cyan] (0.5,-0.5) to[out=90,in=270] (-0.5,0.5);
				\draw[white, line width=1.5mm] (-0.5,-0.5) to[out=90,in=270] (0.5,0.5);
				\draw[ultra thick, black!10!orange] (-0.5,-0.5) to[out=90,in=270] (0.5,0.5);
			\end{tikzpicture}
			\end{array}}{\overset{R_k^+}{\longrightarrow}}\;\begin{array}{c}
		\begin{tikzpicture}[scale=1.3]
			\begin{scope}[shift={(-0.5,0)}]
				\begin{scope}[rotate=45]
				\draw[ultra thick, black!40!cyan] (-0.05,-0.05) -- (0.05,0.05) (-0.05,0.05) -- (0.05,-0.05);
				\end{scope}
			\end{scope}
			\begin{scope}[shift={(0.5,0)}]
				\draw[ultra thick, black!10!orange] (-0.05,-0.05) -- (0.05,0.05) (-0.05,0.05) -- (0.05,-0.05);
			\end{scope}
			\draw[thick] (-2,-0.75) -- (-2,0.75) (-1.75,-0.75) -- (-1.75,0.75) (-1.5,-0.75) -- (-1.5,0.75);
			\draw[ultra thick, black!40!red] (-1.25,0.75) -- (-1.25,0) to[out=270,in=180] (-0.5,-0.75) to[out=0,in=270] (0.25,0) to[out=90,in=180] (0.5,0.25) to[out=0,in=90] (0.75,0) -- (0.75,-0.75); 
			\draw[ultra thick, black!40!green] (-1,0.75) -- (-1,0) to[out=270,in=180] (-0.5,-0.5) to[out=0,in=270] (0,0) to[out=90,in=180] (0.5,0.5) to[out=0,in=90] (1,0) -- (1,-0.75);
			\draw[ultra thick, black!40!blue] (-0.75,0.75) -- (-0.75,0) to[out=270,in=180] (-0.5,-0.25) to[out=0,in=270] (-0.25,0) to[out=90,in=180] (0.5,0.75) to[out=0,in=90] (1.25,0) -- (1.25,-0.75);
			\draw[thick] (2,-0.75) -- (2,0.75) (1.75,-0.75) -- (1.75,0.75) (1.5,-0.75) -- (1.5,0.75);
			\foreach \x/\t in {-2/$a_1$, -1.75/$a_2$, -1.5/$a_3$, -1.25/$b_1'$, -1/$b_2'$, -0.75/$b_3'$, -0.25/$b_3$, 0/$b_2$, 0.25/$b_1$, 0.75/$b_1''$, 1/$b_2''$, 1.25/$b_3''$, 1.5/$c_1$, 1.75/$c_2$, 2/$c_3$}
			{
				\draw[fill=white, draw=\myblue] (\x,0) circle (0.1);
				\node at (\x,0) {\scalebox{0.5}{\t}};
			}
		\end{tikzpicture}
		\end{array}\,,\\
		&R_k^+:\quad \left(\begin{array}{c}
			\left\{{\bf D}_{k-1},\quad {\bf D}_k,\quad {\bf D}_{k+1}\right\}\\
			\bf U\\
			\bf L
		\end{array}\right)\mapsto \left(\begin{array}{c}
		\left\{{\bf D}_{k-1}\cup{\bf D}_k',\quad {\bf D}_k^R,\quad {\bf D}_k''\cup{\bf D}_{k+1}\right\}\\
		{\bf U}\big|_{{\bf D}_k={\bf D}_k'}\cup \, \bigcup_{p\in {\bf D}_k}(p\cap p'')\\
		{\bf L}\big|_{{\bf D}_k={\bf D}_k''}\cup \, \bigcup_{p\in {\bf D}_k}(p\cap p')
		\end{array}\right)\,,\\
		&\begin{array}{c}
			\begin{tikzpicture}[scale=1.3]
				\begin{scope}[shift={(-0.5,0)}]
					\draw[ultra thick, black!10!orange] (-0.05,-0.05) -- (0.05,0.05) (-0.05,0.05) -- (0.05,-0.05);
				\end{scope}
				\begin{scope}[shift={(0.5,0)}]
					\begin{scope}[rotate=45]
					\draw[ultra thick, black!40!cyan] (-0.05,-0.05) -- (0.05,0.05) (-0.05,0.05) -- (0.05,-0.05);
					\end{scope}
				\end{scope}
				\draw[thick] (-1.25,-0.75) -- (-1.25,0.75) (-1,-0.75) -- (-1,0.75) (-0.75,-0.75) -- (-0.75,0.75);
				\draw[ultra thick, black!40!red] (-0.25,-0.75) -- (-0.25,0.75); 
				\draw[ultra thick, black!40!green] (0,-0.75) -- (0,0.75);
				\draw[ultra thick, black!40!blue] (0.25,-0.75) -- (0.25,0.75);
				\draw[thick] (1.25,-0.75) -- (1.25,0.75) (1,-0.75) -- (1,0.75) (0.75,-0.75) -- (0.75,0.75);
				\foreach \x/\t in {-1.25/$a_1$, -1/$a_2$, -0.75/$a_3$, -0.25/$b_1$, 0/$b_2$, 0.25/$b_3$, 0.75/$c_1$, 1/$c_2$, 1.25/$c_3$}
				{
					\draw[fill=white, draw=\myblue] (\x,0) circle (0.1);
					\node at (\x,0) {\scalebox{0.5}{\t}};
				}
			\end{tikzpicture}
		\end{array}\;\underset{\begin{array}{c}
			\begin{tikzpicture}[scale=0.5]
				\draw[ultra thick, black!10!orange] (-0.5,-0.5) to[out=90,in=270] (0.5,0.5);
				\draw[white, line width=1.5mm] (0.5,-0.5) to[out=90,in=270] (-0.5,0.5);
				\draw[ultra thick, black!40!cyan] (0.5,-0.5) to[out=90,in=270] (-0.5,0.5);
			\end{tikzpicture}
			\end{array}}{\overset{R_k^-}{\longrightarrow}}\;\begin{array}{c}
			\begin{tikzpicture}[scale=1.3,yscale=-1]
				\begin{scope}[shift={(-0.5,0)}]
					\begin{scope}[rotate=45]
					\draw[ultra thick, black!40!cyan] (-0.05,-0.05) -- (0.05,0.05) (-0.05,0.05) -- (0.05,-0.05);
					\end{scope}
				\end{scope}
				\begin{scope}[shift={(0.5,0)}]
					\draw[ultra thick, black!10!orange] (-0.05,-0.05) -- (0.05,0.05) (-0.05,0.05) -- (0.05,-0.05);
				\end{scope}
				\draw[thick] (-2,-0.75) -- (-2,0.75) (-1.75,-0.75) -- (-1.75,0.75) (-1.5,-0.75) -- (-1.5,0.75);
				\draw[ultra thick, black!40!red] (-1.25,0.75) -- (-1.25,0) to[out=270,in=180] (-0.5,-0.75) to[out=0,in=270] (0.25,0) to[out=90,in=180] (0.5,0.25) to[out=0,in=90] (0.75,0) -- (0.75,-0.75); 
				\draw[ultra thick, black!40!green] (-1,0.75) -- (-1,0) to[out=270,in=180] (-0.5,-0.5) to[out=0,in=270] (0,0) to[out=90,in=180] (0.5,0.5) to[out=0,in=90] (1,0) -- (1,-0.75);
				\draw[ultra thick, black!40!blue] (-0.75,0.75) -- (-0.75,0) to[out=270,in=180] (-0.5,-0.25) to[out=0,in=270] (-0.25,0) to[out=90,in=180] (0.5,0.75) to[out=0,in=90] (1.25,0) -- (1.25,-0.75);
				\draw[thick] (2,-0.75) -- (2,0.75) (1.75,-0.75) -- (1.75,0.75) (1.5,-0.75) -- (1.5,0.75);
				\foreach \x/\t in {-2/$a_1$, -1.75/$a_2$, -1.5/$a_3$, -1.25/$b_1'$, -1/$b_2'$, -0.75/$b_3'$, -0.25/$b_3$, 0/$b_2$, 0.25/$b_1$, 0.75/$b_1''$, 1/$b_2''$, 1.25/$b_3''$, 1.5/$c_1$, 1.75/$c_2$, 2/$c_3$}
				{
					\draw[fill=white, draw=\myblue] (\x,0) circle (0.1);
					\node at (\x,0) {\scalebox{0.5}{\t}};
				}
			\end{tikzpicture}
		\end{array}\,,\\
		&R_k^-:\quad \left(\begin{array}{c}
			\left\{{\bf D}_{k-1},\quad {\bf D}_k,\quad {\bf D}_{k+1}\right\}\\
			\bf U\\
			\bf L
		\end{array}\right)\mapsto \left(\begin{array}{c}
			\left\{{\bf D}_{k-1}\cup{\bf D}_k',\quad {\bf D}_k^R,\quad {\bf D}_k''\cup{\bf D}_{k+1}\right\}\\
			{\bf U}\big|_{{\bf D}_k={\bf D}_k''}\cup \, \bigcup_{p\in {\bf D}_k}(p\cap p')\\
			{\bf L}\big|_{{\bf D}_k={\bf D}_k'}\cup \, \bigcup_{p\in {\bf D}_k}(p\cap p'')
		\end{array}\right)\,.
	\end{aligned}
\end{equation} 

So we see that braid morphisms $R_k^{\pm}$ correspond to the following simple operations on arcade sets.
Dot subset ${\bf D}_k$ acquires two copies ${\bf D}_k'$ and ${\bf D}_k''$ that are glued from the right to ${\bf D}_{k-1}$ and from the left to ${\bf D}_{k+1}$ respectively.
Arcs are moved accordingly, and we obtain new arcs connecting set ${\bf D}_k$ with ${\bf D}_{k}'$ and ${\bf D}_k''$.
The ordering of dots in the very ${\bf D}_k$ is reversed to ${\bf D}_k^R$ eventually.

Furthermore we will need one more structure introduced on arcs.
We call an arc end point \emph{chipped} if this is the only arc connected to this particular dot.
In the example \eqref{arc_examp} arc $5\cap 6$ is chipped on both ends, whereas arcs $1\cap 3$ and $3\cap4$ are chipped on a single end.
Now again we consider twists $R_k^{\pm}$ and if arcs are chipped at $b_1$ or $c_1$ we introduce extra rules for respective morphisms.
The origin of this rule is the following.
When we consider satellite links their strands go along the strand of the base knot $K$.
From the arc language prospective this looks like their strands do not end on the chipped end, rather go inside, or peel off from, the respective puncture representing a strand of the base knot.
For example,
\begin{equation}
	\begin{array}{c}
		\begin{tikzpicture}
			\draw[ultra thick, burgundy] (1,-1) -- (1,-0.5) to[out=90,in=270] (0,0.5) -- (0,1) (2,-1) -- (2,1);
			\draw[white, line width=1.7mm] (0,-1) -- (0,-0.5) to[out=90,in=270] (1,0.5) -- (1,1);
			\draw[ultra thick, burgundy] (0,-1) -- (0,-0.5) to[out=90,in=270] (1,0.5) -- (1,1);
			\draw[white, line width = 1.2mm] (1.3,1) -- (1.3,0.5) to[out=270,in=90] (0.3,-0.5) to[out=-30,in=210] (2.3,-0.5) -- (2.3,1);
			\draw (1.3,1) -- (1.3,0.5) to[out=270,in=90] (0.3,-0.5) to[out=-30,in=210] (2.3,-0.5) -- (2.3,1);
		\end{tikzpicture}
	\end{array}=\begin{array}{c}
	\begin{tikzpicture}
		\draw[ultra thick, burgundy] (1,-1) -- (1,-0.5) to[out=90,in=270] (0,0.5) -- (0,1) (2,-1) -- (2,1);
		\draw[white, line width=1.7mm] (0,-1) -- (0,-0.5) to[out=90,in=270] (1,0.5) -- (1,1);
		\draw[ultra thick, burgundy] (0,-1) -- (0,-0.5) to[out=90,in=270] (1,0.5) -- (1,1);
		\draw[white, line width = 1.2mm] (1.3,1) -- (1.3,0.5) to[out=-30,in=210] (2.3,0.5) -- (2.3,1);
		\draw (1.3,1) -- (1.3,0.5) to[out=-30,in=210] (2.3,0.5) -- (2.3,1);
	\end{tikzpicture}
	\end{array},\quad 
	R_{12}^+\left[\begin{array}{c}
		\begin{tikzpicture}[scale=0.7]
			\begin{scope}[shift={(0,0)}]
				\draw[ultra thick, orange] (-0.1,-0.1) -- (0.1,0.1) (-0.1,0.1) -- (0.1,-0.1);
			\end{scope}
			\begin{scope}[shift={(1,0)}]
				\draw[ultra thick, orange] (-0.1,-0.1) -- (0.1,0.1) (-0.1,0.1) -- (0.1,-0.1);
			\end{scope}
			\begin{scope}[shift={(2,0)}]
				\draw[ultra thick, orange] (-0.1,-0.1) -- (0.1,0.1) (-0.1,0.1) -- (0.1,-0.1);
			\end{scope}
			\draw[thick] (0.3,0) to[out=270,in=180] (1.3,-0.5) to[out=0,in=270] (2.3,0); 
		\end{tikzpicture}
	\end{array}\right]=\begin{array}{c}
	\begin{tikzpicture}[scale=0.7]
		\begin{scope}[shift={(0,0)}]
			\draw[ultra thick, orange] (-0.1,-0.1) -- (0.1,0.1) (-0.1,0.1) -- (0.1,-0.1);
		\end{scope}
		\begin{scope}[shift={(1,0)}]
			\draw[ultra thick, orange] (-0.1,-0.1) -- (0.1,0.1) (-0.1,0.1) -- (0.1,-0.1);
		\end{scope}
		\begin{scope}[shift={(2,0)}]
			\draw[ultra thick, orange] (-0.1,-0.1) -- (0.1,0.1) (-0.1,0.1) -- (0.1,-0.1);
		\end{scope}
		\draw[thick] (1.3,0) to[out=270,in=180] (1.8,-0.5) to[out=0,in=270] (2.3,0); 
	\end{tikzpicture}
	\end{array}\,.
\end{equation}

We introduce ``chipping indices'' $g_{b,c}$ for the left most dots $b_1$ and $c_1$ as the number of upper arcs connected to the respective dot minus the number of lower arcs ending on this dot.
So the extra rules applied after morphisms $R_k^{\pm}$ are:
\begin{enumerate}
	\item For $R_k^+$ if $g_b=-1$ final ${\bf U}\mapsto {\bf U}\setminus\{b_1\cap b_1''\}$, ${\bf L}\mapsto {\bf L}\setminus\{b_1\cap b_1'\}$.
	\item For $R_k^+$ if $g_c\neq 0$ we add extra point ${\bf D}_k\mapsto c_1'\cup{\bf D}_k$ and an extra arc ${\bf U}\mapsto {\bf U}\cup\{c_1\cap c_1'\}$.
	\item For $R_k^-$ if $g_b=1$ final ${\bf U}\mapsto {\bf U}\setminus\{b_1\cap b_1'\}$, ${\bf L}\mapsto {\bf L}\setminus\{b_1\cap b_1''\}$.
	\item For $R_k^-$ if $g_c\neq 0$ we add extra point ${\bf D}_k\mapsto c_1'\cup{\bf D}_k$ and an extra arc ${\bf L}\mapsto {\bf L}\cup\{c_1\cap c_1'\}$.
\end{enumerate}

This technique is somewhat similar to so called crossing-less matching representation of knot diagrams \cite{2004math.....11015M,Aganagic:2023amh}.

\section{A brief reminder on \texorpdfstring{$U_q(\fs\fu_n)$}{}, RT formalism and brackets}\label{sec:RT}

\subsection{Reshetikhin-Turaev formalism}

To calculate (partially) HOMFLY-PT polynomials we follow the Reshetikhin-Turaev (RT) formalism \cite{RT1,RT2,1112.2654,Mironov:2015aia,Mironov:2015qma,Anokhina:2024lbn,Galakhov:2014sha}.
The calculation is performed for a braided tensor category of $U_q(\fg)$ where different link elements are assigned to functors.
Let us here remind this identification briefly.

For a given link diagram one chooses a time flow direction and orientations on all link components.
Here we assume that all our link diagrams are oriented in such a way that time flows upwards.

To a sheaf of link strands one assigns an element from a tensor power of representations.
Depending on if a strand orientation matches the time flow or is opposite the respective tensor multiplier corresponds to the representation or its complex conjugate:
\begin{equation}
    \begin{array}{c}
	\begin{tikzpicture}
	    \draw[-stealth] (0,0) -- (0,0.5);
	    \draw[stealth-] (0.8,0) -- (0.8,0.5);
	    \draw[-stealth] (1.6,0) -- (1.6,0.5);
	    \draw[stealth-] (2.4,0) -- (2.4,0.5);
	    \draw[-stealth] (3.2,0) -- (3.2,0.5);
	    \node[left] at (0,0.25) {$\scriptstyle Q_1$};
	    \node[left] at (0.8,0.25) {$\scriptstyle Q_2$};
	    \node[left] at (1.6,0.25) {$\scriptstyle Q_3$};
	    \node[left] at (2.4,0.25) {$\scriptstyle Q_4$};
	    \node[left] at (3.2,0.25) {$\scriptstyle Q_5$};
	\end{tikzpicture}
    \end{array}\;\mapsto\; Q_1\otimes \bar Q_2\otimes Q_3\otimes\bar Q_4\otimes Q_5\,.
\end{equation}

To intersections one assigns R-matrix functors:
\begin{equation}
    \begin{array}{c}
	\begin{tikzpicture}[scale=0.5]
	    \draw[-stealth] (0.5,-0.5) to[out=90,in=270] (-0.5,0.5);
	    \draw[white, line width = 0.7mm] (-0.5,-0.5) to[out=90,in=270] (0.5,0.5);
	    \draw[-stealth] (-0.5,-0.5) to[out=90,in=270] (0.5,0.5);
	    \node[left] at (-0.5,-0.5) {$\scriptstyle Q_1$};
	    \node[right] at (0.5,-0.5) {$\scriptstyle Q_2$};
	\end{tikzpicture}
    \end{array}\;\mapsto\;\left(R:\,Q_1\otimes Q_2\to Q_2\otimes Q_1\right),\quad 
    \begin{array}{c}
	\begin{tikzpicture}[scale=0.5, xscale=-1]
	    \draw[-stealth] (0.5,-0.5) to[out=90,in=270] (-0.5,0.5);
	    \draw[white, line width = 0.7mm] (-0.5,-0.5) to[out=90,in=270] (0.5,0.5);
	    \draw[-stealth] (-0.5,-0.5) to[out=90,in=270] (0.5,0.5);
	    \node[right] at (-0.5,-0.5) {$\scriptstyle Q_2$};
	    \node[left] at (0.5,-0.5) {$\scriptstyle Q_1$};
	\end{tikzpicture}
    \end{array}\;\mapsto\;\left(R^{-1}:\,Q_1\otimes Q_2\to Q_2\otimes Q_1\right)\,.
\end{equation}

Turning points, ``caps'' and ``cups'' correspond to the following functors, where $\varnothing$ is a trivial representation:
\begin{equation}
    \begin{array}{c}
	\begin{tikzpicture}[scale=0.5]
	    \draw[-stealth] (-0.5,0) to[out=90,in=180] (0,0.7) to[out=0,in=90] (0.5,0);
	    \node[left] at (-0.5,0) {$\scriptstyle Q$};
	\end{tikzpicture}
    \end{array}\;\mapsto\;(Q\otimes \bar Q\to \varnothing),\quad \begin{array}{c}
	\begin{tikzpicture}[scale=0.5,yscale=-1]
	    \draw[-stealth] (-0.5,0) to[out=90,in=180] (0,0.7) to[out=0,in=90] (0.5,0);
	    \node[left] at (-0.5,0) {$\scriptstyle Q$};
	\end{tikzpicture}
    \end{array}\;\mapsto\;(\varnothing \to Q\otimes \bar Q),
\end{equation}

Additionally one might extend this formalism with trivalent junctions corresponding to $q$-Clebsh-Gordan functors:
\begin{equation}
    \begin{array}{c}
	\begin{tikzpicture}[scale=0.5]
	    \draw[-stealth] (-0.5,-0.5) to[out=90,in=210] (0,0);
	    \draw[-stealth] (0.5,-0.5) to[out=90,in=330] (0,0);
	    \draw[-stealth] (0,0) -- (0,0.5);
	    \node[left] at (-0.5,-0.5) {$\scriptstyle Q_1$};
	    \node[right] at (0.5,-0.5) {$\scriptstyle Q_2$};
	    \node[right] at (0,0.5) {$\scriptstyle Q_3$};
	\end{tikzpicture}
    \end{array}\;\mapsto\;(Q_1\otimes Q_2\to Q_3),\quad \begin{array}{c}
	\begin{tikzpicture}[scale=0.5, yscale=-1]
	    \draw[stealth-] (-0.5,-0.5) to[out=90,in=210] (0,0);
	    \draw[stealth-] (0.5,-0.5) to[out=90,in=330] (0,0);
	    \draw[stealth-] (0,0) -- (0,0.5);
	    \node[left] at (-0.5,-0.5) {$\scriptstyle Q_1$};
	    \node[right] at (0.5,-0.5) {$\scriptstyle Q_2$};
	    \node[right] at (0,0.5) {$\scriptstyle Q_3$};
	\end{tikzpicture}
    \end{array}\;\mapsto\;(Q_3\to Q_1\otimes Q_2)\,.
\end{equation}


\subsection{Algebra \texorpdfstring{$U_q(\fs\fu_n)$}{} and simple reps}

Algebra $U_q(\fs\fu_n)$ is defined by the Cartan matrix $a_{ij}=2\delta_{i,j}-\delta_{i,j+1}-\delta_{i,j-1}$, $i,j=1,\ldots,n-1$ (see e.g. \cite{etingof1998lectures}) and by the following relations:\footnote{We use the following notation for quantum numbers $[n]_q=\frac{q^n-q^{-n}}{q-q^{-1}}$.}
\begin{equation}
	\begin{split}
		q^{h_j}e_iq^{-h_j}=q^{a_{ij}}e_i,\quad q^{h_j}f_iq^{-h_j}=q^{-a_{ij}}f_i,\quad \left[e_i,f_j\right]=\delta_{ij}\frac{q^{h_i}-q^{-h_i}}{q-q^{-1}}\,,\\
		[e_i,e_j]=0,\quad [f_i,f_j]=0,\quad |i-j|>1\,,\\
		e_i^2e_j-[2]_qe_ie_je_i+e_je_i^2=0,\;f_i^2f_j-[2]_qf_if_jf_i+f_jf_i^2=0,\quad |i-j|=1\,.
	\end{split}
\end{equation}

The Hopf structure on this algebra is given by the following co-product structures $\Delta$, $\tilde \Delta$:
\begin{equation}
	\begin{split}
		\Delta(e_i)=e_i\otimes q^{\frac{h_i}{2}}+q^{-\frac{h_i}{2}}\otimes e_i,\quad \Delta(f_i)=f_i\otimes q^{\frac{h_i}{2}}+q^{-\frac{h_i}{2}}\otimes f_i,\quad \Delta(h_i)=h_i\otimes 1+1\otimes h_i\,;\\
		\tilde\Delta(e_i)=e_i\otimes q^{-\frac{h_i}{2}}+q^{\frac{h_i}{2}}\otimes e_i,\quad \tilde\Delta(f_i)=f_i\otimes q^{-\frac{h_i}{2}}+q^{\frac{h_i}{2}}\otimes f_i,\quad \tilde\Delta(h_i)=h_i\otimes 1+1\otimes h_i\,.
	\end{split}
\end{equation}

The R-functor inquired by the RT formalism is constructed $R=P\check R$ from the R-matrix $\check R$, where $P$ permutes tensor factors.
The R-matrix may be calculated in a universal way \cite{khoroshkin1991universal}.
Also, in practice, for small representation it is also convenient to construct it from defining intertwining relations:
\begin{equation}\label{Rmat}
    \check R\circ \Delta=\tilde\Delta\circ\check R\,.
\end{equation}
If appropriate boundary conditions are chosen:
\begin{equation}
	\check R=e^{a^{-1}_{ij}h_i\otimes h_j}\left(1+O(e\otimes f)\right)\,,
\end{equation}
where by $O(e\otimes f)$ we imply a series elements of which correspond to an action by Borel positive generators on the first tensor component and by negative ones on the second component, explicit expression for $\check R$ could be constructed uniquely.

The smallest representations of $U_q(\fs\fu_n)$ of dimension $n$ are fundamental and anti-fundamental ones (for $U_q(\fs\fu_2)$ these reps are isomorphic naturally).
We will use explicitly chosen bases for these representations, so that generators are given by the following matrices (where $E_{i,j}$ is a unit matrix):
\begin{equation}
	\begin{split}
		\mbox{fundamental }\Box: & \quad |1\rangle, \; |2\rangle,\;\ldots,\; |n\rangle\,, \\
		&e_a=E_{a,a+1},\quad f_a=E_{a+1,a},\quad h_a=E_{a,a}-E_{a+1,a+1},\quad a=1,\ldots,n-1\,,\\
		&\mbox{with the highest weight vector }\vec h|1\rangle=(1,0,\ldots,0)|1\rangle\,,\\
		\mbox{anti-fundamental }\bar \Box: & \quad |\bar 1\rangle, \; |\bar 2\rangle,\;\ldots,\; |\bar n\rangle,\\
		&e_a=E_{a+1,a},\quad f_a=E_{a,a+1},\quad h_a=-E_{a,a}+E_{a+1,a+1},\quad a=1,\ldots,n-1\,,\\
		&\mbox{with the highest weight vector }\vec h|\bar n\rangle=(0,\ldots,0,1)|\bar n\rangle\,,\,.
	\end{split}
\end{equation}

For this choice of representations the following R-functors could be calculated:
\begin{equation}\label{Rmatrix}
	\begin{split}
		R=P\check R=\begin{array}{c}
			\begin{tikzpicture}[scale=0.8]
				\draw[thick, -stealth] (-0.5,-0.5) -- (0.5,0.5);
				\draw[thick, -stealth] (-0.1,0.1) -- (-0.5,0.5);
				\draw[thick] (0.1,-0.1) -- (0.5,-0.5);
			\end{tikzpicture}
		\end{array}=q^{-\frac{1}{n}}\left(
		\sum\lm_{i,j}q^{\delta_{ij}}
		\begin{array}{c}
			\begin{tikzpicture}[scale=0.8]
				\draw[thick] (-0.5,-0.5) -- (0.5,0.5) (0.5,-0.5) -- (-0.5,0.5);
				\node[below] at (-0.5,-0.5) {$\scriptstyle i$};
				\node[below] at (0.5,-0.5) {$\scriptstyle j$};
				\node[above] at (-0.5,0.5) {$\scriptstyle j$};
				\node[above] at (0.5,0.5) {$\scriptstyle i$};
			\end{tikzpicture}
		\end{array}+\left(q-q^{-1}\right)
		\sum\lm_{i>j}
		\begin{array}{c}
			\begin{tikzpicture}[scale=0.8]
				\draw[thick] (-0.5,-0.5) -- (0.5,0.5) (0.5,-0.5) -- (-0.5,0.5);
				\node[below] at (-0.5,-0.5) {$\scriptstyle i$};
				\node[below] at (0.5,-0.5) {$\scriptstyle j$};
				\node[above] at (-0.5,0.5) {$\scriptstyle i$};
				\node[above] at (0.5,0.5) {$\scriptstyle j$};
			\end{tikzpicture}
		\end{array}
		\right)\,,\\
		R^{-1}=\check R^{-1}P=\begin{array}{c}
			\begin{tikzpicture}[xscale=-1, scale=0.8]
				\draw[thick, -stealth] (-0.5,-0.5) -- (0.5,0.5);
				\draw[thick, -stealth] (-0.1,0.1) -- (-0.5,0.5);
				\draw[thick] (0.1,-0.1) -- (0.5,-0.5);
			\end{tikzpicture}
		\end{array}=q^{\frac{1}{n}}\left(
		\sum\lm_{i,j}q^{-\delta_{ij}}
		\begin{array}{c}
			\begin{tikzpicture}[scale=0.8]
				\draw[thick] (-0.5,-0.5) -- (0.5,0.5) (0.5,-0.5) -- (-0.5,0.5);
				\node[below] at (-0.5,-0.5) {$\scriptstyle i$};
				\node[below] at (0.5,-0.5) {$\scriptstyle j$};
				\node[above] at (-0.5,0.5) {$\scriptstyle j$};
				\node[above] at (0.5,0.5) {$\scriptstyle i$};
			\end{tikzpicture}
		\end{array}-\left(q-q^{-1}\right)
		\sum\lm_{i<j}
		\begin{array}{c}
			\begin{tikzpicture}[scale=0.8]
				\draw[thick] (-0.5,-0.5) -- (0.5,0.5) (0.5,-0.5) -- (-0.5,0.5);
				\node[below] at (-0.5,-0.5) {$\scriptstyle i$};
				\node[below] at (0.5,-0.5) {$\scriptstyle j$};
				\node[above] at (-0.5,0.5) {$\scriptstyle i$};
				\node[above] at (0.5,0.5) {$\scriptstyle j$};
			\end{tikzpicture}
		\end{array}
		\right)\,.
	\end{split}
\end{equation}
and the following closure functors ``caps'' and ``cups'':
\begin{equation}\label{cupscaps}
	\begin{aligned}
		\begin{array}{c}
			\begin{tikzpicture}[scale=0.8,xscale=-1]
				\draw[thick, postaction={decorate},decoration={markings, mark= at position 0.7 with {\arrow{stealth}}}] (-0.5,0) to[out=270,in=180] (0,-0.4) to[out=0,in=270] (0.5,0);
			\end{tikzpicture}
		\end{array}&=\sum\lm_{k=1}^{n}(-1)^kq^{-k+\frac{n+1}{2}}
		\begin{array}{c}
			\begin{tikzpicture}[scale=0.8]
				\draw[thick] (-0.5,0) to[out=270,in=180] (0,-0.4) to[out=0,in=270] (0.5,0);
				\node[above] at (-0.5,0) {$\scriptstyle k$};
				\node[above] at (0.5,0) {$\scriptstyle \bar k$};
			\end{tikzpicture}
		\end{array}\,,\\
		\begin{array}{c}
			\begin{tikzpicture}[scale=0.8,scale=-1]
				\draw[thick, postaction={decorate},decoration={markings, mark= at position 0.7 with {\arrow{stealth}}}] (-0.5,0) to[out=270,in=180] (0,-0.4) to[out=0,in=270] (0.5,0);
			\end{tikzpicture}
		\end{array}&=\sum\lm_{k=1}^{n}(-1)^kq^{k-\frac{n+1}{2}}
		\begin{array}{c}
			\begin{tikzpicture}[scale=0.8,yscale=-1]
				\draw[thick] (-0.5,0) to[out=270,in=180] (0,-0.4) to[out=0,in=270] (0.5,0);
				\node[below] at (-0.5,0) {$\scriptstyle \bar k$};
				\node[below] at (0.5,0) {$\scriptstyle k$};
			\end{tikzpicture}
		\end{array}\,,\\
		\begin{array}{c}
			\begin{tikzpicture}[scale=0.8]
				\draw[thick, postaction={decorate},decoration={markings, mark= at position 0.7 with {\arrow{stealth}}}] (-0.5,0) to[out=270,in=180] (0,-0.4) to[out=0,in=270] (0.5,0);
			\end{tikzpicture}
		\end{array}&=\sum\lm_{k=1}^{n}(-1)^kq^{k-\frac{n+1}{2}}
		\begin{array}{c}
			\begin{tikzpicture}[scale=0.8]
				\draw[thick] (-0.5,0) to[out=270,in=180] (0,-0.4) to[out=0,in=270] (0.5,0);
				\node[above] at (-0.5,0) {$\scriptstyle \bar k$};
				\node[above] at (0.5,0) {$\scriptstyle k$};
			\end{tikzpicture}
		\end{array}\,,\\
		\begin{array}{c}
			\begin{tikzpicture}[scale=0.8,yscale=-1]
				\draw[thick, postaction={decorate},decoration={markings, mark= at position 0.7 with {\arrow{stealth}}}] (-0.5,0) to[out=270,in=180] (0,-0.4) to[out=0,in=270] (0.5,0);
			\end{tikzpicture}
		\end{array}&=\sum\lm_{k=1}^{n}(-1)^kq^{-k+\frac{n+1}{2}}
		\begin{array}{c}
			\begin{tikzpicture}[scale=0.8,yscale=-1]
				\draw[thick] (-0.5,0) to[out=270,in=180] (0,-0.4) to[out=0,in=270] (0.5,0);
				\node[below] at (-0.5,0) {$\scriptstyle k$};
				\node[below] at (0.5,0) {$\scriptstyle \bar k$};
			\end{tikzpicture}
		\end{array}\,.
	\end{aligned}
\end{equation}
In these expressions indices on graph legs indicate particular vectors of fundamental or anti-fundamental representation respectively flowing in the corresponding channel.
To calculate the resulting HOMFLY polynomial one counts with $q$-weights coloring numbers of the knot diagram edges in the respective vertex state model, see e.g. \cite{Anokhina:2014jza}.
Alternatively, one simply substitutes the knot diagram elements with the formal graph sums and expands the result associatively. 
The resulting graphs contribute with a unit value only if colorings match for all legs across the edges, otherwise the contribution is zero.

Also we would like to consider the adjoint representation $\wedge^2\Box$ constructed in the basis of a tensor square.
This basis consists of the following orthonormal vectors:
\begin{equation}
    \frac{1}{\sqrt{q+q^{-1}}}\left(q^{\frac{1}{2}}|i\rangle\otimes|j\rangle-q^{-\frac{1}{2}}|j\rangle\otimes|i\rangle\right),\quad 1\leq j<i\leq n\,.
\end{equation}
Then we note that R-functors \eqref{Rmatrix} could be expanded as the following graphs where trivalent vertices imply corresponding $q$-Clebsh-Gordan coefficients:
\begin{equation}\label{bracket_gen}
    \begin{aligned}
	&\begin{array}{c}
			\begin{tikzpicture}[scale=0.8]
				\draw[thick, -stealth] (-0.5,-0.5) -- (0.5,0.5);
				\draw[thick, -stealth] (-0.1,0.1) -- (-0.5,0.5);
				\draw[thick] (0.1,-0.1) -- (0.5,-0.5);
			\end{tikzpicture}
	\end{array}=q^{\frac{n-1}{n}}\begin{array}{c}
			\begin{tikzpicture}[scale=0.8]
			    \draw[thick, -stealth] (-0.5,-0.5) to[out=45,in=315] (-0.5,0.5);
				\draw[thick, -stealth] (0.5,-0.5) to[out=135,in=245] (0.5,0.5);
			\end{tikzpicture}
	\end{array}-q^{-\frac{1}{n}}[2]_q\begin{array}{c}
			\begin{tikzpicture}[scale=0.8]
				\draw[thick, -stealth] (0,0.25) -- (-0.5,0.5);
				\draw[thick, -stealth] (0,0.25) -- (0.5,0.5);
				\draw[thick, -stealth] (-0.5,-0.5) -- (0,-0.25);
				\draw[thick, -stealth] (0.5,-0.5) -- (0,-0.25);
				\draw[ultra thick, -stealth,\myblue] (0,-0.25) -- (0,0.25);
				\node[right] at (0,0) {$\scriptstyle \wedge^2\Box$};
			\end{tikzpicture}
		\end{array}\,,\\
	&\begin{array}{c}
			\begin{tikzpicture}[scale=0.8,xscale=-1]
				\draw[thick, -stealth] (-0.5,-0.5) -- (0.5,0.5);
				\draw[thick, -stealth] (-0.1,0.1) -- (-0.5,0.5);
				\draw[thick] (0.1,-0.1) -- (0.5,-0.5);
			\end{tikzpicture}
	\end{array}=q^{-\frac{n-1}{n}}\begin{array}{c}
			\begin{tikzpicture}[scale=0.8]
			    \draw[thick, -stealth] (-0.5,-0.5) to[out=45,in=315] (-0.5,0.5);
				\draw[thick, -stealth] (0.5,-0.5) to[out=135,in=245] (0.5,0.5);
			\end{tikzpicture}
	\end{array}-q^{\frac{1}{n}}[2]_q\begin{array}{c}
			\begin{tikzpicture}[scale=0.8]
				\draw[thick, -stealth] (0,0.25) -- (-0.5,0.5);
				\draw[thick, -stealth] (0,0.25) -- (0.5,0.5);
				\draw[thick, -stealth] (-0.5,-0.5) -- (0,-0.25);
				\draw[thick, -stealth] (0.5,-0.5) -- (0,-0.25);
				\draw[ultra thick, -stealth,\myblue] (0,-0.25) -- (0,0.25);
				\node[right] at (0,0) {$\scriptstyle \wedge^2\Box$};
			\end{tikzpicture}
		\end{array}\,.\\
    \end{aligned}
\end{equation}


\subsection{Algebras \texorpdfstring{$U_q(\fs\fu_3)$}{} and \texorpdfstring{$U_q(\fs\fu_3)$}{}, Kauffman and Kuperberg brackets}
For $U_q(\fs\fu_2)$ $\wedge^2\Box\cong \varnothing$.
Trivial representation $\varnothing$ is a unit with respect to the tensor multiplication, therefore lines in graphs colored with the trivial representation could be simply erased.
Then it is natural to identify $\wedge^2\Box$ channels in \eqref{bracket_gen} with pairs of caps and cups.
Conforming normalizations with \eqref{cupscaps} we arrive to the Kauffman bracket:
\begin{equation}\label{Kauffman}
	\begin{aligned}
		\begin{array}{c}
			\begin{tikzpicture}[scale=0.6]
				\draw[thick, -stealth] (-0.5,-0.5) -- (0.5,0.5);
				\draw[thick, -stealth] (-0.1,0.1) -- (-0.5,0.5);
				\draw[thick] (0.1,-0.1) -- (0.5,-0.5);
			\end{tikzpicture}
		\end{array}=q^{\frac{1}{2}}\begin{array}{c}
			\begin{tikzpicture}[scale=0.6]
				\draw[thick] (-0.5,-0.5) to[out=45,in=315] (-0.5,0.5);
				\draw[thick] (0.5,-0.5) to[out=135,in=245] (0.5,0.5);
			\end{tikzpicture}
		\end{array}+q^{-\frac{1}{2}}\begin{array}{c}
			\begin{tikzpicture}[scale=0.6, rotate=90]
				\draw[thick] (-0.5,-0.5) to[out=45,in=315] (-0.5,0.5);
				\draw[thick] (0.5,-0.5) to[out=135,in=245] (0.5,0.5);
			\end{tikzpicture}
		\end{array},\quad \begin{array}{c}
			\begin{tikzpicture}[scale=0.6,xscale=-1]
				\draw[thick, -stealth] (-0.5,-0.5) -- (0.5,0.5);
				\draw[thick, -stealth] (-0.1,0.1) -- (-0.5,0.5);
				\draw[thick] (0.1,-0.1) -- (0.5,-0.5);
			\end{tikzpicture}
		\end{array}=q^{-\frac{1}{2}}\begin{array}{c}
		\begin{tikzpicture}[scale=0.6]
			\draw[thick] (-0.5,-0.5) to[out=45,in=315] (-0.5,0.5);
			\draw[thick] (0.5,-0.5) to[out=135,in=245] (0.5,0.5);
		\end{tikzpicture}
		\end{array}+q^{\frac{1}{2}}\begin{array}{c}
		\begin{tikzpicture}[scale=0.6, rotate=90]
			\draw[thick] (-0.5,-0.5) to[out=45,in=315] (-0.5,0.5);
			\draw[thick] (0.5,-0.5) to[out=135,in=245] (0.5,0.5);
		\end{tikzpicture}
		\end{array}\,.
	\end{aligned}
\end{equation}
In the final expansions arrows on graphs can be also erased as representations and complex conjugated representations are isomorphic for $U_q(\fs\fu_2)$.

In the case of $U_q(\fs\fu_3)$ there is a simplification $\wedge^2\Box\cong \bar\Box$, and comparing weights we can identify vectors $\bar\Box$ with vectors in the basis of $\Box^{\otimes 2}$:
\begin{equation}
    \begin{aligned}
	& |\bar 1\rangle=\frac{q^{\frac{1}{2}}|2\rangle\otimes|3\rangle-q^{-\frac{1}{2}}|3\rangle\otimes|2\rangle}{\sqrt{q+q^{-1}}},\quad |\bar 2\rangle=\frac{q^{\frac{1}{2}}|1\rangle\otimes|3\rangle-q^{-\frac{1}{2}}|3\rangle\otimes|1\rangle}{\sqrt{q+q^{-1}}},\quad |\bar 3\rangle=\frac{q^{\frac{1}{2}}|1\rangle\otimes|2\rangle-q^{-\frac{1}{2}}|2\rangle\otimes|1\rangle}{\sqrt{q+q^{-1}}}\,.
    \end{aligned}
\end{equation}
Under these circumstances the generic bracket \eqref{bracket_gen} becomes the Kuperberg bracket for $U_q(\fs\fu_3)$:
\begin{equation}\label{Kuperberg}
    \begin{aligned}
	&\begin{array}{c}
			\begin{tikzpicture}[scale=0.8]
				\draw[thick, -stealth] (-0.5,-0.5) -- (0.5,0.5);
				\draw[thick, -stealth] (-0.1,0.1) -- (-0.5,0.5);
				\draw[thick] (0.1,-0.1) -- (0.5,-0.5);
			\end{tikzpicture}
	\end{array}=q^{\frac{2}{3}}\begin{array}{c}
			\begin{tikzpicture}[scale=0.8]
			    \draw[thick, -stealth] (-0.5,-0.5) to[out=45,in=315] (-0.5,0.5);
				\draw[thick, -stealth] (0.5,-0.5) to[out=135,in=245] (0.5,0.5);
			\end{tikzpicture}
	\end{array}-q^{-\frac{1}{3}}[2]_q\begin{array}{c}
			\begin{tikzpicture}[scale=0.8]
				\draw[thick, -stealth] (0,0.25) -- (-0.5,0.5);
				\draw[thick, -stealth] (0,0.25) -- (0.5,0.5);
				\draw[thick, -stealth] (-0.5,-0.5) -- (0,-0.25);
				\draw[thick, -stealth] (0.5,-0.5) -- (0,-0.25);
				\draw[thick, -stealth] (0,0.25) -- (0,-0.25);
				\draw[fill=yellow] (0,-0.25) circle (0.07) (0,0.25) circle (0.07);
			\end{tikzpicture}
		\end{array}\,,\\
	&\begin{array}{c}
			\begin{tikzpicture}[scale=0.8,xscale=-1]
				\draw[thick, -stealth] (-0.5,-0.5) -- (0.5,0.5);
				\draw[thick, -stealth] (-0.1,0.1) -- (-0.5,0.5);
				\draw[thick] (0.1,-0.1) -- (0.5,-0.5);
			\end{tikzpicture}
	\end{array}=q^{-\frac{2}{3}}\begin{array}{c}
			\begin{tikzpicture}[scale=0.8]
			    \draw[thick, -stealth] (-0.5,-0.5) to[out=45,in=315] (-0.5,0.5);
				\draw[thick, -stealth] (0.5,-0.5) to[out=135,in=245] (0.5,0.5);
			\end{tikzpicture}
	\end{array}-q^{\frac{1}{3}}[2]_q\begin{array}{c}
			\begin{tikzpicture}[scale=0.8]
				\draw[thick, -stealth] (0,0.25) -- (-0.5,0.5);
				\draw[thick, -stealth] (0,0.25) -- (0.5,0.5);
				\draw[thick, -stealth] (-0.5,-0.5) -- (0,-0.25);
				\draw[thick, -stealth] (0.5,-0.5) -- (0,-0.25);
				\draw[thick, -stealth] (0,0.25) -- (0,-0.25);
				\draw[fill=yellow] (0,-0.25) circle (0.07) (0,0.25) circle (0.07);
			\end{tikzpicture}
		\end{array}\,,\\
    \end{aligned}
\end{equation}
where vertex $\!\!\begin{array}{c}
	\begin{tikzpicture}
		\draw[fill=yellow] (0,0) circle (0.08);
	\end{tikzpicture}
\end{array}\!\!$ denotes the respective Clebsh-Gordan coefficient insertion.
In a similar manner we identify $\wedge^2\bar\Box\cong \Box$:
\begin{equation}
    \begin{aligned}
	& |1\rangle=\frac{q^{\frac{1}{2}}|\bar 3\rangle\otimes|\bar 2\rangle-q^{-\frac{1}{2}}|\bar 2\rangle\otimes|\bar 3\rangle}{\sqrt{q+q^{-1}}},\quad |2\rangle=\frac{q^{\frac{1}{2}}|\bar 3\rangle\otimes|\bar 1\rangle-q^{-\frac{1}{2}}|\bar 1\rangle\otimes|\bar 3\rangle}{\sqrt{q+q^{-1}}},\quad |3\rangle=\frac{q^{\frac{1}{2}}|\bar 2\rangle\otimes|\bar 1\rangle-q^{-\frac{1}{2}}|\bar 1\rangle\otimes|\bar 2\rangle}{\sqrt{q+q^{-1}}}\,.
    \end{aligned}
\end{equation}
For such $U_q(\fs\fu_3)$ MOY graphs we derive the following two reduction rules allowing to eliminate a bigon and a ``box'':
\begin{equation}
    \begin{array}{c}
	\begin{tikzpicture}[scale=1]
	    \draw[thick,-stealth] (0,-0.5) -- (0,-0.25);
	    \draw[thick,-stealth] (0,0.25) -- (0,0.5);
	    \draw[thick,-stealth] (0,0.25) to[out=210,in=90] (-0.3,0) to[out=270,in=150] (0,-0.25);
	    \begin{scope}[xscale=-1]
	        \draw[thick,-stealth] (0,0.25) to[out=210,in=90] (-0.3,0) to[out=270,in=150] (0,-0.25);
	    \end{scope}
	    \draw[fill=yellow] (0,-0.25) circle (0.05) (0,0.25) circle (0.05);
	\end{tikzpicture}
    \end{array}=\begin{array}{c}
	\begin{tikzpicture}[scale=1]
	    \draw[thick,-stealth] (0,-0.4) -- (0,0.4);
	\end{tikzpicture}
    \end{array},\quad \begin{array}{c}
	\begin{tikzpicture}[scale=0.8]
	    \draw[thick,-stealth] (-0.3,-0.3) -- (-0.3,0.3);
	    \draw[thick,-stealth] (0.3,0.3) -- (-0.3,0.3);
	    \draw[thick,-stealth] (0.3,0.3) -- (0.3,-0.3);
	    \draw[thick,-stealth] (-0.3,-0.3) -- (0.3,-0.3);
	    \draw[thick,-stealth] (-0.3,-0.3) -- (-0.7,-0.7);
	    \draw[thick,-stealth] (-0.7,0.7) -- (-0.3,0.3);
	    \draw[thick,-stealth] (0.3,0.3) -- (0.7,0.7);
	    \draw[thick,-stealth] (0.7,-0.7) -- (0.3,-0.3);
	    \draw[fill=yellow] (-0.3,-0.3) circle (0.07) (-0.3,0.3) circle (0.07) (0.3,0.3) circle (0.07) (0.3,-0.3) circle (0.07);
	\end{tikzpicture}
    \end{array}=\frac{1}{[2]_q^2}\left(\begin{array}{c}
	\begin{tikzpicture}[scale=0.8]
	    \draw[thick,-stealth] (-0.5,0.5) to[out=315,in=45] (-0.5,-0.5);
	    \draw[thick,-stealth] (0.5,-0.5) to[out=135,in=225] (0.5,0.5);
	\end{tikzpicture}
    \end{array}+\begin{array}{c}
	\begin{tikzpicture}[scale=0.8,xscale=-1,rotate=90]
	    \draw[thick,-stealth] (-0.5,0.5) to[out=315,in=45] (-0.5,-0.5);
	    \draw[thick,-stealth] (0.5,-0.5) to[out=135,in=225] (0.5,0.5);
	\end{tikzpicture}
    \end{array}\right)\,.
\end{equation}
Unfortunately, for higher $2m$-gons we \emph{do not find} reducing relations, so we should consider them as new invariant tensors.

To compare these expressions with canonical expressions of the Kuperberg bracket \cite{sikora2005skein,kuperberg1996spiders,2012arXiv1207.0719H} we rescale the trivalent vertex:
\begin{equation}
    \begin{aligned}
	&\begin{array}{c}
	    \begin{tikzpicture}[scale=0.5]
		\draw[thick,-stealth] (0,0) -- (0.866025,-0.5);
		\draw[thick,-stealth] (0,0) -- (-0.866025,-0.5);
		\draw[thick,-stealth] (0,0) -- (0,1);
		\draw[fill=yellow] (0,0) circle (0.1);
	    \end{tikzpicture}
	\end{array}=\frac{1}{\sqrt{-[2]_q}}\begin{array}{c}
	    \begin{tikzpicture}[scale=0.5]
		\draw[thick,-stealth] (0,0) -- (0.866025,-0.5);
		\draw[thick,-stealth] (0,0) -- (-0.866025,-0.5);
		\draw[thick,-stealth] (0,0) -- (0,1);
	    \end{tikzpicture}
	\end{array},\quad
	\begin{array}{c}
	    \begin{tikzpicture}[scale=0.5, yscale=-1]
		\draw[thick,-stealth] (0,0) -- (0.866025,-0.5);
		\draw[thick,-stealth] (0,0) -- (-0.866025,-0.5);
		\draw[thick,-stealth] (0,0) -- (0,1);
		\draw[fill=yellow] (0,0) circle (0.1);
	    \end{tikzpicture}
	\end{array}=\frac{1}{\sqrt{-[2]_q}}\begin{array}{c}
	    \begin{tikzpicture}[scale=0.5, yscale=-1]
		\draw[thick,-stealth] (0,0) -- (0.866025,-0.5);
		\draw[thick,-stealth] (0,0) -- (-0.866025,-0.5);
		\draw[thick,-stealth] (0,0) -- (0,1);
	    \end{tikzpicture}
	\end{array}\,,\\
	&\begin{array}{c}
		\begin{tikzpicture}[scale=0.5]
			\draw[thick,stealth-] (0,0) -- (0.866025,-0.5);
			\draw[thick,stealth-] (0,0) -- (-0.866025,-0.5);
			\draw[thick,stealth-] (0,0) -- (0,1);
			\draw[fill=yellow] (0,0) circle (0.1);
		\end{tikzpicture}
	\end{array}=\frac{1}{\sqrt{-[2]_q}}\begin{array}{c}
		\begin{tikzpicture}[scale=0.5]
			\draw[thick,stealth-] (0,0) -- (0.866025,-0.5);
			\draw[thick,stealth-] (0,0) -- (-0.866025,-0.5);
			\draw[thick,stealth-] (0,0) -- (0,1);
		\end{tikzpicture}
	\end{array},\quad
	\begin{array}{c}
		\begin{tikzpicture}[scale=0.5, yscale=-1]
			\draw[thick,stealth-] (0,0) -- (0.866025,-0.5);
			\draw[thick,stealth-] (0,0) -- (-0.866025,-0.5);
			\draw[thick,stealth-] (0,0) -- (0,1);
			\draw[fill=yellow] (0,0) circle (0.1);
		\end{tikzpicture}
	\end{array}=\frac{1}{\sqrt{-[2]_q}}\begin{array}{c}
		\begin{tikzpicture}[scale=0.5, yscale=-1]
			\draw[thick,stealth-] (0,0) -- (0.866025,-0.5);
			\draw[thick,stealth-] (0,0) -- (-0.866025,-0.5);
			\draw[thick,stealth-] (0,0) -- (0,1);
		\end{tikzpicture}
	\end{array}\,.
    \end{aligned}
\end{equation}
for the new vertex we derive the following relations:
\begin{tcolorbox}
\begin{equation}
\begin{aligned}
    &\begin{array}{c}
			\begin{tikzpicture}[scale=0.8]
				\draw[thick, -stealth] (-0.5,-0.5) -- (0.5,0.5);
				\draw[thick, -stealth] (-0.1,0.1) -- (-0.5,0.5);
				\draw[thick] (0.1,-0.1) -- (0.5,-0.5);
			\end{tikzpicture}
	\end{array}=q^{\frac{2}{3}}\begin{array}{c}
			\begin{tikzpicture}[scale=0.8]
			    \draw[thick, -stealth] (-0.5,-0.5) to[out=45,in=315] (-0.5,0.5);
				\draw[thick, -stealth] (0.5,-0.5) to[out=135,in=245] (0.5,0.5);
			\end{tikzpicture}
	\end{array}+q^{-\frac{1}{3}}\begin{array}{c}
			\begin{tikzpicture}[scale=0.8]
				\draw[thick, -stealth] (0,0.25) -- (-0.5,0.5);
				\draw[thick, -stealth] (0,0.25) -- (0.5,0.5);
				\draw[thick, -stealth] (-0.5,-0.5) -- (0,-0.25);
				\draw[thick, -stealth] (0.5,-0.5) -- (0,-0.25);
				\draw[thick, -stealth] (0,0.25) -- (0,-0.25);
			\end{tikzpicture}
		\end{array},\quad \begin{array}{c}
			\begin{tikzpicture}[scale=0.8,xscale=-1]
				\draw[thick, -stealth] (-0.5,-0.5) -- (0.5,0.5);
				\draw[thick, -stealth] (-0.1,0.1) -- (-0.5,0.5);
				\draw[thick] (0.1,-0.1) -- (0.5,-0.5);
			\end{tikzpicture}
	\end{array}=q^{-\frac{2}{3}}\begin{array}{c}
			\begin{tikzpicture}[scale=0.8]
			    \draw[thick, -stealth] (-0.5,-0.5) to[out=45,in=315] (-0.5,0.5);
				\draw[thick, -stealth] (0.5,-0.5) to[out=135,in=245] (0.5,0.5);
			\end{tikzpicture}
	\end{array}+q^{\frac{1}{3}}\begin{array}{c}
			\begin{tikzpicture}[scale=0.8]
				\draw[thick, -stealth] (0,0.25) -- (-0.5,0.5);
				\draw[thick, -stealth] (0,0.25) -- (0.5,0.5);
				\draw[thick, -stealth] (-0.5,-0.5) -- (0,-0.25);
				\draw[thick, -stealth] (0.5,-0.5) -- (0,-0.25);
				\draw[thick, -stealth] (0,0.25) -- (0,-0.25);
			\end{tikzpicture}
		\end{array}\,,\\
    & \begin{array}{c}
	\begin{tikzpicture}[scale=0.8]
	    \draw[thick,-stealth] (0,-0.7) -- (0,-0.25);
	    \draw[thick,-stealth] (0,0.25) -- (0,0.7);
	    \draw[thick,-stealth] (0,0.25) to[out=210,in=90] (-0.3,0) to[out=270,in=150] (0,-0.25);
	    \begin{scope}[xscale=-1]
	        \draw[thick,-stealth] (0,0.25) to[out=210,in=90] (-0.3,0) to[out=270,in=150] (0,-0.25);
	    \end{scope}
	\end{tikzpicture}
    \end{array}=-[2]_q\begin{array}{c}
	\begin{tikzpicture}[scale=0.8]
	    \draw[thick,-stealth] (0,-0.4) -- (0,0.4);
	\end{tikzpicture}
    \end{array},\quad \begin{array}{c}
	\begin{tikzpicture}[scale=0.8]
	    \draw[thick,-stealth] (-0.3,-0.3) -- (-0.3,0.3);
	    \draw[thick,-stealth] (0.3,0.3) -- (-0.3,0.3);
	    \draw[thick,-stealth] (0.3,0.3) -- (0.3,-0.3);
	    \draw[thick,-stealth] (-0.3,-0.3) -- (0.3,-0.3);
	    \draw[thick,-stealth] (-0.3,-0.3) -- (-0.7,-0.7);
	    \draw[thick,-stealth] (-0.7,0.7) -- (-0.3,0.3);
	    \draw[thick,-stealth] (0.3,0.3) -- (0.7,0.7);
	    \draw[thick,-stealth] (0.7,-0.7) -- (0.3,-0.3);
	\end{tikzpicture}
    \end{array}=\left(\begin{array}{c}
	\begin{tikzpicture}[scale=0.8]
	    \draw[thick,-stealth] (-0.5,0.5) to[out=315,in=45] (-0.5,-0.5);
	    \draw[thick,-stealth] (0.5,-0.5) to[out=135,in=225] (0.5,0.5);
	\end{tikzpicture}
    \end{array}+\begin{array}{c}
	\begin{tikzpicture}[scale=0.8,xscale=-1,rotate=90]
	    \draw[thick,-stealth] (-0.5,0.5) to[out=315,in=45] (-0.5,-0.5);
	    \draw[thick,-stealth] (0.5,-0.5) to[out=135,in=225] (0.5,0.5);
	\end{tikzpicture}
    \end{array}\right)\,.
\end{aligned}
\end{equation}
\end{tcolorbox}


\section{Exercises with Kauffman bracket for \texorpdfstring{$U_q(\fs\fu_2)$}{}}\label{sec:Kauff_ex}

First of all we would like to test the proposed arcade algorithms on a wider sampling of knots.
By tests we imply that we apply the braid group group action on arcades to planarize links and obtain testable relations among them.
Here we consider only the case $\fs\fu_2$ as the respective Kauffman bracket \eqref{Kauffman} is simpler, therefore relations are simpler as well.
Eventually, in each case we derive classical A-polynomials in the quasi-classical limit and compare them to known ones from the literature.

\subsection{Trefoil knot \texorpdfstring{$3_1$}{}}\label{sec:3_1}

First we start with the trefoil knot $3_1$ example of \cite{Galakhov:2024eco}.
The basic link symbols are constructed from links $L$ and $\mu$ (see Fig.\ref{fig:3_1_aux}).
\begin{figure}[ht!]
	\centering
		\begin{tikzpicture}[scale=0.6]
			\begin{scope}[yscale=0.3]
				\draw ([shift={(0:0.8)}]0,0) arc (0:180:0.8);
			\end{scope}
			\begin{scope}[shift={(0,3)}]
				\begin{scope}[yscale=0.3]
					\draw ([shift={(0:0.8)}]0,0) arc (0:180:0.8);
				\end{scope}
			\end{scope}
			\begin{scope}[shift={(0,3)}]
				\begin{scope}[yscale=0.3]
					\draw ([shift={(0:0.3)}]-1.5,0) arc (0:180:0.3);
				\end{scope}
			\end{scope}
			\begin{scope}
				\draw[white, line width = 1.7mm] (0.5,0) to[out=90,in=270] (-0.5,1);
				\draw[ultra thick, burgundy] (0.5,0) to[out=90,in=270] (-0.5,1);
				\draw[white, line width = 1.7mm] (-0.5,0) to[out=90,in=270] (0.5,1);
				\draw[ultra thick, burgundy] (-0.5,0) to[out=90,in=270] (0.5,1);
			\end{scope}
			\begin{scope}[shift={(0,1)}]
				\draw[ultra thick, burgundy] (0.5,0) to[out=90,in=270] (-0.5,1);
				\draw[white, line width = 1.7mm] (-0.5,0) to[out=90,in=270] (0.5,1);
				\draw[ultra thick, burgundy] (-0.5,0) to[out=90,in=270] (0.5,1);
			\end{scope}
			\begin{scope}[shift={(0,2)}]
				\draw[ultra thick, burgundy] (0.5,0) to[out=90,in=270] (-0.5,1);
				\draw[white, line width = 1.7mm] (-0.5,0) to[out=90,in=270] (0.5,1);
				\draw[ultra thick, burgundy] (-0.5,0) to[out=90,in=270] (0.5,1);
			\end{scope}
			\draw[white, line width = 1.7mm] (-0.5,3) to[out=90,in=0] (-1,3.5) to[out=180,in=90] (-1.5,3) -- (-1.5,0) to[out=270,in=180] (-1,-0.5) to[out=0,in=270] (-0.5,0);
			\draw[ultra thick, burgundy] (-0.5,3) to[out=90,in=0] (-1,3.5) to[out=180,in=90] (-1.5,3) -- (-1.5,0) to[out=270,in=180] (-1,-0.5) to[out=0,in=270] (-0.5,0);
			\begin{scope}[xscale=-1]
				\draw[white, line width = 1.7mm] (-0.5,3) to[out=90,in=0] (-1,3.5) to[out=180,in=90] (-1.5,3) -- (-1.5,0) to[out=270,in=180] (-1,-0.5) to[out=0,in=270] (-0.5,0);
				\draw[ultra thick, burgundy] (-0.5,3) to[out=90,in=0] (-1,3.5) to[out=180,in=90] (-1.5,3) -- (-1.5,0) to[out=270,in=180] (-1,-0.5) to[out=0,in=270] (-0.5,0);
			\end{scope}
			\begin{scope}[yscale=0.3]
				\draw[white, line width = 1mm] ([shift={(180:0.8)}]0,0) arc (180:360:0.8);
				\draw ([shift={(180:0.8)}]0,0) arc (180:360:0.8);
			\end{scope}
			\begin{scope}[shift={(0,3)}]
				\begin{scope}[yscale=0.3]
					\draw[white, line width = 1mm] ([shift={(180:0.8)}]0,0) arc (180:360:0.8);
					\draw ([shift={(180:0.8)}]0,0) arc (180:360:0.8);
				\end{scope}
			\end{scope}
			\begin{scope}[shift={(0,3)}]
				\begin{scope}[yscale=0.3]
					\draw[white, line width = 1mm] ([shift={(180:0.3)}]-1.5,0) arc (180:360:0.3);
					\draw ([shift={(180:0.3)}]-1.5,0) arc (180:360:0.3);
				\end{scope}
			\end{scope}
			\node[right] at (1.5,0) {$\scriptstyle L'$};
			\node[right] at (1.5,3) {$\scriptstyle L$};
			\node[left] at (-1.8,3) {$\scriptstyle \mu$};
		\end{tikzpicture}
	\caption{Our conventions for auxiliary links for knot $3_1$ diagram.}\label{fig:3_1_aux}
\end{figure}

First we consider the spin shift operator.
The planarization procedure for its cups is depicted in Fig.~\ref{fig:3_1} (column (a)).
Further, we  simply close it with two caps and obtain the following planarized link that is decomposed over link symbols with the help of the Kauffman bracket:
\begin{equation}
\lambda \Psi_{3_1}(j)=\Psi_{3_1}\left(j+\frac{1}{2}\right)+\Psi_{3_1}\left(j-\frac{1}{2}\right)=\begin{array}{c}
	\begin{tikzpicture}[scale=0.5]
		\foreach \i in {1,...,4}
		{\begin{scope}[shift={(\i,0)}]
				\draw[ultra thick, orange] (-0.1,-0.1) -- (0.1,0.1) (-0.1,0.1) -- (0.1,-0.1);
		\end{scope}}
		\foreach \a/\b in {2.6/1.2,2.6/1.,2.975/0.225,2.95/0.45,2.025/0.225}
		{\draw[thick] ([shift={(0:\b)}]\a,0) arc (0:180:\b);}
		\foreach \a/\b in {2.4/1.2,2.4/1.,2.175/0.575,2.15/0.35}
		{\draw[thick] ([shift={(180:\b)}]\a,0) arc (180:360:\b);}
		\foreach \a/\b in {1.725/0.525,3.5/0.3}
		{
			\draw[line width=1.2mm,white] ([shift={(180:\b)}]\a,0) arc (180:360:\b);
			\draw[thick,\myblue] ([shift={(180:\b)}]\a,0) arc (180:360:\b);
		}
	\end{tikzpicture}
\end{array}=q^{\frac{3}{2}}\mu  \left(2 \mu ^2-3\right) \langle 1\rangle+q^{-\frac{3}{2}}\mu  \left(\mu ^2 q^2+1\right)\langle L\rangle+q^{-\frac{1}{2}}\mu\langle L^2\rangle \,.
\end{equation}
This expression coincides with \cite[eq. (3.12)]{Galakhov:2024eco}.

\begin{figure}[ht!]
	\centering
	\begin{tikzpicture}
		\draw[ultra thick, burgundy] (-3,0) to[out=270,in=180] (-2.5,-0.5) to[out=0,in=270] (-2,0) (-1,0) to[out=270,in=180] (-0.5,-0.5) to[out=0,in=270] (0,0);
		\foreach \i in {0, ..., 2} {
			\begin{scope}[shift={(0,1.5*\i)}]
				\draw[ultra thick, burgundy] (-3,0) -- (-3,1.5) (-1,0) to[out=90,in=270] (-2,1.5) (0,0) -- (0,1.5);
				\draw[line width=1.7mm,white] (-2,0) to[out=90,in=270] (-1,1.5);
				\draw[ultra thick, burgundy] (-2,0) to[out=90,in=270] (-1,1.5);
			\end{scope}
		}
		\draw[ultra thick, burgundy] (-3,4.5) to[out=90,in=180] (-2.5,5) to[out=0,in=90] (-2,4.5) (-1,4.5) to[out=90,in=180] (-0.5,5) to[out=0,in=90] (0,4.5);
		\foreach \i in {0, ..., 3} {
			\begin{scope}[shift={(0,1.5 * \i)}]
				\draw[dashed] (-4,0) -- (0,0) (0.5,0) -- (1,0) (4.5,0) -- (5,0);
			\end{scope}
		}
		\foreach \i  in {0, ..., 2} {
			\begin{scope}[shift={(0,1.5 * \i)}]
				\draw[thick, -stealth] (-3.5,0) -- (-3.5,1.5);
				\node[left] at (-3.5,0.75) {$R_2^+$};
			\end{scope}
		}
		\begin{scope}[shift={(1,0)}]
			\begin{scope}[xscale=0.7,yscale=0.5]
				\foreach \i in {1,...,4}
				{\begin{scope}[shift={(\i,0)}]
						\draw[ultra thick, orange] (-0.05,-0.05) -- (0.05,0.05) (-0.05,0.05) -- (0.05,-0.05);
				\end{scope}}
				\foreach \a/\b in {}
				{\draw ([shift={(0:\b)}]\a,0) arc (0:180:\b);}
				\foreach \a/\b in {2./0.5,4./0.5}
				{\draw ([shift={(180:\b)}]\a,0) arc (180:360:\b);}
			\end{scope}
		\end{scope}
		\begin{scope}[shift={(1,1.5)}]
			\begin{scope}[xscale=0.7,yscale=0.5]
				\foreach \i in {1,...,4}
				{\begin{scope}[shift={(\i,0)}]
						\draw[ultra thick, orange] (-0.05,-0.05) -- (0.05,0.05) (-0.05,0.05) -- (0.05,-0.05);
				\end{scope}}
				\foreach \a/\b in {3.08333/0.583333}
				{\draw ([shift={(0:\b)}]\a,0) arc (0:180:\b);}
				\foreach \a/\b in {2.41667/0.916667,4.08333/0.416667}
				{\draw ([shift={(180:\b)}]\a,0) arc (180:360:\b);}
			\end{scope}
		\end{scope}
		\begin{scope}[shift={(1,3)}]
			\begin{scope}[xscale=0.7,yscale=0.5]
				\foreach \i in {1,...,4}
				{\begin{scope}[shift={(\i,0)}]
						\draw[ultra thick, orange] (-0.05,-0.05) -- (0.05,0.05) (-0.05,0.05) -- (0.05,-0.05);
				\end{scope}}
				\foreach \a/\b in {2.70833/1.04167,2.95833/0.291667,2.91667/0.583333}
				{\draw ([shift={(0:\b)}]\a,0) arc (0:180:\b);}
				\foreach \a/\b in {2.41667/1.08333,4.125/0.375,2.16667/0.5}
				{\draw ([shift={(180:\b)}]\a,0) arc (180:360:\b);}
			\end{scope}
		\end{scope}
		\begin{scope}[shift={(1,4.5)}]
			\begin{scope}[xscale=0.7,yscale=0.5]
				\foreach \i in {1,...,4}
				{\begin{scope}[shift={(\i,0)}]
						\draw[ultra thick, orange] (-0.05,-0.05) -- (0.05,0.05) (-0.05,0.05) -- (0.05,-0.05);
				\end{scope}}
				\foreach \a/\b in {2.6/1.2,2.6/1.,2.975/0.225,2.95/0.45,2.025/0.225}
				{\draw ([shift={(0:\b)}]\a,0) arc (0:180:\b);}
				\foreach \a/\b in {2.4/1.2,4.15/0.35,2.4/1.,2.175/0.575,2.15/0.35}
				{\draw ([shift={(180:\b)}]\a,0) arc (180:360:\b);}
			\end{scope}
		\end{scope}
		\begin{scope}[shift={(4,0)}]
			\begin{scope}[shift={(1,0)}]
				\begin{scope}[xscale=0.7,yscale=0.5]
					\foreach \i in {1,...,4}
					{\begin{scope}[shift={(\i,0)}]
							\draw[ultra thick, orange] (-0.05,-0.05) -- (0.05,0.05) (-0.05,0.05) -- (0.05,-0.05);
					\end{scope}}
					\foreach \a/\b in {2./1.5,2./0.5}
					{\draw  ([shift={(0:\b)}]\a,0) arc (0:180:\b);}
					\foreach \a/\b in {3./0.5}
					{\draw  ([shift={(180:\b)}]\a,0) arc (180:360:\b);}
				\end{scope}
			\end{scope}
			\begin{scope}[shift={(1,1.5)}]
				\begin{scope}[xscale=0.7,yscale=0.5]
					\foreach \i in {1,...,4}
					{\begin{scope}[shift={(\i,0)}]
							\draw[ultra thick, orange] (-0.05,-0.05) -- (0.05,0.05) (-0.05,0.05) -- (0.05,-0.05);
					\end{scope}}
					\foreach \a/\b in {1.5/1.}
					{\draw  ([shift={(0:\b)}]\a,0) arc (0:180:\b);}
					\foreach \a/\b in {2./0.5}
					{\draw  ([shift={(180:\b)}]\a,0) arc (180:360:\b);}
				\end{scope}
			\end{scope}
			\begin{scope}[shift={(1,3)}]
				\begin{scope}[xscale=0.7,yscale=0.5]
					\foreach \i in {1,...,4}
					{\begin{scope}[shift={(\i,0)}]
							\draw[ultra thick, orange] (-0.05,-0.05) -- (0.05,0.05) (-0.05,0.05) -- (0.05,-0.05);
					\end{scope}}
					\foreach \a/\b in {1.08333/0.583333,3./0.5}
					{\draw  ([shift={(0:\b)}]\a,0) arc (0:180:\b);}
					\foreach \a/\b in {2.41667/1.08333,2.08333/0.416667}
					{\draw  ([shift={(180:\b)}]\a,0) arc (180:360:\b);}
				\end{scope}
			\end{scope}
			\begin{scope}[shift={(1,4.5)}]
				\begin{scope}[xscale=0.7,yscale=0.5]
					\foreach \i in {1,...,4}
					{\begin{scope}[shift={(\i,0)}]
							\draw[ultra thick, orange] (-0.05,-0.05) -- (0.05,0.05) (-0.05,0.05) -- (0.05,-0.05);
					\end{scope}}
					\foreach \a/\b in {1./0.5,2.70833/0.958333,2.91667/0.416667}
					{\draw  ([shift={(0:\b)}]\a,0) arc (0:180:\b);}
					\foreach \a/\b in {2.45833/1.20833,2.41667/0.916667,2.125/0.375}
					{\draw  ([shift={(180:\b)}]\a,0) arc (180:360:\b);}
				\end{scope}
			\end{scope}
		\end{scope}
		\node[below] at (2.5,-0.5) {(a)};
		\node[below] at (6.5,-0.5) {(b)};
		\node[below] at (-2.5,-0.5) {$\scriptstyle A$};
		\node[below] at (-0.5,-0.5) {$\scriptstyle B$};
		\node[below] at (-2.5,5) {$\scriptstyle C$};
		\node[below] at (-0.5,5) {$\scriptstyle D$};
	\end{tikzpicture}
	\caption{Braid group action \eqref{arc_braid} on arcades for the braid of $3_1$.}\label{fig:3_1}
\end{figure}

Next we would like to deduce a relation for generators $\langle L^p\rangle$ indicating that the link basis is actually 2d over coefficient field $\IC[\mu]$.
To do so consider link $L'$ and planarize it to the upper level in two ways: directly and with a twist \eqref{cupmove} through cup $A$.
The direct planarization transports simply $L'$ to $L$.
The planarization process for an open part of twisted $L'$ is depicted in Fig.~\ref{fig:3_1} (column (b)).
The result is twisted back through cap $C$.
Eventually, we find that these manipulations lead to the following relation:
\begin{equation}\label{3_1_L}
	\langle L\rangle=\begin{array}{c}
		\begin{tikzpicture}[scale=0.5]
			\foreach \i in {1,...,4}
			{\begin{scope}[shift={(\i,0)}]
					\draw[ultra thick, orange] (-0.1,-0.1) -- (0.1,0.1) (-0.1,0.1) -- (0.1,-0.1);
			\end{scope}}
			\foreach \a/\b in {3./0.333333,2.70833/0.958333}
			{\draw[thick] ([shift={(0:\b)}]\a,0) arc (0:180:\b);}
			\foreach \a/\b in {2.45833/1.20833,2.41667/0.916667,2.20833/0.458333}
			{\draw[thick] ([shift={(180:\b)}]\a,0) arc (180:360:\b);}
			\foreach \a/\b in {1.79167/0.541667}
			{
				\draw[line width=0.7mm,white] ([shift={(0:\b)}]\a,0) arc (0:180:\b);
				\draw[thick,\myblue] ([shift={(0:\b)}]\a,0) arc (0:180:\b);
			}
			\foreach \a/\b in {1.91667/0.416667}
			{
				\draw[line width=0.7mm,white] ([shift={(180:\b)}]\a,0) arc (180:360:\b);
				\draw[thick,\myblue] ([shift={(180:\b)}]\a,0) arc (180:360:\b);
			}
		\end{tikzpicture}
	\end{array}=q^{-1}\left(\mu ^2 q^2-q^2-1\right)\langle 1\rangle+\mu^2\langle L\rangle+q^{-1}\langle L^2\rangle\,.
\end{equation}
This expression is equivalent to \cite[eq. (3.7)]{Galakhov:2024eco}.

Expression \eqref{3_1_L} indicates that \emph{any} link $\CO$ can be rewritten as a linear combination of basis elements:
\begin{equation}
	\langle \CO\rangle_{3_1}=c_1(\mu)\; \langle 1\rangle_{3_1}+c_2(\mu) \; \langle L\rangle_{3_1}\,,
\end{equation}
where $c_i$ are polynomials in operator $\mu$ with rational coefficients in $q^{\frac{1}{2}}$.

\subsection{Cinquefoil knot \texorpdfstring{$5_1$}{}}

The next to the simplest torus knot is the cinquefoil knot $5_1$.
It is quite similar to the trefoil knot differing by the amount of twists in the middle braid.
We simply generalize all the notions introduced for $3_1$.
Respective braid actions on the arcs are depicted in Fig.~\ref{fig:5_1}.

\begin{figure}[ht!]
\centering
\begin{tikzpicture}
    \draw[ultra thick, burgundy] (-3,0) to[out=270,in=180] (-2.5,-0.5) to[out=0,in=270] (-2,0) (-1,0) to[out=270,in=180] (-0.5,-0.5) to[out=0,in=270] (0,0);
    \foreach \i in {0, ..., 4} {
	\begin{scope}[shift={(0,1.5*\i)}]
	    \draw[ultra thick, burgundy] (-3,0) -- (-3,1.5) (-1,0) to[out=90,in=270] (-2,1.5) (0,0) -- (0,1.5);
	    \draw[line width=1.7mm,white] (-2,0) to[out=90,in=270] (-1,1.5);
	    \draw[ultra thick, burgundy] (-2,0) to[out=90,in=270] (-1,1.5);
	\end{scope}
    }
    \draw[ultra thick, burgundy] (-3,7.5) to[out=90,in=180] (-2.5,8) to[out=0,in=90] (-2,7.5) (-1,7.5) to[out=90,in=180] (-0.5,8) to[out=0,in=90] (0,7.5);
    \foreach \i in {0, ..., 5} {
	\begin{scope}[shift={(0,1.5 * \i)}]
	    \draw[dashed] (-4,0) -- (0,0) (0.5,0) -- (1,0) (4.5,0) -- (5,0);
	\end{scope}
    }
    \foreach \i  in {0, ..., 4} {
	\begin{scope}[shift={(0,1.5 * \i)}]
	\draw[thick, -stealth] (-3.5,0) -- (-3.5,1.5);
	    \node[left] at (-3.5,0.75) {$R_2^+$};
	\end{scope}
    }
    \begin{scope}[shift={(1,0)}]
	\begin{scope}[xscale=0.7,yscale=0.5]
\foreach \i in {1,...,4}
{\begin{scope}[shift={(\i,0)}]
\draw[ultra thick, orange] (-0.05,-0.05) -- (0.05,0.05) (-0.05,0.05) -- (0.05,-0.05);
\end{scope}}
\foreach \a/\b in {}
{\draw ([shift={(0:\b)}]\a,0) arc (0:180:\b);}
\foreach \a/\b in {2./0.5,4./0.5}
{\draw ([shift={(180:\b)}]\a,0) arc (180:360:\b);}
	\end{scope}
    \end{scope}
    \begin{scope}[shift={(1,1.5)}]
	\begin{scope}[xscale=0.7,yscale=0.5]
\foreach \i in {1,...,4}
{\begin{scope}[shift={(\i,0)}]
\draw[ultra thick, orange] (-0.05,-0.05) -- (0.05,0.05) (-0.05,0.05) -- (0.05,-0.05);
\end{scope}}
\foreach \a/\b in {3.08333/0.583333}
{\draw ([shift={(0:\b)}]\a,0) arc (0:180:\b);}
\foreach \a/\b in {2.41667/0.916667,4.08333/0.416667}
{\draw ([shift={(180:\b)}]\a,0) arc (180:360:\b);}
	\end{scope}
    \end{scope}
    \begin{scope}[shift={(1,3)}]
	\begin{scope}[xscale=0.7,yscale=0.5]
\foreach \i in {1,...,4}
{\begin{scope}[shift={(\i,0)}]
\draw[ultra thick, orange] (-0.05,-0.05) -- (0.05,0.05) (-0.05,0.05) -- (0.05,-0.05);
\end{scope}}
\foreach \a/\b in {2.70833/1.04167,2.95833/0.291667,2.91667/0.583333}
{\draw ([shift={(0:\b)}]\a,0) arc (0:180:\b);}
\foreach \a/\b in {2.41667/1.08333,4.125/0.375,2.16667/0.5}
{\draw ([shift={(180:\b)}]\a,0) arc (180:360:\b);}
	\end{scope}
    \end{scope}
    \begin{scope}[shift={(1,4.5)}]
	\begin{scope}[xscale=0.7,yscale=0.5]
\foreach \i in {1,...,4}
{\begin{scope}[shift={(\i,0)}]
\draw[ultra thick, orange] (-0.05,-0.05) -- (0.05,0.05) (-0.05,0.05) -- (0.05,-0.05);
\end{scope}}
\foreach \a/\b in {2.6/1.2,2.6/1.,2.975/0.225,2.95/0.45,2.025/0.225}
{\draw ([shift={(0:\b)}]\a,0) arc (0:180:\b);}
\foreach \a/\b in {2.4/1.2,4.15/0.35,2.4/1.,2.175/0.575,2.15/0.35}
{\draw ([shift={(180:\b)}]\a,0) arc (180:360:\b);}
	\end{scope}
    \end{scope}
    \begin{scope}[shift={(1,6)}]
	\begin{scope}[xscale=0.7,yscale=0.5]
\foreach \i in {1,...,4}
{\begin{scope}[shift={(\i,0)}]
\draw[ultra thick, orange] (-0.05,-0.05) -- (0.05,0.05) (-0.05,0.05) -- (0.05,-0.05);
\end{scope}}
\foreach \a/\b in {2.94286/0.342857,2.59524/1.2619,2.91429/0.514286,2.60714/1.10714,2.97143/0.171429,2.61905/0.952381,2.01667/0.183333}
{\draw ([shift={(0:\b)}]\a,0) arc (0:180:\b);}
\foreach \a/\b in {2.44048/1.27381,4.17857/0.321429,2.45238/1.11905,2.46429/0.964286,2.13333/0.466667,2.11667/0.283333,3.04286/0.242857}
{\draw ([shift={(180:\b)}]\a,0) arc (180:360:\b);}
	\end{scope}
    \end{scope}
    \begin{scope}[shift={(1,7.5)}]
	\begin{scope}[xscale=0.7,yscale=0.5]
\foreach \i in {1,...,4}
{\begin{scope}[shift={(\i,0)}]
\draw[ultra thick, orange] (-0.05,-0.05) -- (0.05,0.05) (-0.05,0.05) -- (0.05,-0.05);
\end{scope}}
\foreach \a/\b in {2.9375/0.4375,2.5625/1.3125,2.95833/0.291667,2.5625/1.1875,2.5625/1.0625,2.97917/0.145833,2.5625/0.9375,2.02083/0.145833,2.04167/0.291667}
{\draw ([shift={(0:\b)}]\a,0) arc (0:180:\b);}
\foreach \a/\b in {2.4375/1.3125,4.1875/0.3125,2.4375/1.1875,2.4375/1.0625,2.4375/0.9375,3.04167/0.208333,2.14583/0.520833,2.125/0.375,2.10417/0.229167}
{\draw ([shift={(180:\b)}]\a,0) arc (180:360:\b);}
	\end{scope}
    \end{scope}
    \begin{scope}[shift={(4,0)}]
    \begin{scope}[shift={(1,0)}]
	\begin{scope}[xscale=0.7,yscale=0.5]
\foreach \i in {1,...,4}
{\begin{scope}[shift={(\i,0)}]
		\draw[ultra thick, orange] (-0.05,-0.05) -- (0.05,0.05) (-0.05,0.05) -- (0.05,-0.05);
\end{scope}}
\foreach \a/\b in {2./1.5,2./0.5}
{\draw  ([shift={(0:\b)}]\a,0) arc (0:180:\b);}
\foreach \a/\b in {3./0.5}
{\draw  ([shift={(180:\b)}]\a,0) arc (180:360:\b);}
	\end{scope}
    \end{scope}
    \begin{scope}[shift={(1,1.5)}]
	\begin{scope}[xscale=0.7,yscale=0.5]
\foreach \i in {1,...,4}
{\begin{scope}[shift={(\i,0)}]
		\draw[ultra thick, orange] (-0.05,-0.05) -- (0.05,0.05) (-0.05,0.05) -- (0.05,-0.05);
\end{scope}}
\foreach \a/\b in {1.5/1.}
{\draw  ([shift={(0:\b)}]\a,0) arc (0:180:\b);}
\foreach \a/\b in {2./0.5}
{\draw  ([shift={(180:\b)}]\a,0) arc (180:360:\b);}
	\end{scope}
    \end{scope}
    \begin{scope}[shift={(1,3)}]
	\begin{scope}[xscale=0.7,yscale=0.5]
\foreach \i in {1,...,4}
{\begin{scope}[shift={(\i,0)}]
		\draw[ultra thick, orange] (-0.05,-0.05) -- (0.05,0.05) (-0.05,0.05) -- (0.05,-0.05);
\end{scope}}
\foreach \a/\b in {1.08333/0.583333,3./0.5}
{\draw  ([shift={(0:\b)}]\a,0) arc (0:180:\b);}
\foreach \a/\b in {2.41667/1.08333,2.08333/0.416667}
{\draw  ([shift={(180:\b)}]\a,0) arc (180:360:\b);}
	\end{scope}
    \end{scope}
    \begin{scope}[shift={(1,4.5)}]
	\begin{scope}[xscale=0.7,yscale=0.5]
\foreach \i in {1,...,4}
{\begin{scope}[shift={(\i,0)}]
		\draw[ultra thick, orange] (-0.05,-0.05) -- (0.05,0.05) (-0.05,0.05) -- (0.05,-0.05);
\end{scope}}
\foreach \a/\b in {1./0.5,2.70833/0.958333,2.91667/0.416667}
{\draw  ([shift={(0:\b)}]\a,0) arc (0:180:\b);}
\foreach \a/\b in {2.45833/1.20833,2.41667/0.916667,2.125/0.375}
{\draw  ([shift={(180:\b)}]\a,0) arc (180:360:\b);}
	\end{scope}
    \end{scope}
    \begin{scope}[shift={(1,6)}]
	\begin{scope}[xscale=0.7,yscale=0.5]
\foreach \i in {1,...,4}
{\begin{scope}[shift={(\i,0)}]
		\draw[ultra thick, orange] (-0.05,-0.05) -- (0.05,0.05) (-0.05,0.05) -- (0.05,-0.05);
\end{scope}}
\foreach \a/\b in {0.95/0.45,2.675/1.075,2.65/0.85,2.875/0.375}
{\draw  ([shift={(0:\b)}]\a,0) arc (0:180:\b);}
\foreach \a/\b in {2.475/1.275,2.45/1.05,2.425/0.825,2.15/0.35}
{\draw  ([shift={(180:\b)}]\a,0) arc (180:360:\b);}
	\end{scope}
    \end{scope}
    \begin{scope}[shift={(1,7.5)}]
	\begin{scope}[xscale=0.7,yscale=0.5]
\foreach \i in {1,...,4}
{\begin{scope}[shift={(\i,0)}]
		\draw[ultra thick, orange] (-0.05,-0.05) -- (0.05,0.05) (-0.05,0.05) -- (0.05,-0.05);
\end{scope}}
\foreach \a/\b in {0.916667/0.416667,2.65/1.15,2.63333/0.966667,2.61667/0.783333,2.85/0.35}
{\draw  ([shift={(0:\b)}]\a,0) arc (0:180:\b);}
\foreach \a/\b in {2.48333/1.31667,2.46667/1.13333,2.45/0.95,2.43333/0.766667,2.16667/0.333333}
{\draw  ([shift={(180:\b)}]\a,0) arc (180:360:\b);}
	\end{scope}
    \end{scope}
    \end{scope}
\end{tikzpicture}
    \caption{Braid group action \eqref{arc_braid} on arcades for the braid of $5_1$.}\label{fig:5_1}
\end{figure}

From Fig.~\ref{fig:5_1} we derive the following relations for shift operators defined in \cite{Galakhov:2024eco}:
\begin{equation}
\begin{aligned}
	&\lambda\Psi_{5_1}(j)=\Psi_{5_1}\left(j+\frac{1}{2}\right)+\Psi_{5_1}\left(j-\frac{1}{2}\right)=\begin{array}{c}
		\begin{tikzpicture}[scale=0.5]
			\foreach \i in {1,...,4}
			{\begin{scope}[shift={(\i,0)}]
					\draw[ultra thick, orange] (-0.05,-0.05) -- (0.05,0.05) (-0.05,0.05) -- (0.05,-0.05);
			\end{scope}}
			\foreach \a/\b in {2.9375/0.4375,2.5625/1.3125,2.95833/0.291667,2.5625/1.1875,2.5625/1.0625,2.97917/0.145833,2.5625/0.9375,2.02083/0.145833,2.04167/0.291667}
			{\draw ([shift={(0:\b)}]\a,0) arc (0:180:\b);}
			\foreach \a/\b in {2.4375/1.3125,2.4375/1.1875,2.4375/1.0625,2.4375/0.9375,3.04167/0.208333,2.14583/0.520833,2.125/0.375,2.10417/0.229167}
			{\draw ([shift={(180:\b)}]\a,0) arc (180:360:\b);}
			\foreach \a/\b in {1.64583/0.520833,3.5/0.375}
			{\draw[line width = 0.7mm,white] ([shift={(180:\b)}]\a,0) arc (180:360:\b);}
			\foreach \a/\b in {1.64583/0.520833,3.5/0.375}
			{\draw[thick,\myblue] ([shift={(180:\b)}]\a,0) arc (180:360:\b);}
		\end{tikzpicture}
	\end{array}=\\
    &=\frac{\mu  \left(\mu ^2+2 \mu ^4 q^6-6 \mu ^2 q^6+3 q^6-2 \mu ^4 q^4+6 \mu ^2 q^4-2 q^4-\mu ^2 q^2\right)}{q^{7/2}}\langle 1\rangle+\\
    &+\frac{\mu  \left(2 \mu ^4 q^6-4 \mu ^2 q^6+q^6-2 \mu ^4 q^4+4 \mu ^2 q^4+q^4-4 \mu ^2 q^2+2 q^2+3\right)}{q^{9/2}}\langle L\rangle+\\
    &+\frac{\mu  \left(-3 \mu ^2+2 \mu ^4 q^4-4 \mu ^2 q^4+q^4-\mu ^2 q^2+2 q^2\right)}{q^{7/2}}\langle L^2\rangle+\frac{\mu  \left(\mu ^4 q^4-\mu ^2 q^4-q^4+2 \mu ^2 q^2-q^2-4\right)}{q^{9/2}}\langle L^3\rangle+\\
    &+\frac{\mu  \left(\mu ^2+\mu ^2 q^2-q^2\right)}{q^{7/2}}\langle L^4\rangle+\frac{\mu }{q^{9/2}}\langle L^5\rangle\,.
\end{aligned}
\end{equation}

Now using another column in Fig.~\ref{fig:5_1} we derive a constraint for symbols $\langle L^p\rangle$:
\begin{equation}
\begin{aligned}
	&\langle L\rangle =\begin{array}{c}
		\begin{tikzpicture}[scale=0.5]
			\foreach \i in {1,...,4}
			{\begin{scope}[shift={(\i,0)}]
					\draw[ultra thick, orange] (-0.05,-0.05) -- (0.05,0.05) (-0.05,0.05) -- (0.05,-0.05);
			\end{scope}}
			\foreach \a/\b in {2.93333/0.266667,2.65/1.15,2.63333/0.966667,2.61667/0.783333}
			{\draw  ([shift={(0:\b)}]\a,0) arc (0:180:\b);}
			\foreach \a/\b in {2.48333/1.31667,2.46667/1.13333,2.45/0.95,2.43333/0.766667,2.25/0.416667}
			{\draw  ([shift={(180:\b)}]\a,0) arc (180:360:\b);}
			\foreach \a/\b in {1.75/0.583333}
			{
				\draw[line width=0.7mm, white] ([shift={(0:\b)}]\a,0) arc (0:180:\b);
				\draw[thick, \myblue] ([shift={(0:\b)}]\a,0) arc (0:180:\b);}
			\foreach \a/\b in {1.83333/0.5}
			{
				\draw[line width=0.7mm, white] ([shift={(180:\b)}]\a,0) arc (180:360:\b);
				\draw[thick, \myblue] ([shift={(180:\b)}]\a,0) arc (180:360:\b);}
		\end{tikzpicture}
	\end{array}=\\
    &=\frac{\mu ^2 q^4-2 \mu ^2 q^2+q^2+1}{q^3}\langle 1\rangle+\frac{-2 \mu ^2+2 \mu ^2 q^2}{q^2}\langle L\rangle+\frac{2 \mu ^2 q^2-q^2-3}{q^3}\langle L^2\rangle+\frac{\mu ^2}{q^2}\langle L^3\rangle+\frac{1}{q^3}\langle L^4\rangle\,.
\end{aligned}
\end{equation}

We could easily check the validity of relations presented in this subsection for the following generic spin-$j$ formulas obtained directly from the RT formalism:
\begin{equation}
    \begin{aligned}
	& \mu \Psi_{5_1}(j)=-\left(q^{2j+1}+q^{-2j-1}\right)\Psi_{5_1}(j)\,,\\
	& \langle L^p\rangle=\sum\lm_{k=0}^{2j}(-1)^kq^{5\left(k(1-k)+4 jk-2j^2\right)}\left[2(2j-k)+1\right]_q\times\left(-q^{4j-2k+1}-q^{-4j+2k-1}\right)^p\,.
    \end{aligned}
\end{equation}

We describe this equation quasi-classically by considering a limit $q\to 1$, $\mu\to m+m^{-1}$, $\lambda\to l+l^{-1}$, $\langle L^p\rangle\to x^p$.
Then the relation for $\langle L^p\rangle$ links factorizes:
\begin{equation}
\frac{(x+1) \left(x^2+x-1\right) \left(x m^2+m^4+1\right)}{m^2} = 0\,.
\end{equation}

For an ``interesting'' root of this equation $x=-m^2-m^{-2}$ we have the following factorization of the shift operator equation:
\begin{equation}
\frac{\left(l+m^5\right) \left(l m^5+1\right)}{l m^5}=0\,.
\end{equation}
Clearly one of the roots contributes to the canonical A-polynomial $A_{5_1}(l,m)=l+m^5$ (cf. \cite[Table 2]{Gukov:2003na}).

\subsection{Figure-eight knot \texorpdfstring{$4_1$}{}}

In the case of the figure-eight knot symbols are again generated by $\mu$ and $L$ (see Fig.\ref{fig:4_1_aux}).
\begin{figure}[ht!]
	\centering
	\begin{tikzpicture}[scale=0.6]
		\draw[ultra thick, burgundy] (-1.5,0) to[out=270,in=180]  (-1,-0.5) to[out=0,in=270] (-0.5,0) (0.5,0) to[out=270,in=180] (1,-0.5) to[out=0,in=270] (1.5,0);
		\begin{scope}[yscale=0.3] 
		\draw ([shift={(0:0.8)}]0,0) arc (0:180:0.8);
		\end{scope}
		\draw[line width=1.5mm, white] (0.5,0) to[out=90,in=270] (-0.5,1);
		\draw[ultra thick, burgundy] (-1.5,0) -- (-1.5,1) (1.5,0) -- (1.5,1) (0.5,0) to[out=90,in=270] (-0.5,1);
		\draw[line width=1.5mm, white] (-0.5,0) to[out=90,in=270] (0.5,1);
		\draw[ultra thick, burgundy] (-0.5,0) to[out=90,in=270] (0.5,1);
		\begin{scope}[yscale=0.3]
			\draw[line width=1.2mm, white] ([shift={(180:0.8)}]0,0) arc (180:360:0.8);
			\draw ([shift={(180:0.8)}]0,0) arc (180:360:0.8);
		\end{scope}
		\begin{scope}[shift={(0,1)}]
			\draw[ultra thick, burgundy] (-1.5,0) -- (-1.5,1) (1.5,0) -- (1.5,1) (0.5,0) to[out=90,in=270] (-0.5,1);
			\draw[line width=1.5mm, white] (-0.5,0) to[out=90,in=270] (0.5,1);
			\draw[ultra thick, burgundy] (-0.5,0) to[out=90,in=270] (0.5,1);
		\end{scope}
		\begin{scope}[shift={(0,2)}]
			\draw[ultra thick, burgundy] (-1.5,0) -- (-1.5,1) (-0.5,0) -- (-0.5,1) (0.5,0) to[out=90,in=270] (1.5,1);
			\draw[line width=1.5mm, white] (1.5,0) to[out=90,in=270] (0.5,1);
			\draw[ultra thick, burgundy] (1.5,0) to[out=90,in=270] (0.5,1);
		\end{scope}
		\begin{scope}[shift={(0,3)}]
			\draw[ultra thick, burgundy] (-1.5,0) -- (-1.5,1) (-0.5,0) -- (-0.5,1) (0.5,0) to[out=90,in=270] (1.5,1);
			\draw[line width=1.5mm, white] (1.5,0) to[out=90,in=270] (0.5,1);
			\draw[ultra thick, burgundy] (1.5,0) to[out=90,in=270] (0.5,1);
		\end{scope}
		\begin{scope}[shift={(0,4)}]
			\begin{scope}[yscale=0.3]
			\draw ([shift={(0:0.3)}]-1.5,0) arc (0:180:0.3);
			\draw ([shift={(0:0.8)}]1.0,0) arc (0:180:0.8);
			\end{scope}
		\end{scope}
		\begin{scope}[shift={(0,4)}]
			\draw[line width=1.5mm, white] (-0.5,0) to[out=90,in=180] (0,0.5) to[out=0,in=90] (0.5,0) (-1.5,0) to[out=90,in=180] (0,1) to[out=0,in=90] (1.5,0);
			\draw[ultra thick, burgundy] (-0.5,0) to[out=90,in=180] (0,0.5) to[out=0,in=90] (0.5,0) (-1.5,0) to[out=90,in=180] (0,1) to[out=0,in=90] (1.5,0);
		\end{scope}
		\begin{scope}[shift={(0,4)}]
			\begin{scope}[yscale=0.3]
				\draw[line width=1.2mm, white] ([shift={(180:0.3)}]-1.5,0) arc (180:360:0.3);
				\draw[line width=1.2mm, white] ([shift={(180:0.8)}]1.0,0) arc (180:360:0.8);
				\draw ([shift={(180:0.3)}]-1.5,0) arc (180:360:0.3);
				\draw ([shift={(180:0.8)}]1.0,0) arc (180:360:0.8);
			\end{scope}
		\end{scope}
		\node[left] at (-1.8,4) {$\mu$};
		\node[right] at (1.8,4) {$L$};
		\node[right] at (1.8,0) {$L'$};
	\end{tikzpicture}
	\caption{Our conventions for auxiliary links for knot $3_1$ diagram.}\label{fig:4_1_aux}
\end{figure}
We could have chosen another collection of loops around two strands $L'$ as a generator instead of $L$, yet they are related, and $L$ would be for us more convenient. 

First of all we consider the spin shift operator $\lambda$.
To decompose the respective link over the symbol basis first we peel off cups $A$ and $B$ depicted in Fig.~\ref{fig:Fig8} as arcs.
Further we evolve them with the braid word $R_3^-R_3^-R_2^+R_2^+$ corresponding to the figure-eight knot as it was prescribed in Sec.~\ref{sec:arcade}, this evolution process is depicted in a series of diagrams in Fig.~\ref{fig:Fig8}(column (a)).
Eventually, we close the link with two caps $C$ and $D$ depicted in Fig.~\ref{fig:Fig8}.

\begin{figure}[ht!]
	\centering
	\begin{tikzpicture}
		\draw[ultra thick, burgundy] (-3,0) to[out=270,in=180] (-2.5,-0.5) to[out=0,in=270] (-2,0) (-1,0) to[out=270,in=180] (-0.5,-0.5) to[out=0,in=270] (0,0);
		\begin{scope}
			\draw[ultra thick, burgundy] (-3,0) -- (-3,1.5) (0,0) -- (0,1.5) (-1,0) to[out=90,in=270] (-2,1.5);
			\draw[line width=1.7mm, white] (-2,0) to[out=90,in=270] (-1,1.5);
			\draw[ultra thick, burgundy] (-2,0) to[out=90,in=270] (-1,1.5);
		\end{scope}
		\begin{scope}[shift={(0,1.5)}]
			\draw[ultra thick, burgundy] (-3,0) -- (-3,1.5) (0,0) -- (0,1.5) (-1,0) to[out=90,in=270] (-2,1.5);
			\draw[line width=1.7mm, white] (-2,0) to[out=90,in=270] (-1,1.5);
			\draw[ultra thick, burgundy] (-2,0) to[out=90,in=270] (-1,1.5);
		\end{scope}
		\begin{scope}[shift={(0,3)}]
			\draw[ultra thick, burgundy] (-3,0) -- (-3,1.5) (-2,0) -- (-2,1.5) (-1,0) to[out=90,in=270] (0,1.5);
			\draw[line width=1.7mm, white] (0,0) to[out=90,in=270] (-1,1.5);
			\draw[ultra thick, burgundy] (0,0) to[out=90,in=270] (-1,1.5);
		\end{scope}
		\begin{scope}[shift={(0,4.5)}]
			\draw[ultra thick, burgundy] (-3,0) -- (-3,1.5) (-2,0) -- (-2,1.5) (-1,0) to[out=90,in=270] (0,1.5);
			\draw[line width=1.7mm, white] (0,0) to[out=90,in=270] (-1,1.5);
			\draw[ultra thick, burgundy] (0,0) to[out=90,in=270] (-1,1.5);
		\end{scope}
		\begin{scope}[shift={(0,6)}]
			\draw[ultra thick, burgundy] (-2,0) to[out=90,in=180] (-1.5,0.5) to[out=0,in=90] (-1,0) (-3,0) to[out=90,in=180] (-1.5,1) to[out=0,in=90] (0,0);
		\end{scope}
		\foreach \i in {0, ..., 4} {
			\draw[dashed] (-4,1.5*\i) -- (0.5,1.5*\i) (4,1.5*\i) -- (4.5,1.5*\i);
		}
		\foreach \i/\s in {0/$R_2^+$, 1/$R_2^+$, 2/$R_3^-$, 3/$R_3^-$} {
			\draw[-stealth] (-3.5,1.5*\i) -- (-3.5,1.5*\i+1.5);
			\node[left] at (-3.5,1.5*\i+0.75) {\s};
		}
		\node[below] at (2.5,-0.5) {(a)};
		\node[below] at (6.5,-0.5) {(b)};
		\node[below] at (10.5,-0.5) {(c)};
		\node[below] at (-2.5,-0.5) {$\scriptstyle A$};
		\node[below] at (-0.5,-0.5) {$\scriptstyle B$};
		\node[below] at (-1.5,6.5) {$\scriptstyle C$};
		\node[below] at (-1.5,7) {$\scriptstyle D$};
		\begin{scope}[shift={(-0.5,0)}]
		\begin{scope}[shift={(1,0)}]
			\begin{scope}[xscale=0.7,yscale=0.5]
				\foreach \i in {1,...,4}
				{\begin{scope}[shift={(\i,0)}]
						\draw[ultra thick, orange] (-0.05,-0.05) -- (0.05,0.05) (-0.05,0.05) -- (0.05,-0.05);
				\end{scope}}
				\foreach \a/\b in {}
				{\draw ([shift={(0:\b)}]\a,0) arc (0:180:\b);}
				\foreach \a/\b in {2./0.5,4./0.5}
				{\draw ([shift={(180:\b)}]\a,0) arc (180:360:\b);}
			\end{scope}
		\end{scope}
		\begin{scope}[shift={(1,1.5)}]
			\begin{scope}[xscale=0.7,yscale=0.5]
				\foreach \i in {1,...,4}
				{\begin{scope}[shift={(\i,0)}]
						\draw[ultra thick, orange] (-0.05,-0.05) -- (0.05,0.05) (-0.05,0.05) -- (0.05,-0.05);
				\end{scope}}
				\foreach \a/\b in {3.08333/0.583333}
				{\draw ([shift={(0:\b)}]\a,0) arc (0:180:\b);}
				\foreach \a/\b in {2.41667/0.916667,4.08333/0.416667}
				{\draw ([shift={(180:\b)}]\a,0) arc (180:360:\b);}
			\end{scope}
		\end{scope}
		\begin{scope}[shift={(1,3)}]
			\begin{scope}[xscale=0.7,yscale=0.5]
				\foreach \i in {1,...,4}
				{\begin{scope}[shift={(\i,0)}]
						\draw[ultra thick, orange] (-0.05,-0.05) -- (0.05,0.05) (-0.05,0.05) -- (0.05,-0.05);
				\end{scope}}
				\foreach \a/\b in {2.70833/1.04167,2.95833/0.291667,2.91667/0.583333}
				{\draw ([shift={(0:\b)}]\a,0) arc (0:180:\b);}
				\foreach \a/\b in {2.41667/1.08333,4.125/0.375,2.16667/0.5}
				{\draw ([shift={(180:\b)}]\a,0) arc (180:360:\b);}
			\end{scope}
		\end{scope}
		\begin{scope}[shift={(1,4.5)}]
			\begin{scope}[xscale=0.7,yscale=0.5]
				\foreach \i in {1,...,4}
				{\begin{scope}[shift={(\i,0)}]
						\draw[ultra thick, orange] (-0.05,-0.05) -- (0.05,0.05) (-0.05,0.05) -- (0.05,-0.05);
				\end{scope}}
				\foreach \a/\b in {3.325/0.925,3.20833/1.54167,3.35/1.15,3.175/0.575,3.15/0.35}
				{\draw ([shift={(0:\b)}]\a,0) arc (0:180:\b);}
				\foreach \a/\b in {1.96667/0.633333,4.125/0.625,4.125/0.375,2.03333/0.366667,3.025/0.225}
				{\draw ([shift={(180:\b)}]\a,0) arc (180:360:\b);}
			\end{scope}
		\end{scope}
		\begin{scope}[shift={(1,6)}]
			\begin{scope}[xscale=0.7,yscale=0.5]
				\foreach \i in {1,...,4}
				{\begin{scope}[shift={(\i,0)}]
						\draw[ultra thick, orange] (-0.05,-0.05) -- (0.05,0.05) (-0.05,0.05) -- (0.05,-0.05);
				\end{scope}}
				\foreach \a/\b in {3.45238/1.11905,3.2619/1.59524,3.44048/1.27381,3.46429/0.964286,3.13333/0.466667,3.11667/0.283333,4.04286/0.242857}
				{\draw ([shift={(0:\b)}]\a,0) arc (0:180:\b);}
				\foreach \a/\b in {1.91667/0.583333,3.94286/0.342857,3.91429/0.514286,2./0.333333,3.97143/0.171429,3.7619/1.09524,3.77381/0.940476,3.88571/0.685714}
				{\draw ([shift={(180:\b)}]\a,0) arc (180:360:\b);}
			\end{scope}
		\end{scope}
		\end{scope}
	\begin{scope}[shift={(3.5,0)}]
	\begin{scope}[shift={(1,0)}]
		\begin{scope}[xscale=0.7,yscale=0.5]
			\foreach \i in {1,...,4}
			{\begin{scope}[shift={(\i,0)}]
					\draw[ultra thick, orange] (-0.05,-0.05) -- (0.05,0.05) (-0.05,0.05) -- (0.05,-0.05);
			\end{scope}}
			\foreach \a/\b in {2.5/1.}
			{\draw ([shift={(0:\b)}]\a,0) arc (0:180:\b);}
			\foreach \a/\b in {2.5/1.}
			{\draw ([shift={(180:\b)}]\a,0) arc (180:360:\b);}
		\end{scope}
	\end{scope}
	\begin{scope}[shift={(1,1.5)}]
		\begin{scope}[xscale=0.7,yscale=0.5]
			\foreach \i in {1,...,4}
			{\begin{scope}[shift={(\i,0)}]
					\draw[ultra thick, orange] (-0.05,-0.05) -- (0.05,0.05) (-0.05,0.05) -- (0.05,-0.05);
			\end{scope}}
			\foreach \a/\b in {2.5/1.}
			{\draw ([shift={(0:\b)}]\a,0) arc (0:180:\b);}
			\foreach \a/\b in {2.5/1.}
			{\draw ([shift={(180:\b)}]\a,0) arc (180:360:\b);}
		\end{scope}
	\end{scope}
	\begin{scope}[shift={(1,3)}]
		\begin{scope}[xscale=0.7,yscale=0.5]
			\foreach \i in {1,...,4}
			{\begin{scope}[shift={(\i,0)}]
					\draw[ultra thick, orange] (-0.05,-0.05) -- (0.05,0.05) (-0.05,0.05) -- (0.05,-0.05);
			\end{scope}}
			\foreach \a/\b in {2.5/1.}
			{\draw ([shift={(0:\b)}]\a,0) arc (0:180:\b);}
			\foreach \a/\b in {2.5/1.}
			{\draw ([shift={(180:\b)}]\a,0) arc (180:360:\b);}
		\end{scope}
	\end{scope}
	\begin{scope}[shift={(1,4.5)}]
		\begin{scope}[xscale=0.7,yscale=0.5]
			\foreach \i in {1,...,4}
			{\begin{scope}[shift={(\i,0)}]
					\draw[ultra thick, orange] (-0.05,-0.05) -- (0.05,0.05) (-0.05,0.05) -- (0.05,-0.05);
			\end{scope}}
			\foreach \a/\b in {3./1.5,3./0.5}
			{\draw ([shift={(0:\b)}]\a,0) arc (0:180:\b);}
			\foreach \a/\b in {2./0.5,4./0.5}
			{\draw ([shift={(180:\b)}]\a,0) arc (180:360:\b);}
		\end{scope}
	\end{scope}
	\begin{scope}[shift={(1,6)}]
		\begin{scope}[xscale=0.7,yscale=0.5]
			\foreach \i in {1,...,4}
			{\begin{scope}[shift={(\i,0)}]
					\draw[ultra thick, orange] (-0.05,-0.05) -- (0.05,0.05) (-0.05,0.05) -- (0.05,-0.05);
			\end{scope}}
			\foreach \a/\b in {3.08333/1.58333,3.33333/1.,3.08333/0.416667}
			{\draw ([shift={(0:\b)}]\a,0) arc (0:180:\b);}
			\foreach \a/\b in {1.91667/0.416667,3.66667/1.,3.91667/0.416667}
			{\draw ([shift={(180:\b)}]\a,0) arc (180:360:\b);}
		\end{scope}
	\end{scope}
	\end{scope}
	\begin{scope}[shift={(7.5,0)}]
	\begin{scope}[shift={(1,0)}]
		\begin{scope}[xscale=0.7,yscale=0.5]
			\foreach \i in {1,...,4}
			{\begin{scope}[shift={(\i,0)}]
					\draw[ultra thick, orange] (-0.05,-0.05) -- (0.05,0.05) (-0.05,0.05) -- (0.05,-0.05);
			\end{scope}}
			\foreach \a/\b in {2.41667/1.08333,2.08333/0.416667}
			{\draw ([shift={(0:\b)}]\a,0) arc (0:180:\b);}
			\foreach \a/\b in {3./0.5}
			{\draw ([shift={(180:\b)}]\a,0) arc (180:360:\b);}
		\end{scope}
	\end{scope}
	\begin{scope}[shift={(1,1.5)}]
		\begin{scope}[xscale=0.7,yscale=0.5]
			\foreach \i in {1,...,4}
			{\begin{scope}[shift={(\i,0)}]
					\draw[ultra thick, orange] (-0.05,-0.05) -- (0.05,0.05) (-0.05,0.05) -- (0.05,-0.05);
			\end{scope}}
			\foreach \a/\b in {1.91667/0.583333}
			{\draw ([shift={(0:\b)}]\a,0) arc (0:180:\b);}
			\foreach \a/\b in {2.08333/0.416667}
			{\draw ([shift={(180:\b)}]\a,0) arc (180:360:\b);}
		\end{scope}
	\end{scope}
	\begin{scope}[shift={(1,3)}]
		\begin{scope}[xscale=0.7,yscale=0.5]
			\foreach \i in {1,...,4}
			{\begin{scope}[shift={(\i,0)}]
					\draw[ultra thick, orange] (-0.05,-0.05) -- (0.05,0.05) (-0.05,0.05) -- (0.05,-0.05);
			\end{scope}}
			\foreach \a/\b in {1.5/0.25,3./0.5}
			{\draw ([shift={(0:\b)}]\a,0) arc (0:180:\b);}
			\foreach \a/\b in {2.5/1.,2.125/0.375}
			{\draw ([shift={(180:\b)}]\a,0) arc (180:360:\b);}
		\end{scope}
	\end{scope}
	\begin{scope}[shift={(1,4.5)}]
		\begin{scope}[xscale=0.7,yscale=0.5]
			\foreach \i in {1,...,4}
			{\begin{scope}[shift={(\i,0)}]
					\draw[ultra thick, orange] (-0.05,-0.05) -- (0.05,0.05) (-0.05,0.05) -- (0.05,-0.05);
			\end{scope}}
			\foreach \a/\b in {1.5/0.25,3.41667/1.08333,3.08333/0.416667}
			{\draw ([shift={(0:\b)}]\a,0) arc (0:180:\b);}
			\foreach \a/\b in {2.08333/0.583333,2.04167/0.291667,4./0.5}
			{\draw ([shift={(180:\b)}]\a,0) arc (180:360:\b);}
		\end{scope}
	\end{scope}
	\begin{scope}[shift={(1,6)}]
		\begin{scope}[xscale=0.7,yscale=0.5]
			\foreach \i in {1,...,4}
			{\begin{scope}[shift={(\i,0)}]
					\draw[ultra thick, orange] (-0.05,-0.05) -- (0.05,0.05) (-0.05,0.05) -- (0.05,-0.05);
			\end{scope}}
			\foreach \a/\b in {1.5/0.25,3.45833/1.20833,3.41667/0.916667,3.125/0.375}
			{\draw ([shift={(0:\b)}]\a,0) arc (0:180:\b);}
			\foreach \a/\b in {2./0.5,2./0.25,3.70833/0.958333,3.91667/0.416667}
			{\draw ([shift={(180:\b)}]\a,0) arc (180:360:\b);}
		\end{scope}
	\end{scope}
	\end{scope}
	\end{tikzpicture}
	\caption{Braid group action \eqref{arc_braid} on arcades for the braid of $4_1$.}\label{fig:Fig8}
\end{figure}

In practice, for convenience we have modified the resulting arcs slightly, so the links differ by the first Reidemeister move leading to a $q$-monomial coefficient:
\begin{equation}\label{4_1_lambda}
\begin{aligned}
	&\left(-q^{-\frac{3}{2}}\right)\lambda\Psi_{4_1}(j)=\begin{array}{c}
		\begin{tikzpicture}[scale=0.5]
			\foreach \i in {1,...,4}
			{\begin{scope}[shift={(\i,0)}]
					\draw[ultra thick, orange] (-0.05,-0.05) -- (0.05,0.05) (-0.05,0.05) -- (0.05,-0.05);
			\end{scope}}
			\foreach \a/\b in {3.41667/1.08333,3.29167/1.54167,3.41667/1.25,3.41667/0.916667,3.13333/0.466667,3.11667/0.283333,3.98333/0.183333,1.375/0.125}
			{\draw[thick] ([shift={(0:\b)}]\a,0) arc (0:180:\b);}
			\foreach \a/\b in {2./0.5,3.88333/0.283333,3.86667/0.466667,2.04167/0.291667,3.75/1.08333,3.75/0.916667,3.85/0.65}
			{\draw[thick] ([shift={(180:\b)}]\a,0) arc (180:360:\b);}
			\foreach \a/\b in {2.525/1.275,2.68333/0.516667}
			{
				\draw[line width=1.5mm,white] ([shift={(180:\b)}]\a,0) arc (180:360:\b);
				\draw[ultra thick,\myblue] ([shift={(180:\b)}]\a,0) arc (180:360:\b);
			}
		\end{tikzpicture}
	\end{array}=\\
	&=\frac{\mu ^2 \left(-\mu ^2+2 q^6+2 \mu ^2 q^4+q^4+2 \mu ^2 q^2-2 q^2-2\right)}{q^{5/2}}\langle 1\rangle+\frac{\mu ^2 \left(2 q^4+\mu ^2 q^2-1\right)}{q^{3/2}}\langle L\rangle+\mu ^2 q^{3/2}\langle L^2\rangle+\\
	&+\frac{\left(q^2-1\right) \left(\mu ^2+q^2-1\right)}{q^{7/2}}\langle T_1\rangle+\frac{\mu ^2 \left(q^4-1\right)}{q^{7/2}}\langle T_2\rangle+\frac{q^6+\mu ^2 q^4-q^2}{q^{7/2}}\langle T_3\rangle+\mu ^2 q^{5/2}\langle T_4\rangle+q^{9/2}\langle T_5\rangle\,,
\end{aligned}
\end{equation}
where we have introduced new link symbols:
\begin{subequations}
\begin{equation}
	\langle T_1\rangle=\begin{array}{c}
		\begin{tikzpicture}[scale=0.5]
			\foreach \i in {1,...,4}
			{\begin{scope}[shift={(\i,0)}]
					\draw[ultra thick, orange] (-0.05,-0.05) -- (0.05,0.05) (-0.05,0.05) -- (0.05,-0.05);
			\end{scope}}
			\foreach \a/\b in {3./1.5,3./0.5}
			{\draw[thick] ([shift={(0:\b)}]\a,0) arc (0:180:\b);}
			\foreach \a/\b in {2./0.5,4./0.5}
			{\draw[thick] ([shift={(180:\b)}]\a,0) arc (180:360:\b);}
		\end{tikzpicture}
	\end{array}=\begin{array}{c}
	\begin{tikzpicture}[scale=0.5]
		\foreach \i in {1,...,4}
		{\begin{scope}[shift={(\i,0)}]
				\draw[ultra thick, orange] (-0.05,-0.05) -- (0.05,0.05) (-0.05,0.05) -- (0.05,-0.05);
		\end{scope}}
		\foreach \a/\b in {3.08333/1.41667,3.08333/0.583333}
		{\draw[thick] ([shift={(0:\b)}]\a,0) arc (0:180:\b);}
		\foreach \a/\b in {4.08333/0.416667,1.5/0.166667}
		{\draw[thick] ([shift={(180:\b)}]\a,0) arc (180:360:\b);}
		\foreach \a/\b in {2.33333/1.}
		{
			\draw[line width=1.5mm,white] ([shift={(0:\b)}]\a,0) arc (0:180:\b);
			\draw[ultra thick, \myblue] ([shift={(0:\b)}]\a,0) arc (0:180:\b);
		}
		\foreach \a/\b in {2.91667/0.416667}
		{
			\draw[line width=1.5mm,white] ([shift={(180:\b)}]\a,0) arc (180:360:\b);
			\draw[ultra thick, \myblue] ([shift={(180:\b)}]\a,0) arc (180:360:\b);
		}
	\end{tikzpicture}
	\end{array}=-q\mu ^2\langle 1\rangle -q^2\langle L\rangle\,,
\end{equation}
\begin{equation}
	\langle T_2\rangle=\begin{array}{c}
		\begin{tikzpicture}[scale=0.5]
			\foreach \i in {1,...,4}
			{\begin{scope}[shift={(\i,0)}]
					\draw[ultra thick, orange] (-0.05,-0.05) -- (0.05,0.05) (-0.05,0.05) -- (0.05,-0.05);
			\end{scope}}
			\foreach \a/\b in {3.08333/1.58333,3.33333/1.,3.08333/0.416667}
			{\draw[thick] ([shift={(0:\b)}]\a,0) arc (0:180:\b);}
			\foreach \a/\b in {1.91667/0.416667,3.66667/1.,3.91667/0.416667}
			{\draw[thick] ([shift={(180:\b)}]\a,0) arc (180:360:\b);}
		\end{tikzpicture}
	\end{array}=\begin{array}{c}
	\begin{tikzpicture}[scale=0.5]
		\foreach \i in {1,...,4}
		{\begin{scope}[shift={(\i,0)}]
				\draw[ultra thick, orange] (-0.05,-0.05) -- (0.05,0.05) (-0.05,0.05) -- (0.05,-0.05);
		\end{scope}}
		\foreach \a/\b in {3.16667/1.5,3.33333/1.,3.16667/0.5}
		{\draw[thick] ([shift={(0:\b)}]\a,0) arc (0:180:\b);}
		\foreach \a/\b in {3.66667/1.,4./0.333333,1.5/0.166667}
		{\draw[thick] ([shift={(180:\b)}]\a,0) arc (180:360:\b);}
		\foreach \a/\b in {2.33333/1.}
		{
			\draw[line width=1.5mm,white] ([shift={(0:\b)}]\a,0) arc (0:180:\b);
			\draw[ultra thick, \myblue] ([shift={(0:\b)}]\a,0) arc (0:180:\b);
		}
		\foreach \a/\b in {2.83333/0.5}
		{
			\draw[line width=1.5mm,white] ([shift={(180:\b)}]\a,0) arc (180:360:\b);
			\draw[ultra thick, \myblue] ([shift={(180:\b)}]\a,0) arc (180:360:\b);
		}
	\end{tikzpicture}
	\end{array}=q \left(-2 \mu ^2+q^2+1\right)\langle 1\rangle+\langle L\rangle-q^3\langle L^2\rangle\,,
\end{equation}
\begin{equation}
	\langle T_3\rangle=\begin{array}{c}
		\begin{tikzpicture}[scale=0.5]
			\foreach \i in {1,...,4}
			{\begin{scope}[shift={(\i,0)}]
					\draw[ultra thick, orange] (-0.05,-0.05) -- (0.05,0.05) (-0.05,0.05) -- (0.05,-0.05);
			\end{scope}}
			\foreach \a/\b in {3.125/1.625,3.375/1.125,3.375/0.875,3.125/0.375}
			{\draw[thick] ([shift={(0:\b)}]\a,0) arc (0:180:\b);}
			\foreach \a/\b in {1.875/0.375,3.625/1.125,3.625/0.875,3.875/0.375}
			{\draw[thick] ([shift={(180:\b)}]\a,0) arc (180:360:\b);}
		\end{tikzpicture}
	\end{array}=\begin{array}{c}
	\begin{tikzpicture}[scale=0.5]
		\foreach \i in {1,...,4}
		{\begin{scope}[shift={(\i,0)}]
				\draw[ultra thick, orange] (-0.05,-0.05) -- (0.05,0.05) (-0.05,0.05) -- (0.05,-0.05);
		\end{scope}}
		\foreach \a/\b in {3.20833/1.54167,3.375/1.125,3.375/0.875,3.20833/0.458333}
		{\draw[thick] ([shift={(0:\b)}]\a,0) arc (0:180:\b);}
		\foreach \a/\b in {3.625/1.125,3.625/0.875,3.95833/0.291667,1.5/0.166667}
		{\draw[thick] ([shift={(180:\b)}]\a,0) arc (180:360:\b);}
		\foreach \a/\b in {2.33333/1.}
		{
			\draw[line width=1.5mm,white] ([shift={(0:\b)}]\a,0) arc (0:180:\b);
			\draw[ultra thick, \myblue] ([shift={(0:\b)}]\a,0) arc (0:180:\b);
		}
		\foreach \a/\b in {2.79167/0.541667}
		{
			\draw[line width=1.5mm,white] ([shift={(180:\b)}]\a,0) arc (180:360:\b);
			\draw[ultra thick, \myblue] ([shift={(180:\b)}]\a,0) arc (180:360:\b);
		}
	\end{tikzpicture}
	\end{array}=\begin{array}{l}\mu ^2 q \left(q^2-2\right)\langle 1\rangle+q^2 \left(-2 \mu ^2+2 q^2+1\right)\langle L\rangle+\\+\mu ^2 \left(-q^3\right)\langle L^2\rangle-q^4\langle L^3\rangle\,,\end{array}
\end{equation}
\begin{equation}
	\langle T_4\rangle=\begin{array}{c}
		\begin{tikzpicture}[scale=0.5]
			\foreach \i in {1,...,4}
			{\begin{scope}[shift={(\i,0)}]
					\draw[ultra thick, orange] (-0.05,-0.05) -- (0.05,0.05) (-0.05,0.05) -- (0.05,-0.05);
			\end{scope}}
			\foreach \a/\b in {3.15/1.65,3.4/1.2,3.4/1.,3.4/0.8,3.15/0.35}
			{\draw[thick] ([shift={(0:\b)}]\a,0) arc (0:180:\b);}
			\foreach \a/\b in {1.85/0.35,3.6/1.2,3.6/1.,3.6/0.8,3.85/0.35}
			{\draw[thick] ([shift={(180:\b)}]\a,0) arc (180:360:\b);}
		\end{tikzpicture}
	\end{array}=\begin{array}{c}
	\begin{tikzpicture}[scale=0.5]
		\foreach \i in {1,...,4}
		{\begin{scope}[shift={(\i,0)}]
				\draw[ultra thick, orange] (-0.05,-0.05) -- (0.05,0.05) (-0.05,0.05) -- (0.05,-0.05);
		\end{scope}}
		\foreach \a/\b in {3.23333/1.56667,3.4/1.2,3.4/1.,3.4/0.8,3.23333/0.433333}
		{\draw[thick] ([shift={(0:\b)}]\a,0) arc (0:180:\b);}
		\foreach \a/\b in {3.6/1.2,3.6/1.,3.6/0.8,3.93333/0.266667,1.5/0.166667}
		{\draw[thick] ([shift={(180:\b)}]\a,0) arc (180:360:\b);}
		\foreach \a/\b in {2.33333/1.}
		{
			\draw[line width=1.5mm,white] ([shift={(0:\b)}]\a,0) arc (0:180:\b);
			\draw[ultra thick, \myblue] ([shift={(0:\b)}]\a,0) arc (0:180:\b);
		}
		\foreach \a/\b in {2.76667/0.566667}
		{
			\draw[line width=1.5mm,white] ([shift={(180:\b)}]\a,0) arc (180:360:\b);
			\draw[ultra thick, \myblue] ([shift={(180:\b)}]\a,0) arc (180:360:\b);
		}
	\end{tikzpicture}
	\end{array}=\begin{array}{l}-q \left(2 \mu ^2+q^4-2 \mu ^2 q^2+q^2\right)\langle 1 \rangle+2 \mu ^2 (q-1) q^2 (q+1)\langle L\rangle+\\ +q^3 \left(-2 \mu ^2+3 q^2+1\right)\langle L^2\rangle+\mu ^2 \left(-q^4\right)\langle L^3\rangle-q^5\langle L^4\rangle\,,\end{array}
\end{equation}
\begin{equation}
	\langle T_5\rangle=\begin{array}{c}
		\begin{tikzpicture}[scale=0.5]
			\foreach \i in {1,...,4}
			{\begin{scope}[shift={(\i,0)}]
					\draw[ultra thick, orange] (-0.05,-0.05) -- (0.05,0.05) (-0.05,0.05) -- (0.05,-0.05);
			\end{scope}}
			\foreach \a/\b in {3.16667/1.66667,3.41667/1.25,3.41667/1.08333,3.41667/0.916667,3.41667/0.75,3.16667/0.333333}
			{\draw[thick] ([shift={(0:\b)}]\a,0) arc (0:180:\b);}
			\foreach \a/\b in {1.83333/0.333333,3.58333/1.25,3.58333/1.08333,3.58333/0.916667,3.58333/0.75,3.83333/0.333333}
			{\draw[thick] ([shift={(180:\b)}]\a,0) arc (180:360:\b);}
		\end{tikzpicture}
	\end{array}=\begin{array}{c}
	\begin{tikzpicture}[scale=0.5]
		\foreach \i in {1,...,4}
		{\begin{scope}[shift={(\i,0)}]
				\draw[ultra thick, orange] (-0.05,-0.05) -- (0.05,0.05) (-0.05,0.05) -- (0.05,-0.05);
		\end{scope}}
		\foreach \a/\b in {3.25/1.58333,3.41667/1.25,3.41667/1.08333,3.41667/0.916667,3.41667/0.75,3.25/0.416667}
		{\draw[thick] ([shift={(0:\b)}]\a,0) arc (0:180:\b);}
		\foreach \a/\b in {3.58333/1.25,3.58333/1.08333,3.58333/0.916667,3.58333/0.75,3.91667/0.25,1.5/0.166667}
		{\draw[thick] ([shift={(180:\b)}]\a,0) arc (180:360:\b);}
		\foreach \a/\b in {2.33333/1.}
		{
			\draw[line width=1.5mm,white] ([shift={(0:\b)}]\a,0) arc (0:180:\b);
			\draw[ultra thick, \myblue] ([shift={(0:\b)}]\a,0) arc (0:180:\b);
		}
		\foreach \a/\b in {2.75/0.583333}
		{
			\draw[line width=1.5mm,white] ([shift={(180:\b)}]\a,0) arc (180:360:\b);
			\draw[ultra thick, \myblue] ([shift={(180:\b)}]\a,0) arc (180:360:\b);
		}
	\end{tikzpicture}
	\end{array}=\begin{array}{l}-q\mu ^2 \left(q^4-2 q^2+2\right)\langle 1\rangle-q^2 \left(2 \mu ^2+3 q^4-4 \mu ^2 q^2+2 q^2\right)\langle L\rangle+\\ +\mu ^2 q^3 \left(3 q^2-2\right)\langle L^2\rangle+q^4 \left(1-2 \mu ^2+4 q^2\right)\langle L^3\rangle -\\ -q^5\mu ^2 \langle L^4\rangle-q^6\langle L^5\rangle\,.\end{array}
\end{equation}
\end{subequations}
We should note that flattened links $T_k$ being represented as arcades are related by a simple relation $T_{k+1}=R_3^-T_k$.
To relate those links to $L^p$ in the above expressions we hopped the loop around the second puncture to a blue thick loop through cap $C$ in Fig.~\ref{fig:Fig8} and then re-expanded again using the Kauffman bracket.

Next we consider a self-relation among symbols $L^p$.
To do so we rewrite symbol $L'$ in two equivalent forms:
\begin{equation}\label{L_prime}
	\langle L'\rangle=\begin{array}{c}
		\begin{tikzpicture}[scale=0.5]
			\foreach \i in {0,...,3} 
			{
				\begin{scope}[shift={(\i,0)}]
					\draw[orange, ultra thick] (-0.05,-0.05) -- (0.05, 0.05) (0.05,-0.05) -- (-0.05, 0.05);
				\end{scope}
			}
			\draw[thick] (0.5,0) to[out=90,in=180] (1.5,0.5) to[out=0,in=90] (2.5,0);
			\begin{scope}[yscale=-1]
				\draw[thick] (0.5,0) to[out=90,in=180] (1.5,0.5) to[out=0,in=90] (2.5,0);
			\end{scope}
		\end{tikzpicture}
	\end{array}=-q^{\frac{3}{2}}\begin{array}{c}
	\begin{tikzpicture}[scale=0.5]
		\foreach \i in {0,...,3} 
		{
			\begin{scope}[shift={(\i,0)}]
				\draw[orange, ultra thick] (-0.05,-0.05) -- (0.05, 0.05) (0.05,-0.05) -- (-0.05, 0.05);
			\end{scope}
		}
		\draw[thick] (0,-0.5) to[out=180,in=270] (-0.5,0) to[out=90,in=180] (0,0.3) -- (1,0.3) to[out=0, in=180] (2,-0.5) to[out=0,in=270] (2.5,0) to[out=90,in=0] (2,0.6) -- (0.7,0.6);
		\draw[line width=1.2mm,white] (0.7,0.6) to[out=180,in=0] (0,-0.5);
		\draw[thick] (0.7,0.6) to[out=180,in=0] (0,-0.5);
		\draw[thick, dashed, burgundy] (0.7,0) -- (0.7,1);
	\end{tikzpicture}
	\end{array}\,,
\end{equation}
where we used the first Reidemeister  move once.
Then we pull both links upwards through the knot by acting morphisms word $R_3^-R_3^-R_2^+R_2^+$ on both sides.
The result of this action is displayed in Fig.~\ref{fig:Fig8} (b) and (c) respectively.
Simple link $L'$ pulled upwards becomes link $T_2$ on the upper layer, whereas for the twisted link $L'$ we could follow only the action on the loop to the right of the dashed line in \eqref{L_prime}.
Then we again hop the loop from the first strand via cap $D$ in Fig.~\ref{fig:Fig8}, this move is accompanied by the other first Reidemeister move.
As the result we derive the following relation between link symbols:
\begin{equation}\label{4_1_self}
	q^{-3}\langle T_2\rangle=\begin{array}{c}
		\begin{tikzpicture}[scale=0.8]
			\foreach \i in {1,...,4}
			{\begin{scope}[shift={(\i,0)}]
					\draw[ultra thick, orange] (-0.05,-0.05) -- (0.05,0.05) (-0.05,0.05) -- (0.05,-0.05);
			\end{scope}}
			\foreach \a/\b in {1.75/0.125,3.5/1.25,3.5/1.,3.04167/0.291667,3.95833/0.291667,1.375/0.125}
			{\draw[thick] ([shift={(0:\b)}]\a,0) arc (0:180:\b);}
			\foreach \a/\b in {2.125/0.375,2.0625/0.1875,3.75/1.,3.91667/0.583333,1.25/0.125,1.625/0.125}
			{\draw[thick] ([shift={(180:\b)}]\a,0) arc (180:360:\b);}
			\foreach \a/\b in {1.5/0.125,2.45833/1.20833,2.6875/1.5625}
			{
				\draw[line width=0.7mm,white] ([shift={(180:\b)}]\a,0) arc (180:360:\b);
				\draw[thick,\myblue] ([shift={(180:\b)}]\a,0) arc (180:360:\b);
			}
		\end{tikzpicture}
	\end{array}=-\frac{\mu ^2 \left(q^2-1\right)^2}{q^2}\langle 1\rangle+\left(-2 q^3+2 \mu ^2 q-q\right)\langle L\rangle+\mu ^2 q^2\langle L^2\rangle+q^3\langle L^3\rangle\,.
\end{equation}

To derive the classical A-polynomial we consider the quasi-classical limit $q\to 1$, $\mu\to m+m^{-1}$, $\lambda\to l+l^{-1}$, $\langle L^p\rangle\to x^p$ of \eqref{4_1_self} and \eqref{4_1_lambda}:
\begin{equation}
	\begin{aligned}
		& (x+2) \left(m^4 x+m^4+m^2 x^2+m^2 x+m^2+x+1\right)=0\,,\\
		& l^2 m^4+l m^8 x^3+3 l m^8 x^2+l m^8 x-2 l m^8+2 l m^6 x^4+7 l m^6 x^3+6 l m^6 x^2-2 l m^6 x-4 l m^6+l m^4 x^5+\\
		&+4 l m^4 x^4+7 l m^4 x^3+6 l m^4 x^2-l m^4 x-4 l m^4+2 l m^2 x^4+7 l m^2 x^3+6 l m^2 x^2-2 l m^2 x-4 l m^2+l x^3+\\ 
		&+3 l x^2+l x-2 l+m^4=0\,.
	\end{aligned}
\end{equation}
To exclude the $x$-variable from these expressions we consider an ``interesting'' root of the first equation:
\begin{equation}
	x=\frac{-m^4-m^2-1\pm\sqrt{m^8-2 m^6-m^4-2 m^2+1}}{2 m^2}\,,
\end{equation}
and substitute it in the second equation to derive the A-polynomial of the figure-eight knot (cf. \cite[Table 2]{Gukov:2003na}):
\begin{equation}
	A_{4_1}(l,m)=l^2 m^4+l \left(-m^8+m^6+2 m^4+m^2-1\right)+m^4=0\,.
\end{equation}


\section{Towards colored A-polynomials: trefoil knot \texorpdfstring{$3_1$}{} for \texorpdfstring{$\fs\fu_3$}{}}\label{sec:su(3)}

As usual \cite{Moore:1988qv,RT1,RT2} in the case of $\fs\fu_3$ HOMFLY-PT polynomials are treated as a wave function $\Psi(x)$ of the 3d Chern-Simons theory on the knot complement.
The choice of the polarization (``$x$''-variable) is not unique in the quantum theory.
We assume that the natural dependence of the HOMFLY-PT polynomial of a representation $R$ coloring the knotted Wilson loop reflects exactly the dependence of the wave function on some self-adjoint operator eigenvalue (later we will see how this operator is constructed).
For the case of $\fs\fu_3$ all irreps $R$ are modules of the highest weight vector, here we denote the highest weight as $(w_1,w_2)$.

For these wave functions we define the action of operators corresponding to A-cycle and B-cycle Wilson line insertions at the torus body boundary.
Starting with $\fs\fu_3$ in comparison to $\fs\fu_2$ the representation conjugation defining the direction of the lines on knot diagrams matters, so in general we acquire four distinct Wilson line operators acting on knot wave functions:
\begin{equation}\label{operators}
	\begin{aligned}
		&\mu_+\Psi_K(w_1,w_2)=\begin{array}{c}
			\begin{tikzpicture}
				\draw[burgundy, ultra thick] (0,-0.3) -- (0,0);
				\begin{scope}[yscale=0.5]
					\draw ([shift={(0:0.25)}]0,0) arc (0:180:0.25);
					\draw[white, line width=0.7mm] ([shift={(180:0.25)}]0,0) arc (180:360:0.25);
					\draw[postaction={decorate},decoration={markings, mark= at position 0.8 with {\arrow{stealth}}}] ([shift={(180:0.25)}]0,0) arc (180:360:0.25);
				\end{scope}
				\draw[white, line width=1.5mm] (0,0) -- (0,0.4);
				\draw[burgundy, ultra thick, -stealth] (0,0) -- (0,0.4);
			\end{tikzpicture}
		\end{array} =\left(q^{2+\frac{4}{3}w_1+\frac{2}{3}w_2}+q^{\frac{2}{3}(w_2-w_1)}+q^{-2-\frac{2}{3}w_1-\frac{4}{3}w_2}\right)  \Psi_K(w_1,w_2)\,,\\
		&\mu_-\Psi_K(w_1,w_2)=\begin{array}{c}
		\begin{tikzpicture}
			\draw[burgundy, ultra thick] (0,-0.3) -- (0,0);
			\begin{scope}[yscale=0.5]
				\draw ([shift={(0:0.25)}]0,0) arc (0:180:0.25);
				\draw[white, line width=0.7mm] ([shift={(180:0.25)}]0,0) arc (180:360:0.25);
				\draw[postaction={decorate},decoration={markings, mark= at position 0.2 with {\arrowreversed{stealth}}}] ([shift={(180:0.25)}]0,0) arc (180:360:0.25);
			\end{scope}
			\draw[white, line width=1.5mm] (0,0) -- (0,0.4);
			\draw[burgundy, ultra thick, -stealth] (0,0) -- (0,0.4);
		\end{tikzpicture}
		\end{array} =\left(q^{2+\frac{4}{3}w_2+\frac{2}{3}w_1}+q^{\frac{2}{3}(w_1-w_2)}+q^{-2-\frac{2}{3}w_2-\frac{4}{3}w_1}\right)\Psi_K(w_1,w_2),\\
		&\lambda_+\Psi_K(R)=\begin{array}{c}
			\begin{tikzpicture}
				\draw[burgundy, ultra thick, -stealth] (0,-0.25) -- (0,0.25);
				\draw[-stealth] (0.25,-0.25) -- (0.25,0.25);
			\end{tikzpicture}
		\end{array} =\Psi_K(R\otimes \Box)=\Psi_K(w_1+1,w_2)+\Psi_K(w_1-1,w_2+1)+\Psi_K(w_1,w_2-1)\,,\\ &\lambda_-\Psi_K(R)=\begin{array}{c}
		\begin{tikzpicture}
			\draw[burgundy, ultra thick, -stealth] (0,-0.25) -- (0,0.25);
			\draw[-stealth] (0.25,0.25) -- (0.25,-0.25);
		\end{tikzpicture}
		\end{array}=\Psi_K(R\otimes \bar\Box)=\Psi_K(w_1,w_2+1)+\Psi_K(w_1+1,w_2-1)+\Psi_K(w_1-1,w_2)\,,
	\end{aligned}
\end{equation}
where we assume representation $R$ is of a rather generic form.

In this section we consider the case of the trefoil knot $3_1$.
So in what follows to lighten up a bit notations we assume that all the equivalences between symbols are valid \emph{only} being applied under sandwiching them with brackets $\langle \cdot\rangle_{3_1}$.

In this case we introduce the following basic link symbols:
\begin{equation}
\begin{aligned}
	&L_+=\begin{array}{c}
		\begin{tikzpicture}[scale=0.5]
			\foreach \i in {0, ..., 3}
			{
				\begin{scope}[shift={(\i,0)}]
					\draw[orange, ultra thick]  (-0.1,-0.1) -- (0.1,0.1) (-0.1,0.1) -- (0.1,-0.1);
				\end{scope}
			}
			\draw [-stealth] (1.5,-0.5) to[out=0,in=270] (2.5,0) to[out=90,in=0] (1.5,0.5) to[out=180,in=90] (0.5,0) to[out=270,in=180] (1.5,-0.5);
		\end{tikzpicture}
	\end{array},\quad L_-=\begin{array}{c}
	\begin{tikzpicture}[scale=0.5]
		\foreach \i in {0, ..., 3}
		{
			\begin{scope}[shift={(\i,0)}]
				\draw[orange, ultra thick]  (-0.1,-0.1) -- (0.1,0.1) (-0.1,0.1) -- (0.1,-0.1);
			\end{scope}
		}
		\draw [stealth-] (1.5,-0.5) to[out=0,in=270] (2.5,0) to[out=90,in=0] (1.5,0.5) to[out=180,in=90] (0.5,0) to[out=270,in=180] (1.5,-0.5);
	\end{tikzpicture}
	\end{array}\,,\\
	&\Theta_{+}=\begin{array}{c}
		\begin{tikzpicture}[scale=0.5]
			\foreach \i in {0, ..., 3}
			{
				\begin{scope}[shift={(\i,0)}]
					\draw[orange, ultra thick]  (-0.1,-0.1) -- (0.1,0.1) (-0.1,0.1) -- (0.1,-0.1);
				\end{scope}
			}
			\draw [-stealth] (1.5,0.4) to[out=150,in=90] (0.5,0) to[out=270,in=210] (1.5,-0.4);
			\draw [-stealth] (1.5,0.4) to[out=30,in=90] (2.5,0) to[out=270,in=330] (1.5,-0.4);
			\draw [-stealth] (1.5,0.4) -- (1.5,-0.4);
		\end{tikzpicture}
	\end{array},\quad \Theta_{-}=\begin{array}{c}
		\begin{tikzpicture}[scale=0.5]
			\foreach \i in {0, ..., 3}
			{
				\begin{scope}[shift={(\i,0)}]
					\draw[orange, ultra thick]  (-0.1,-0.1) -- (0.1,0.1) (-0.1,0.1) -- (0.1,-0.1);
				\end{scope}
			}
			\draw [stealth-] (1.5,0.4) to[out=150,in=90] (0.5,0) to[out=270,in=210] (1.5,-0.4);
			\draw [stealth-] (1.5,0.4) to[out=30,in=90] (2.5,0) to[out=270,in=330] (1.5,-0.4);
			\draw [stealth-] (1.5,0.4) -- (1.5,-0.4);
		\end{tikzpicture}
	\end{array}\,.
\end{aligned}
\end{equation}

We will find in what follows also the following link symbols:
\begin{equation}
    \begin{aligned}
	&X_+=\begin{array}{c}
		\begin{tikzpicture}[scale=0.5]
			\draw[-stealth] (0.866025,0.5) -- (0,1);
			\draw[-stealth] (-0.866025,0.5) -- (0,1);
			\draw[-stealth] (0.866025,0.5) -- (0.866025,-0.5);
			\draw[-stealth] (-0.866025,0.5) -- (-0.866025,-0.5);
			\draw[-stealth] (0,-1) -- (0.866025,-0.5);
			\draw[-stealth] (0,-1) -- (-0.866025,-0.5);
			\begin{scope}[shift={(-1.5,0)}]
				\draw[ultra thick,orange] (-0.1,-0.1) -- (0.1,0.1) (-0.1,0.1) -- (0.1,-0.1);
			\end{scope}
			\begin{scope}[shift={(1.5,0)}]
				\draw[ultra thick,orange] (-0.1,-0.1) -- (0.1,0.1) (-0.1,0.1) -- (0.1,-0.1);
			\end{scope}
			\draw[-stealth] (-0.866025,0.5) to[out=150,in=90] (-2,0) to[out=270,in=210] (-0.866025,-0.5);
			\draw[-stealth] (0,1.4) to[out=150,in=90] (-2.3,0) to[out=270,in=210] (0,-1.4);
			\begin{scope}[xscale=-1]
				\draw[-stealth] (-0.866025,0.5) to[out=150,in=90] (-2,0) to[out=270,in=210] (-0.866025,-0.5);
				\draw[-stealth] (0,1.4) to[out=150,in=90] (-2.3,0) to[out=270,in=210] (0,-1.4);
			\end{scope}
			\draw[-stealth] (0,1.4) -- (0,1);
			\draw[-stealth] (0,-1) -- (0,-1.4);
			\begin{scope}[shift={(-2.7,0)}]
				\draw[ultra thick,orange] (-0.1,-0.1) -- (0.1,0.1) (-0.1,0.1) -- (0.1,-0.1);
			\end{scope}
			\begin{scope}[shift={(2.7,0)}]
				\draw[ultra thick,orange] (-0.1,-0.1) -- (0.1,0.1) (-0.1,0.1) -- (0.1,-0.1);
			\end{scope}
		\end{tikzpicture}
	\end{array},\quad X_-=\begin{array}{c}
	\begin{tikzpicture}[scale=0.5]
		\draw[stealth-] (0.866025,0.5) -- (0,1);
		\draw[stealth-] (-0.866025,0.5) -- (0,1);
		\draw[stealth-] (0.866025,0.5) -- (0.866025,-0.5);
		\draw[stealth-] (-0.866025,0.5) -- (-0.866025,-0.5);
		\draw[stealth-] (0,-1) -- (0.866025,-0.5);
		\draw[stealth-] (0,-1) -- (-0.866025,-0.5);
		\begin{scope}[shift={(-1.5,0)}]
			\draw[ultra thick,orange] (-0.1,-0.1) -- (0.1,0.1) (-0.1,0.1) -- (0.1,-0.1);
		\end{scope}
		\begin{scope}[shift={(1.5,0)}]
			\draw[ultra thick,orange] (-0.1,-0.1) -- (0.1,0.1) (-0.1,0.1) -- (0.1,-0.1);
		\end{scope}
		\draw[stealth-] (-0.866025,0.5) to[out=150,in=90] (-2,0) to[out=270,in=210] (-0.866025,-0.5);
		\draw[stealth-] (0,1.4) to[out=150,in=90] (-2.3,0) to[out=270,in=210] (0,-1.4);
		\begin{scope}[xscale=-1]
			\draw[stealth-] (-0.866025,0.5) to[out=150,in=90] (-2,0) to[out=270,in=210] (-0.866025,-0.5);
			\draw[stealth-] (0,1.4) to[out=150,in=90] (-2.3,0) to[out=270,in=210] (0,-1.4);
		\end{scope}
		\draw[stealth-] (0,1.4) -- (0,1);
		\draw[stealth-] (0,-1) -- (0,-1.4);
		\begin{scope}[shift={(-2.7,0)}]
			\draw[ultra thick,orange] (-0.1,-0.1) -- (0.1,0.1) (-0.1,0.1) -- (0.1,-0.1);
		\end{scope}
		\begin{scope}[shift={(2.7,0)}]
			\draw[ultra thick,orange] (-0.1,-0.1) -- (0.1,0.1) (-0.1,0.1) -- (0.1,-0.1);
		\end{scope}
	\end{tikzpicture}
	\end{array}\,,\\
	& Y_+=\begin{array}{c}
	    \begin{tikzpicture}[scale=0.5]
		\begin{scope}[shift={(-1,0)}]
		    \draw[orange, ultra thick] (-0.1,-0.1) -- (0.1,0.1) (-0.1,0.1) -- (0.1,-0.1);
		\end{scope}
		\begin{scope}[shift={(1,0)}]
		    \draw[orange, ultra thick] (-0.1,-0.1) -- (0.1,0.1) (-0.1,0.1) -- (0.1,-0.1);
		\end{scope}
		\begin{scope}[shift={(-2.5,0)}]
		    \draw[orange, ultra thick] (-0.1,-0.1) -- (0.1,0.1) (-0.1,0.1) -- (0.1,-0.1);
		\end{scope}
		\begin{scope}[shift={(2.5,0)}]
		    \draw[orange, ultra thick] (-0.1,-0.1) -- (0.1,0.1) (-0.1,0.1) -- (0.1,-0.1);
		\end{scope}
		\draw[-stealth] (-0.4,0) to[out=120,in=0] (-1,0.5) to[out=180,in=60] (-1.6,0);
		\begin{scope}[yscale=-1]
		    \draw[-stealth] (-0.4,0) to[out=120,in=0] (-1,0.5) to[out=180,in=60] (-1.6,0);
		\end{scope}
		\begin{scope}[shift={(2,0)}]
		    \draw[-stealth] (-0.4,0) to[out=120,in=0] (-1,0.5) to[out=180,in=60] (-1.6,0);
		    \begin{scope}[yscale=-1]
			\draw[-stealth] (-0.4,0) to[out=120,in=0] (-1,0.5) to[out=180,in=60] (-1.6,0);
		    \end{scope}
		\end{scope}
		\draw[-stealth] (-0.4,0) -- (0.4,0);
		\draw[-stealth] (-2,0) -- (-1.6,0);
		\draw[-stealth] (1.6,0) -- (2,0);
		\draw[stealth-] (2,0) to[out=60,in=0] (0,1) to[out=180,in=120] (-2,0);
		\begin{scope}[yscale=-1]
		    \draw[stealth-] (2,0) to[out=60,in=0] (0,1) to[out=180,in=120] (-2,0);
		\end{scope}
	    \end{tikzpicture}
	\end{array},\quad Y_-=\begin{array}{c}
	    \begin{tikzpicture}[scale=0.5,rotate=180]
		\begin{scope}[shift={(-1,0)}]
		    \draw[orange, ultra thick] (-0.1,-0.1) -- (0.1,0.1) (-0.1,0.1) -- (0.1,-0.1);
		\end{scope}
		\begin{scope}[shift={(1,0)}]
		    \draw[orange, ultra thick] (-0.1,-0.1) -- (0.1,0.1) (-0.1,0.1) -- (0.1,-0.1);
		\end{scope}
		\begin{scope}[shift={(-2.5,0)}]
		    \draw[orange, ultra thick] (-0.1,-0.1) -- (0.1,0.1) (-0.1,0.1) -- (0.1,-0.1);
		\end{scope}
		\begin{scope}[shift={(2.5,0)}]
		    \draw[orange, ultra thick] (-0.1,-0.1) -- (0.1,0.1) (-0.1,0.1) -- (0.1,-0.1);
		\end{scope}
		\draw[-stealth] (-0.4,0) to[out=120,in=0] (-1,0.5) to[out=180,in=60] (-1.6,0);
		\begin{scope}[yscale=-1]
		    \draw[-stealth] (-0.4,0) to[out=120,in=0] (-1,0.5) to[out=180,in=60] (-1.6,0);
		\end{scope}
		\begin{scope}[shift={(2,0)}]
		    \draw[-stealth] (-0.4,0) to[out=120,in=0] (-1,0.5) to[out=180,in=60] (-1.6,0);
		    \begin{scope}[yscale=-1]
			\draw[-stealth] (-0.4,0) to[out=120,in=0] (-1,0.5) to[out=180,in=60] (-1.6,0);
		    \end{scope}
		\end{scope}
		\draw[-stealth] (-0.4,0) -- (0.4,0);
		\draw[-stealth] (-2,0) -- (-1.6,0);
		\draw[-stealth] (1.6,0) -- (2,0);
		\draw[stealth-] (2,0) to[out=60,in=0] (0,1) to[out=180,in=120] (-2,0);
		\begin{scope}[yscale=-1]
		    \draw[stealth-] (2,0) to[out=60,in=0] (0,1) to[out=180,in=120] (-2,0);
		\end{scope}
	    \end{tikzpicture}
	\end{array}\,.
    \end{aligned}
\end{equation}
However we can easily compute they are related to already pre-defined symbols in the following way:
\begin{equation}\label{XY}
	\begin{aligned}
	    & \Theta_+^2=X_++[2]_q\times \Theta_-,\quad \Theta_-^2=X_-+[2]_q\times\Theta_+\,,\\
	    & \Theta_+\Theta_-=[3]_q+2\mu_+\mu_-+\mu_-^2L_++\mu_+^2L_-+L_-L_++q^{-1}Y_++qY_-\,,\\
	    & \Theta_-\Theta_+=[3]_q+2\mu_+\mu_-+\mu_-^2L_++\mu_+^2L_-+L_-L_++q Y_++q^{-1}Y_-\,.
	\end{aligned}
\end{equation}

Here we should stress that the decomposition method over MOY diagrams of flattened link turns out to be \emph{not universal} for $\fs\fu_{n\geq 3}$.
Here we see two types of indecomposable graphs $X_{\pm}$, $Y_{\pm}$ that happen to appear as ``powers'' of links $\Theta_\pm$ we have chosen as elementary.
Unfortunately, at the moment we are not aware of how the problem of classification for indecomposable graphs should be solved in general, neither of how to construct relations between indecomposable graphs and repeated elementary links.

In the same way as we computed shift operators and self-link relations in Sec.~\ref{sec:3_1} using the arcade representations from Fig.~\ref{fig:3_1} we calculate these relations this case of $\fs\fu_3$ by introducing directions on the diagrams and applying the Kuperberg bracket instead of the Kauffman one:
\begin{equation}\label{lambda_+}
	\begin{aligned}
    &\lambda_+=\begin{array}{c}
		\begin{tikzpicture}[scale=0.5]
\foreach \i in {1,...,4}
{\begin{scope}[shift={(\i,0)}]
\draw[ultra thick, orange] (-0.1,-0.1) -- (0.1,0.1) (-0.1,0.1) -- (0.1,-0.1);
\end{scope}}
\foreach \a/\b in {2.6/1.2,2.6/1.,2.975/0.225,2.95/0.45,2.025/0.225}
{\draw[thick] ([shift={(0:\b)}]\a,0) arc (0:180:\b);}
\foreach \a/\b in {2.4/1.2,2.4/1.,2.175/0.575,2.15/0.35}
{\draw[thick] ([shift={(180:\b)}]\a,0) arc (180:360:\b);}
			\foreach \a/\b/\s in {1.725/0.525/stealth-,3.5/0.3/-stealth}
			{
				\draw[line width=1.2mm,white] ([shift={(180:\b)}]\a,0) arc (180:360:\b);
				\draw[thick,\myblue,\s] ([shift={(180:\b)}]\a,0) arc (180:360:\b);
			}
		\end{tikzpicture}
	\end{array}=\mu _+^2 \mu _- L_+-\mu _- L_-+\mu _+^2 L_--\mu _+ L_++\mu _+ L_+ q^2+\Theta _- \mu _+ L_+ q+\frac{\mu _-^3}{q^2}-\\
	&-\frac{\mu _+ \mu _-}{q^2}+\frac{\Theta _- \mu _+ \mu _-}{q}+\frac{\Theta _+ \mu _+ \mu _-}{q}+\Theta _+ q-\frac{\Theta _+}{q}+X_-\,.
\end{aligned}
\end{equation}

\begin{equation}\label{lambda_-}
	\begin{aligned}
    &\lambda_-=\begin{array}{c}
		\begin{tikzpicture}[scale=0.5]
\foreach \i in {1,...,4}
{\begin{scope}[shift={(\i,0)}]
\draw[ultra thick, orange] (-0.1,-0.1) -- (0.1,0.1) (-0.1,0.1) -- (0.1,-0.1);
\end{scope}}
\foreach \a/\b in {2.6/1.2,2.6/1.,2.975/0.225,2.95/0.45,2.025/0.225}
{\draw[thick] ([shift={(0:\b)}]\a,0) arc (0:180:\b);}
\foreach \a/\b in {2.4/1.2,2.4/1.,2.175/0.575,2.15/0.35}
{\draw[thick] ([shift={(180:\b)}]\a,0) arc (180:360:\b);}
			\foreach \a/\b/\s in {1.725/0.525/-stealth,3.5/0.3/stealth-}
			{
				\draw[line width=1.2mm,white] ([shift={(180:\b)}]\a,0) arc (180:360:\b);
				\draw[thick,\myblue,\s] ([shift={(180:\b)}]\a,0) arc (180:360:\b);
			}
		\end{tikzpicture}
	\end{array}=\mu _-^2 \mu _+ L_--\mu _+ L_+-\mu _- L_-+\mu _-^2 L_++\mu _- L_- q^2+\Theta _+ \mu _- L_- q+\frac{\mu _+^3}{q^2}-\\
	&-\frac{\mu _- \mu _+}{q^2}+\frac{\Theta _- \mu _- \mu _+}{q}+\frac{\Theta _+ \mu _- \mu _+}{q}+\Theta _- q-\frac{\Theta _-}{q}+X_+\,.
	\end{aligned}
\end{equation}

\begin{equation}\label{L_+}
    L_+=\begin{array}{c}
		\begin{tikzpicture}[scale=0.5]
\foreach \i in {1,...,4}
{\begin{scope}[shift={(\i,0)}]
\draw[ultra thick, orange] (-0.1,-0.1) -- (0.1,0.1) (-0.1,0.1) -- (0.1,-0.1);
\end{scope}}
\foreach \a/\b in {3./0.333333,2.70833/0.958333}
{\draw[thick] ([shift={(0:\b)}]\a,0) arc (0:180:\b);}
\foreach \a/\b in {2.45833/1.20833,2.41667/0.916667,2.20833/0.458333}
{\draw[thick] ([shift={(180:\b)}]\a,0) arc (180:360:\b);}
\foreach \a/\b in {1.79167/0.541667}
{
    \draw[line width=0.7mm,white] ([shift={(0:\b)}]\a,0) arc (0:180:\b);
    \draw[thick,\myblue,-stealth] ([shift={(0:\b)}]\a,0) arc (0:180:\b);
}
\foreach \a/\b in {1.91667/0.416667}
{
    \draw[line width=0.7mm,white] ([shift={(180:\b)}]\a,0) arc (180:360:\b);
    \draw[thick,\myblue] ([shift={(180:\b)}]\a,0) arc (180:360:\b);
}
		\end{tikzpicture}
	\end{array}=\mu _-+\mu _+ \mu _- L_++\mu _+ L_-+\Theta _+ L_+ q+\frac{\Theta _- \mu _-}{q}\,.
\end{equation}

\begin{equation}\label{L_-}
    L_-=\begin{array}{c}
		\begin{tikzpicture}[scale=0.5]
\foreach \i in {1,...,4}
{\begin{scope}[shift={(\i,0)}]
\draw[ultra thick, orange] (-0.1,-0.1) -- (0.1,0.1) (-0.1,0.1) -- (0.1,-0.1);
\end{scope}}
\foreach \a/\b in {3./0.333333,2.70833/0.958333}
{\draw[thick] ([shift={(0:\b)}]\a,0) arc (0:180:\b);}
\foreach \a/\b in {2.45833/1.20833,2.41667/0.916667,2.20833/0.458333}
{\draw[thick] ([shift={(180:\b)}]\a,0) arc (180:360:\b);}
\foreach \a/\b in {1.79167/0.541667}
{
    \draw[line width=0.7mm,white] ([shift={(0:\b)}]\a,0) arc (0:180:\b);
    \draw[thick,\myblue] ([shift={(0:\b)}]\a,0) arc (0:180:\b);
}
\foreach \a/\b in {1.91667/0.416667}
{
    \draw[line width=0.7mm,white] ([shift={(180:\b)}]\a,0) arc (180:360:\b);
    \draw[thick,\myblue,stealth-] ([shift={(180:\b)}]\a,0) arc (180:360:\b);
}
		\end{tikzpicture}
	\end{array}=\mu _++\mu _- L_++\mu _- \mu _+ L_-+\Theta _- L_- q+\frac{\Theta _+ \mu _+}{q}\,.
\end{equation}

\begin{equation}\label{Theta_+}
    \Theta_+=\begin{array}{c}
		\begin{tikzpicture}[scale=0.5]
\foreach \i in {1,...,4}
{\begin{scope}[shift={(\i,0)}]
\draw[ultra thick, orange] (-0.1,-0.1) -- (0.1,0.1) (-0.1,0.1) -- (0.1,-0.1);
\end{scope}}
\foreach \a/\b in {3./0.333333,2.70833/0.958333}
{\draw[thick] ([shift={(0:\b)}]\a,0) arc (0:180:\b);}
\foreach \a/\b in {2.45833/1.20833,2.41667/0.916667,2.20833/0.458333}
{\draw[thick] ([shift={(180:\b)}]\a,0) arc (180:360:\b);}
\foreach \a/\b in {1.79167/0.541667}
{
    \draw[line width=0.7mm,white] ([shift={(0:\b)}]\a,0) arc (0:180:\b);
    \draw[thick,\myblue,-stealth] ([shift={(0:\b)}]\a,0) arc (0:180:\b);
}
\foreach \a/\b in {1.91667/0.416667}
{
    \draw[line width=0.7mm,white] ([shift={(180:\b)}]\a,0) arc (180:360:\b);
    \draw[thick,\myblue] ([shift={(180:\b)}]\a,0) arc (180:360:\b);
}
\draw[thick,\myblue] (1.25,0) -- (1.5,0);
		\end{tikzpicture}
	\end{array}=L_- L_+ q+\frac{\mu _- \mu _+}{q}+Y_+\,.
\end{equation}

\begin{equation}\label{Theta_-}
    \Theta_-=\begin{array}{c}
		\begin{tikzpicture}[scale=0.5]
\foreach \i in {1,...,4}
{\begin{scope}[shift={(\i,0)}]
\draw[ultra thick, orange] (-0.1,-0.1) -- (0.1,0.1) (-0.1,0.1) -- (0.1,-0.1);
\end{scope}}
\foreach \a/\b in {3./0.333333,2.70833/0.958333}
{\draw[thick] ([shift={(0:\b)}]\a,0) arc (0:180:\b);}
\foreach \a/\b in {2.45833/1.20833,2.41667/0.916667,2.20833/0.458333}
{\draw[thick] ([shift={(180:\b)}]\a,0) arc (180:360:\b);}
\foreach \a/\b in {1.79167/0.541667}
{
    \draw[line width=0.7mm,white] ([shift={(0:\b)}]\a,0) arc (0:180:\b);
    \draw[thick,\myblue] ([shift={(0:\b)}]\a,0) arc (0:180:\b);
}
\foreach \a/\b in {1.91667/0.416667}
{
    \draw[line width=0.7mm,white] ([shift={(180:\b)}]\a,0) arc (180:360:\b);
    \draw[thick,\myblue,stealth-] ([shift={(180:\b)}]\a,0) arc (180:360:\b);
}
\draw[thick,\myblue] (1.25,0) -- (1.5,0);
		\end{tikzpicture}
	\end{array}=L_- L_+ q+\frac{\mu _- \mu _+}{q}+Y_-\,.
\end{equation}

Here we would like to argue that diagrams like $\Theta_+$, $X_+$ and $Y_+$ that after rotation by $\pi$ become $\Theta_-$, $X_-$ and $Y_-$ respectively are indistinguishable in fact from their counterparts:
\begin{equation}
	\Theta_{\pm}=\Theta,\quad  X_{\pm}=X,\quad  Y_{\pm}=Y\,.
\end{equation}
Here we spell the argument solely for $\Theta_+=\Theta_-=\Theta$ implying in the other cases arguments are identical.
Thinking of $\Theta_{\pm}$ as surface operators \cite{Galakhov:2014aha} we observe that as an action on a WZW block $\Theta_+$ is conjugated with respect to $\Theta_-$ by the R-functor:
\begin{equation}
	R\circ \Theta_+=\Theta_-\circ R, \quad \begin{array}{c}
		\begin{tikzpicture}[scale=0.5]
			\draw[burgundy, ultra thick] (1,0) -- (1,-0.5) (2,0) -- (2,-0.5);
			\begin{scope}[yscale=0.6]
			\draw [white, line width=0.7mm] (1.5,0.4) to[out=150,in=90] (0.5,0) to[out=270,in=210] (1.5,-0.4);
			\draw [white, line width=0.7mm] (1.5,0.4) to[out=30,in=90] (2.5,0) to[out=270,in=330] (1.5,-0.4);
			\draw [white, line width=0.7mm] (1.5,0.4) -- (1.5,-0.4);
			\draw [-stealth] (1.5,0.4) to[out=150,in=90] (0.5,0) to[out=270,in=210] (1.5,-0.4);
			\draw [-stealth] (1.5,0.4) to[out=30,in=90] (2.5,0) to[out=270,in=330] (1.5,-0.4);
			\draw [-stealth] (1.5,0.4) -- (1.5,-0.4);
			\end{scope}
			\draw[white,line width=1.2mm] (2,0) to[out=90,in=270] (1,1);
			\draw[burgundy, ultra thick] (2,0) to[out=90,in=270] (1,1);
			\draw[white,line width=1.2mm] (1,0) to[out=90,in=270] (2,1);
			\draw[burgundy, ultra thick] (1,0) to[out=90,in=270] (2,1);
		\end{tikzpicture}
	\end{array}=\begin{array}{c}
	\begin{tikzpicture}[scale=0.5]
		\draw[white,line width=1.2mm] (2,0) to[out=90,in=270] (1,1);
		\draw[burgundy, ultra thick] (2,0) to[out=90,in=270] (1,1);
		\draw[white,line width=1.2mm] (1,0) to[out=90,in=270] (2,1);
		\draw[burgundy, ultra thick] (1,0) to[out=90,in=270] (2,1);
		\begin{scope}[shift={(0,1)}]
			\begin{scope}[yscale=-0.6]
				\draw [white, line width=0.7mm] (1.5,0.4) to[out=150,in=90] (0.5,0) to[out=270,in=210] (1.5,-0.4);
				\draw [white, line width=0.7mm] (1.5,0.4) to[out=30,in=90] (2.5,0) to[out=270,in=330] (1.5,-0.4);
				\draw [white, line width=0.7mm] (1.5,0.4) -- (1.5,-0.4);
				\draw [-stealth] (1.5,0.4) to[out=150,in=90] (0.5,0) to[out=270,in=210] (1.5,-0.4);
				\draw [-stealth] (1.5,0.4) to[out=30,in=90] (2.5,0) to[out=270,in=330] (1.5,-0.4);
				\draw [-stealth] (1.5,0.4) -- (1.5,-0.4);
			\end{scope}
		\end{scope}
		\draw[white,line width=1.2mm] (1,1) -- (1,1.5) (2,1) -- (2,1.5);
		\draw[burgundy, ultra thick] (1,1) -- (1,1.5) (2,1) -- (2,1.5);
	\end{tikzpicture}
	\end{array}\,.
\end{equation}
Then $\Theta_+$ and $\Theta_-$ commute with the R-functor squared $R^2$.
We expect that the $R$-functor is diagonalized in the isotypical components of the tensor square.
This implies that for $\Theta_{\pm}$ isotypical components are eigen spaces as well.
In this basis, where $R$ is diagonal, $R$ will commute with $\Theta_{\pm}$ too if $R^2$ does.
So we derive:
\begin{equation}
	\Theta_+=R^{-1}\Theta_-R = \Theta_-=:\Theta\,.
\end{equation}

Clearly, relations \eqref{lambda_+} -- \eqref{Theta_-} together with \eqref{XY} would be enough to exclude all the symbols to leave only relations $A_{k=1,2}(\lambda_+,\lambda_-,\mu_+,\mu_-)=0$ -- A-polynomials -- for Wilson line \emph{operators} living in the Hilbert space on the inflated knot boundary, if operators and symbols were simple classical variables.
However, unfortunately, a derivation of actual quantum relations, quantum A-polynomials, requires much more work that is a bit more artistic than algorithmic as we have already seen in \cite{Galakhov:2024eco} for the mere case of quantum A-polynomial of trefoil knot in $\fs\fu_2$.
Yet in the classical derivation we would like to see a strong indication that the quantum story works as well.

In the rest of this section we would like to discus  the quasi-classical limit of the presented relations $q\to 1$ when operators and symbols become ordinary commutative variables -- classical expectation values of respective Wilson loops and links.
We take this limit with a simultaneous limit of large weights $w_i\to \infty$, so that exponentiated operators and shift operators acquire vacuum values:
\begin{equation}
	\begin{aligned}
	&q^{\frac{2}{3}w_1}\Psi_{3_1}(w_1,w_2)=:m_1\Psi_{3_1}(w_1,w_2),\quad q^{\frac{2}{3}w_2}\Psi_{3_1}(w_1,w_2)=:m_2\Psi_{3_1}(w_1,w_2)\,,\\ &\Psi_{3_1}(w_1+1,w_2)=:l_1\Psi_{3_1}(w_1,w_2),\quad 
	\Psi_{3_1}(w_1,w_2+1)=:l_1\Psi_{3_1}(w_1,w_2)\,.
	\end{aligned}
\end{equation}
Under these notations we find:
\begin{equation}
	\begin{aligned}
	&\mu_+=m_2 m_1^2+\frac{1}{m_1 m_2^2}+\frac{m_2}{m_1},\quad \mu_-=m_1 m_2^2+\frac{1}{m_2 m_1^2}+\frac{m_1}{m_2}\,,\\
	&\lambda_+=l_1+\frac{1}{l_2}+\frac{l_2}{l_1},\quad \lambda_-=\frac{l_1}{l_2}+l_2+\frac{1}{l_1}\,.
	\end{aligned}
\end{equation}
For the symbols we define the following expectation values:
\begin{equation}
	L_+\to z_1,\quad L_-\to z_2,\quad \Theta\to \theta\,. 
\end{equation}

Then constraints become in the quasi-classical limit (we do not perform all the substitutions as equations become rather bulky):
\begin{equation}\label{classical}
	\begin{aligned}
		&\lambda_+=(\theta -2) \theta +\mu _-^3+\mu _- \left((2 \theta -1) \mu _++\mu _+^2 z_1-z_2\right)+\theta  \mu _+ z_1+\mu _+^2 z_2\,,\\
		&\lambda_-=(\theta -2) \theta +\mu _+^3+\mu _- \left((2 \theta -1) \mu _++\theta  z_2\right)-\mu _+ z_1+\mu _-^2 \left(\mu _+ z_2+z_1\right)\,,\\
		&\theta^2=2 \theta +\mu _-^2 z_1+\mu _+^2 z_2-z_1 z_2+3\,,\\
		&z_1=\mu _- \left(\theta +\mu _+ z_1+1\right)+\theta  z_1+\mu _+ z_2\,,\\
		&z_2=\mu _+ \left(\theta +\mu _- z_2+1\right)+\theta  z_2+\mu _- z_1\,.
	\end{aligned}
\end{equation}
Equations \eqref{classical} have a finite yet a large amount of roots.
And at this moment we do not have at hands an effective mechanism to distinguish actual roots appearing on actual integration cycles in the Chern-Simons path integral  from fake, spurious ones.

Let us discuss further few particular roots of the above equations:
\begin{equation}\label{one-root}
	\begin{aligned}
		&z_1=m_2^2 m_1^4+\frac{m_2^2}{m_1^2}+\frac{1}{m_1^2 m_2^4},\;z_2=m_1^2 m_2^4+\frac{m_1^2}{m_2^2}+\frac{1}{m_1^4 m_2^2},\;\theta=-m_2^3 m_1^3-m_1^3-m_2^3-\frac{1}{m_1^3}-\frac{1}{m_2^3}-\frac{1}{m_1^3 m_2^3}\,,\\
		&l_1= m_1^6 m_2^3,\quad l_2= m_1^3 m_2^6\,.
	\end{aligned}
\end{equation}
The RT formalism allows us to represent this particular Chern-Simons path integral as a sum $\sum_Q \rho_Q^3\; {\rm dim}_qQ$, where the summation runs over all irreducible isotypical components of $R^{\otimes 2}$, and $\rho_Q$ and ${\rm dim}_qQ$ are an eigen value of the R-matrix and a quantum dimension respectively.
Expectation values for symbols $z_1$ and $z_2$ due to relations \eqref{operators} measure exactly which irrep channel $Q=(2w_1,2w_2)$ contributes to this particular branch of the Chern-Simons path integral in the most significant way.

Now consider another root:
\begin{equation}\label{one-root2}
	\begin{aligned}
		&z_1=\frac{m_2^2}{m_1^2}+\frac{m_1}{m_2},\; z_2=\frac{m_1^2}{m_2^2}+\frac{m_2}{m_1},\; \theta=-m_1^3-m_2^3-\frac{1}{m_1^3}-\frac{1}{m_2^3}-1\,,\\
		&l_1=- m_1^6 m_2^3,\quad l_2=- m_1^3 m_2^6\,.
	\end{aligned}
\end{equation}
For this root we are unable to assign such a clear meaning to expectation values of links $z_{1,2}$, however relations between the shift and spin operators are quite reminiscent of relation $l=-m^3$ that is the root of the classical A-polynomial of the trefoil knot in the case of $\fs\fu_2$ (see \cite{Galakhov:2024eco}).
In this case we would like to construct an asymptotic of the wave function for this branch.
We expect the following scaling of quantities:
\begin{equation}
	w_i=\frac{u_i}{\hbar},\quad q=e^{\hbar}, \quad \Psi_{3_1}=\exp\left(\frac{W(u_1,u_2)}{\hbar}\right),\quad \hbar\to 0\,.
\end{equation}
In this case we observe:
\begin{equation}
	m_1=e^{\frac{2}{3}u_1},\quad m_2=e^{\frac{2}{3}u_2},\quad l_1=\exp\left(\p_{u_1}W\right),\quad l_2=\exp\left(\p_{u_2}W\right)\,.
\end{equation}
So we find for the wave function eikonal the following expression:
\begin{equation}
	W(u_1,u_2)=2(u_1^2+u_1u_2+u_2^2)+\pi\I(u_1+u_2)\,.
\end{equation}

\section{Conclusion}

In this note we have continued a development of the program proposed in \cite{Galakhov:2024eco} to interpret (quantum) A-polynomials of knots as relations among links colored by small representations and ``decorating'' knots.
The key idea on this route is to ``planarize'' decorating links and simplify them maximally by using brackets allowing to untangle these links partially.

Here we have performed two steps:
\begin{itemize}
	\item We have introduced the braid group action on arcades allowing one to planarize arbitrary links in a systematic algorithmic manner.
	\item We have considered an example of reducing planarized links for the gauge algebra $\fs\fu_3$ and showed that there is a closed system of multiplicative relations for link symbols.
	This allows to derive immediate relations for commutative quasi-classical variables leading to classical analogs of A-polynomials for $\fs\fu_3$.
\end{itemize}

To conclude this note we would like to highlight further directions in this program and other problems needing a solution:
\begin{itemize}
	\item It would be nice to derive estimates for degrees of emerging equations and dimensions of at least multiplicative bases of link symbol variables.
	As we have seen in this problem parameter $n$ in $\fs\fu_n$ is highly involved making this question rather non-trivial already for $\fs\fu_3$.
	\item The majorly new example to the canonical discussion of A-polynomials we considered in this note is $\fs\fu_3$.
	This example turns out to lie in many senses on a boundary separating its neighbors $\fs\fu_2$ and $\fs\fu_4$:
	\begin{enumerate}
		\item The moduli space of connections on the torus tubular boundary has two canonical coordinates and two canonical momenta.
		Simultaneously, for $\fs\fu_3$ in comparison to $\fs\fu_2$ the fundamental and anti-fundamental representations are not isomorphic, so we could use as quantum operators corresponding to \emph{two} coordinates Hopf link operators in these representations, and for \emph{two} momenta \emph{two} cabling links respectively.
		Starting with $\fs\fu_4$ more independent operators are required.
		\item The Kuperberg bracket maps planarized links into MOY diagrams with a single type of edges.
		For $\fs\fu_2$ there where no vertices and internal edges, in the result MOY diagrams were reduced to non-intersecting links in this case.
		Starting with $\fs\fu_4$ MOY diagrams would have more than one type of edges.
		Yet we hope that by coloring auxiliary decorating links by only antisymmetric tensor powers $\wedge^k\Box$ as in $n$-webs used for foam categorification of $U_q(\fs\fl_n)$ \cite{2012arXiv1212.6076L,Chun:2015gda} one would be able to tame the growth of needed MOY diagram edge colorings.
		Moreover, Hopf and cabling link operators colored with $\wedge^k\Box$, $k=1,\ldots,n-1$ should be enough to cover symplectic coordinates of the moduli space.
		Also it would be interesting to compare these approaches to recursion equations following from skein modules \cite{Ekholm:2020csl,Ekholm:2024ceb}.
		\item For $\fs\fu_3$ relations among MOY diagrams are simpler, eliminations of bigons and boxes are allowed.
		Despite this fact we were quite lucky to guess relations allowing to re-express new non-contractible MOY diagrams \eqref{XY} as simple polynomials of simpler diagrams.
	\end{enumerate}
	All these issues indicate that a lot of work should be done before deciding on a possibility to extend these methods of deriving quantum A-polynomial equations to even $\fs\fu_3$, not to mention $\fs\fu_{n\geq 4}$ or other algebras beyond Dynkin A-series.
	On this route switching to picture-valued link invariants and skein relations with topological coefficients in the fashion of \cite{akimova2020labels,ManNik}.
	\item To simplify and abbreviate our considerations we concentrated on reducing relations among links to respective quasi-classical $q\to 1$ limits.
	The example of \cite{Galakhov:2024eco} indicates that to arrive to the quantum A-polynomials one is required to consider simply more relations for more links, and the bases of links that can not be untangled are finite -- an analog of the Gr\"obner basis in the case of multiplicative rings.
	Nevertheless, a systematic way to enumerate all necessary relations in the quantum case is required.
	In principle, one might expect this problem to be as difficult as the derivation process of Ward identities from a raw path integral expression might be, see e.g. \cite{Mishnyakov:2024rmb}, since the usual division, power root operations, and the fundamental theorem of algebra might not work properly in the ring of quantum operators.
\end{itemize}

\section*{Acknowledgments}

We would like to thank Vassily Manturov for intriguing discussions. 
The  work of D.G. is supported by by the RSF grant 24-71-10058.




\bibliographystyle{utphys}
\bibliography{biblio}

\end{document}